\documentclass[twocolumn]{aastex7}

\usepackage{hyperref}
\usepackage{graphicx}
\usepackage{amsmath}
\usepackage{amssymb}
\usepackage{txfonts}
\usepackage{ulem}
\usepackage{refs}  
\usepackage{booktabs}   
\usepackage{morefloats}
\usepackage{color}
\usepackage{colortbl} 
\usepackage{natbib}
\usepackage{lipsum}
\usepackage{wrapfig}
\usepackage{glossaries}
\usepackage{gensymb}
\usepackage{multirow}
\usepackage{caption}
\usepackage{enumitem}
\usepackage{subfig}
\usepackage{orcidlink}
\glsdisablehyper

\bibpunct{(}{)}{;}{a}{}{,}

\begin{document}

\shortauthors{Ludwig et al.}

\title{The Stripped-Star Ultraviolet Magellanic Cloud Survey (SUMS): 

The UV Photometric Catalog and Stripped Star Candidate Selection}

\correspondingauthor{B.~Ludwig}

\author[0000-0003-0857-2989]{B.~Ludwig}
\affiliation{Department of Astronomy and Astrophysics, University of Toronto, 50 St. George Street, Toronto, Ontario, M5S 3H4, Canada}
\affiliation{Institute of Astronomy, KU Leuven, Celestijnenlaan 200D, 3001, Leuven, Belgium }
\email[show]{bethany.ludwig@kuleuven.be}

\author[0000-0001-7081-0082]{M.~R.~Drout}
\affiliation{Department of Astronomy and Astrophysics, University of Toronto, 50 St. George Street, Toronto, Ontario, M5S 3H4, Canada}
\email{maria.drout@utoronto.ca}

\author[0000-0002-6960-6911]{Y.~G\"{o}tberg}
\affiliation{Institute of Science and Technology Austria (ISTA), Am Campus 1, 3400 Klosterneuburg, Austria}
\email{ylva.gotberg@ist.ac.at}

\author[0000-0002-1172-0754]{D.~Lang}
\affiliation{Perimeter Institute for Theoretical Physics, 31 Caroline St N, Waterloo, Ontario N2L 2Y5, Canada}
\email{dlang@perimeterinstitute.ca}

\author[0000-0002-5522-0217]{A.~Laroche}
\affiliation{Department of Astronomy and Astrophysics, University of Toronto, 50 St. George Street, Toronto, Ontario, M5S 3H4, Canada}
\email{alex.laroche@mail.utoronto.ca}

\definecolor{UV}{RGB}{115,80,185}
\newcommand{\BL}[1]{{\color{blue} #1}}
\newcommand{\MRD}[1]{{\color{blue} #1}}
\newcommand{\red}[1]{{\color{red} #1}}
\newcommand{\YG}[1]{{\color{purple} #1}}
\newcommand{\rev}[1]{{\bf #1}}

\newcommand{\NSourcesLMC}{471,504}
\newcommand{\NSourcesSMC}{263,358}
\newcommand{\NSources}{734,862}
\newcommand{\NQremain}{13,343}
\newcommand{\NQremainLMC}{8,898}
\newcommand{\NQremainSMC}{4,445}
\newcommand{\NSEDCutLost}{8,159} 
\newcommand{\NSEDCutRemain}{5,184} 
\newcommand{\NSEDCutRemainLMC}{3,130} 
\newcommand{\NSEDCutRemainSMC}{2,054} 
\newcommand{\NAdditionalblueincmds}{2,518} 
\newcommand{\NAdditionalfluxfrac}{1,783}
\newcommand{\NAdditionalRemain}{883}
\newcommand{\NAdditionalRemainLMC}{571}
\newcommand{\NAdditionalRemainSMC}{312}
\newcommand{\NNoGaiaMatch}{62}
\newcommand{\NNoGaiaMatchPercent}{7\%} 
\newcommand{\NGaiaMatch}{821}
\newcommand{\NGaiaGOFOverThree}{176}
\newcommand{\NGaiaParallaxCut}{19}
\newcommand{\NGaiaParallaxCutPercent}{$\sim$2\%}
\newcommand{\NGaiaPMCut}{44}
\newcommand{\NGaiaPMCutPercent}{5\%}
\newcommand{\NGaiaMembersPercent}{$\sim$92\%} 
\newcommand{\NFinalLMC}{522}
\newcommand{\NFinalSMC}{298}
\newcommand{\NFinal}{820}
\newcommand{\NVBELMC}{78}
\newcommand{\NVBGLMC}{235}
\newcommand{\NBELMC}{91}
\newcommand{\NBGLMC}{118}
\newcommand{\NVBESMC}{27}
\newcommand{\NVBGSMC}{76}
\newcommand{\NBESMC}{91}
\newcommand{\NBGSMC}{104}
\newcommand{\NAVZEROLMC}{79}
\newcommand{\NAVZEROSMC}{26}
\newcommand{\NAVPONELMC}{144}
\newcommand{\NAVPONESMC}{87}
\newcommand{\NAVPTWENTYTWOLMC}{262}
\newcommand{\NAVPTHREELMC}{383}
\newcommand{\Nsimbad}{59}
\newcommand{\Nsimbadspectype}{6}
\newcommand{\NSEDBrightCut}{192}
\newcommand{\NSEDBrightCutPercent}{$\sim$23\%}
\newcommand{\NSEDChiCut}{183}
\newcommand{\NSEDChiCutPerent}{$\sim$22\%}
\newcommand{\NSEDPlotted}{303}
\newcommand{\NSEDFracErrTotal}{250}
\newcommand{\NSEDFracErrLMC}{158}
\newcommand{\NSEDFracErrSMC}{92}
\newcommand{\NSEDIsoTotal}{53}
\newcommand{\NSEDIsoLMC}{39} 
\newcommand{\NSEDIsoSMC}{14}
\newcommand{\NSEDIsoPercent}{$\sim$12\%}
\newcommand{\Nsummacsubtotalimages}{2,636}
\newcommand{\Nsummactotalimages}{2,420} 
\newcommand{\NsummactotalimagesLMC}{1,846} 
\newcommand{\NsummactotalimagesSMC}{574} 
\newcommand{\Nsummactracking}{216}
\newcommand{\NsummactrackingSMC}{82}
\newcommand{\NsummactrackingLMC}{134} 
\newcommand{\Nsummactrackingpercent}{8} 

\newacronym[plural=CMDs,firstplural=Color Magnitude Diagrams (CMDs)]{CMD}{CMD}{Color Magnitude Diagram}
\newacronym[plural=SEDs,firstplural=Spectral Energy Distribution (SEDs)]{SED}{SED}{Spectral Energy Distribution}
\newacronym[plural=RSGs,firstplural=Red Supegiants (RSGs)]{RSG}{RSG}{Red Supergiant}
\newacronym{ZAMS}{ZAMS}{Zero Age Main Sequence}
\newacronym{MS}{MS}{Main Sequence}
\newacronym{AGB}{AGB}{Asymptotic Giant Branch}
\newacronym[plural=PNe,firstplural=Planetary Nebulae (PNe)]{PN}{PN}{Planetary Nebula}
\newacronym{WR}{WR}{Wolf Rayet}
\newacronym[plural=WDs,firstplural=White Dwarfs (WDs)]{WD}{WD}{White Dwarf}
\newacronym[plural=SNe,firstplural=Supernovae (SNe)]{SN}{SN}{Supernova}
\newacronym{NS}{NS}{Neutron Star}
\newacronym{PSF}{PSF}{Point-Spread Function}
\newacronym{SUMS}{SUMS}{The Stripped-Star Ultraviolet Magellanic Clouds Survey}
\newacronym{MCPS}{MCPS}{Magellanic Clouds Photometric Survey}
\newacronym{SUMaC}{SUMaC}{\emph{Swift} Ultraviolet Survey of the Magellanic Clouds}
\newcommand{\Av}{A$_{\rm{V}}$}
\newcommand{\Tractor}{\emph{Tractor}}
\newcommand{\theTractor}{\emph{the~Tractor}}
\newcommand{\TheTractor}{\emph{The~Tractor}}
\newcommand{\theTractors}{\emph{the~Tractor's}}
\newcommand{\Msun}{\ensuremath{\,\text{M}_\odot}}
\newcommand{\appendixspacer}{\medskip}

\defcitealias{Drout2023}{DGL+23}

\begin{abstract}
Most massive stars {($\sim$8-25\Msun)} interact with a binary companion during their lifetimes. These interactions can remove the hydrogen-rich envelope, producing intermediate-mass ($\sim$2-8\Msun) and helium-rich stars. These ``stripped stars'' are predicted to emit predominantly in the ultraviolet (UV) and can therefore be identified via a UV excess---provided they are not outshone by their companion. However, despite their importance to binary evolution, supernovae, and ionizing feedback, few stripped stars have been confirmed. This is likely due to the scarcity of wide-field, high angular-resolution, UV surveys of stellar populations with reliable distances and extinction estimates. To address this, we present the Stripped-Star Ultraviolet Magellanic Clouds Survey (SUMS) catalog. We use \theTractor\ forward modeling software to perform PSF photometry on \Nsummactotalimages{} \emph{Swift}-UVOT images of the LMC and SMC. The resulting catalog contains \NSources{} sources in three UV filters to a depth of $\sim 20$ Vega mag. We perform validation tests on the photometry pipeline and highlight the catalog's broad applicability. We then identify sources with excess UV light compared to main-sequence stars and apply a series of quality cuts. From this, we identify \NFinalLMC{} candidate stripped stars in the LMC and \NFinalSMC{} in the SMC. We assess the potential contamination from other UV excess systems and argue the dominant uncertainty to be dust: early main-sequence stars can mimic the colors of stripped star binaries when extinction is overcorrected. This survey lays the groundwork for the first systematic census of stripped stars and opens new windows into binary evolution and massive star populations.
\end{abstract}


\section{Introduction}\label{sec:intro}

One of the most common predicted outcomes of binary evolution are stars that lose their hydrogen envelopes via interaction with a companion star. The transfer of mass from a donor to an accretor star can proceed either via Roche lobe overflow \citep{Kippenhahn1967, Paczynski1967} or common envelope evolution {\citep{paczynski_common_1976}}, and current binary population synthesis models predict that $\gtrsim$30\% of massive stars will undergo some level of ``stripping'' \citep[e.g.][]{Sana2012}. 
Many evolutionary models predict that these stripped stars should spend a majority of their core-helium burning phase ($\sim$10\% of the stellar lifetime) as hot and compact stars that emit most of their light in the extreme ultraviolet \citep[e.g.][]{Gotberg2018}. However, in some cases if only partial stripping occurs, they may remain at cooler temperatures \citep[e.g.][]{Klencki2022,Ramachandran2023,Ramachandran2024}.    

The stripping of a hydrogen-rich envelope by a companion can, in principle, occur across any mass range. At the low-mass end ($\lesssim 2,\Msun$; \citealt{Gotberg2018, Heber2016}), binary-stripped stars resemble subdwarfs, while at the high-mass end ($\gtrsim 10,\Msun$; \citealt{Crowther2007}), they resemble Wolf-Rayet stars—objects that are also thought to be capable of shedding their envelopes through strong stellar winds, independent of binary interaction.
In between these two extremes{, massive stars between 8-25\Msun{} that lose their envelopes through binary interaction evolve into 
intermediate mass ($\sim$2-8 M$_\odot$) stripped stars \citep{podsiadlowski_presupernova_1992, Gotberg2018}.}
In this manuscript, we refer to these intermediate mass stripped stars as simply \emph{``stripped stars''}{, however} as we cannot determine precise masses here, we extend this mass regime to broadly include binary stripped stars between $\sim$1 and 8 M$_\odot$.

These stripped stars are of particular interest as potential progenitors of hydrogen-poor core-collapse supernovae {\citep{podsiadlowski_presupernova_1992,de_donder_relative_1998,Smith2011,Drout2011,yoon_nature_2012,Eldridge2013,Tauris2015,eldridge_disappearance_2016,folatelli_disappearance_2016,Lyman2016}}, contributors to cosmic reionization {\citep{van_bever_effect_1999,Ma2016,Stanway2016,rosdahl_sphinx_2018,Gotberg2020A,Secunda2020}}, and could themselves emit gravitational waves in the mHz regime detectable by the upcoming Laser Interferometer Space Antenna (LISA) \citep{Nelemans2004,Gotberg2020B,Wu2020}. {Additionally, these stars likely represent a necessary evolutionary stage in the formation of compact object binaries that eventually merge through gravitational wave emission. Models for double neutron star mergers  require at least two distinct phases of envelope stripping \citep{Tauris2017,vigna-gomez_formation_2018}. Constraining the demographics of this population is therefore critical, as merger rate predictions remain highly uncertain and depend sensitively on poorly constrained aspects of binary evolution such as mass transfer efficiency, common-envelope evolution, and natal kicks \citep{belczynski_binary_2018,mandel_binary_2021,Broekgaarden2022,mandel_rates_2022}.}

\cite{Gotberg2018} predicted that some stripped stars may be identifiable through the additional UV light they contribute to their binary systems if its companion does not outshine it. \citet[][hereafter DGL+23]{Drout2023} then further refined these predictions using a grid of (hot) core-helium burning stripped stars combined with \gls{MS} stars. They found that stripped stars are expected to appear bluewards of the \gls{ZAMS} in UV-optical \glspl{CMD} if they have either \gls{MS} companions with masses below $\sim$8 \Msun\ or compact object companions.
Given that stripped stars of these masses would likely have had more than $\sim$6 \Msun\ of their H-rich envelope removed \citep{Gotberg2018}, this identification method therefore favors systems that (i) have gone through non-conservative mass transfer, (ii) have compact object companions, or (iii) were disrupted from their natal binaries and are now stripped isolated stars \citep[e.g. as in][]{Renzo2019}. 
While there are considerable uncertainties, including the efficiency of binary mass transfer and the outcomes of common envelope evolution, \citetalias{Drout2023} provide an order-of-magnitude estimate that approximately 10-25\%\footnote{This estimate is based on the number of systems in fiducial binary populations that have either undergone common envelope evolution or contain compact object companions.} of systems containing a stripped star may be detectable via excess UV light in their \glspl{SED}.

It is with this theoretical understanding of where a subset of stripped star binaries should exist in UV-optical \glspl{CMD} that we pursue a survey designed to identify hot stripped stars in the Magellanic Clouds. We have termed this effort the Stripped-Star Ultraviolet Magellanic Clouds Survey (SUMS). We have selected the Magellanic Clouds as an ideal first laboratory as: (i) distances are known (ii) the line-of-sight extinction is relatively low (iii) large swaths of stars can be studied and (iv) their subsolar metallicities will enable us to explore the effects of metallicity on binary populations. 

Despite their importance, predicted ubiquity, and detectability, stripped stars evaded direct observation for decades. This was almost certainly an observational bias, as publicly available UV photometric catalogs with sufficient depth, resolution, and coverage that overlap with massive stars have been relatively sparse. In particular, the Magellanic Clouds are both large, covering multiple square degrees on the sky, and feature densely crowded regions. GALEX, for example, has imaged portions of the Clouds, but its resolution (FWHM $\sim$5.3$\arcsec$) is insufficient to resolve individual stars in crowded regions \citep{Simons2014}. Similarly, the Ultraviolet Imaging Telescope (UIT) has offered UV photometry of the Clouds, but with partial coverage and a resolution of FWHM $\sim$3.4$\arcsec$ \citep{Parker1998,Cornett1997,Cornett1994}. The Optical/UV Monitor onboard the XMM Newton Satellite (XMM-OM) has an improved resolution of FWHM $\sim$ 2.5$\arcsec$, however the photometry published from this mission to date \citep{Page2012} relies on aperture photometry and is therefore subject to contamination from nearby sources. Finally, while the Hubble Space Telescope (HST) has provided high-resolution photometry of individual regions of the Magellanic Clouds \citep[e.g.][]{Sabbi2016,Milone2023,Petia2017}, its limited field of view restricts global coverage. 

It was with these considerations in mind that we opted to perform new \gls{PSF} photometry on a set of UV images taken of the Magellanic Clouds with the UV-optical telescope (UVOT) onboard the \emph{Swift} satellite, \citep{Gehrels2004,Roming2005,Poole2008} with the ultimate goal of identifying candidate stripped stars. \emph{Swift-}UVOT has a spatial resolution of $\sim$2.5$\arcsec$ and performed imaging of over 14 square degrees in the Magellanic Clouds in three UV filters between 2010 and 2013 \citep{Siegel2014,Hagen2017}. However, no point source catalog from this data has yet been published. 

We note that since we began our work, the UV Imaging Telescope (UVIT) on board the Astrosat mission \citep{Gehrels2004,Tandon2017} has also conducted a survey of the Magellanic Clouds and recently released a catalog of UV photometry for the Small Magellanic Cloud \citep{Devaraj2023,Hota2024} with state-of-the-art resolution ($\sim$1.4$\arcsec$). This data, along with that from upcoming missions such as UVEX \citep{Kulkarni2021}, will also be invaluable for the study of post-interaction binaries.

Earlier versions of the photometric pipeline presented in this paper were used in \citetalias{Drout2023} to both (i) identify a set of targets with potential UV excess for spectroscopic follow-up and (ii) provide the final set of photometry for the 25 stripped star candidates they discuss in detail. These objects--identified as part of the broader SUMS effort--were shown to have luminosities, colors, and spectral morphologies consistent with expectations for binaries containing intermediate mass stripped stars. The spectroscopic sample could be grouped into three categories: (i) stripped star dominated systems that appeared to host 
stripped stars with no evidence for a luminous companion, (ii) systems showing spectroscopic features from both a stripped star and B-type MS companion, and (iii) systems that resembled B-type stars spectroscopically but exhibited UV excesses in their \glspl{SED}. This diversity is consistent with expectations for binaries, where the stripped star and companion contribute varying fractions of the total light. Subsequent modeling of a subset of these systems (\citetalias{Drout2023}; \citealt{Gotberg2023}) revealed high temperatures (50–100 kK), small radii (0.5–1 $R_\odot$), intermediate luminosities ($\log (L/L_\odot) \sim 3$–5), and hydrogen-depleted surfaces (X$_\mathrm{H} \sim$ 0–0.3)---all consistent with the long-lived helium-core burning phase of intermediate mass stripped stars. Interestingly, though, their winds appear to be weaker than expected \citep{Vink2017,Krtivcka2016,Nugis2000}.

These results highlight the effectiveness of our UV photometry and selection method in identifying stripped stars. To assess whether their observed prevalence matches theoretical expectations, however, a larger and more systematic search is needed. In this paper, we apply our finalized UV photometric pipeline to the full \emph{Swift}-UVOT survey of the Magellanic Clouds and identify a broad sample of candidate stripped star binaries based on their \glspl{SED}. To support ongoing and future studies, and to serve as a community resource, we publicly release the SUMS UV catalog, candidate sample, and data processing pipeline on GitHub\footnote{\url{https://github.com/AstroLudwig/SUMS_UVPhotometricCatalog}}
 under the CC BY 4.0 and MIT licenses, respectively, with version 1.0 permanently archived on Zenodo (\!\dataset[DOI: 10.5281/zenodo.17551743]{https://doi.org/10.5281/zenodo.17551743}; \citealt{sums_github}).

This manuscript is organized as follows: In Section \ref{sec:Data}, we describe the UV and optical data from which our catalog is derived. Section \ref{sec:Photometry} outlines and validates the photometric pipeline used to perform PSF photometry on UV point sources in the \emph{Swift-}UVOT images. We present the structure of the resulting SUMS catalog in Section \ref{sec:catalog}.
In Section \ref{sec:Candidates}, we identify \NFinal{} candidate stripped star systems in the Magellanic Clouds and explore their properties in Section \ref{sec:properties}. Section \ref{sec:otherblue} discusses possible contaminants-including foreground and background sources-and other UV-excess systems that may fall within our selection criteria. Finally, Section \ref{sec:summary} provides a summary and outlines future research directions.
 
\begin{figure*}
    \centering
    \includegraphics[width=0.92\textwidth]{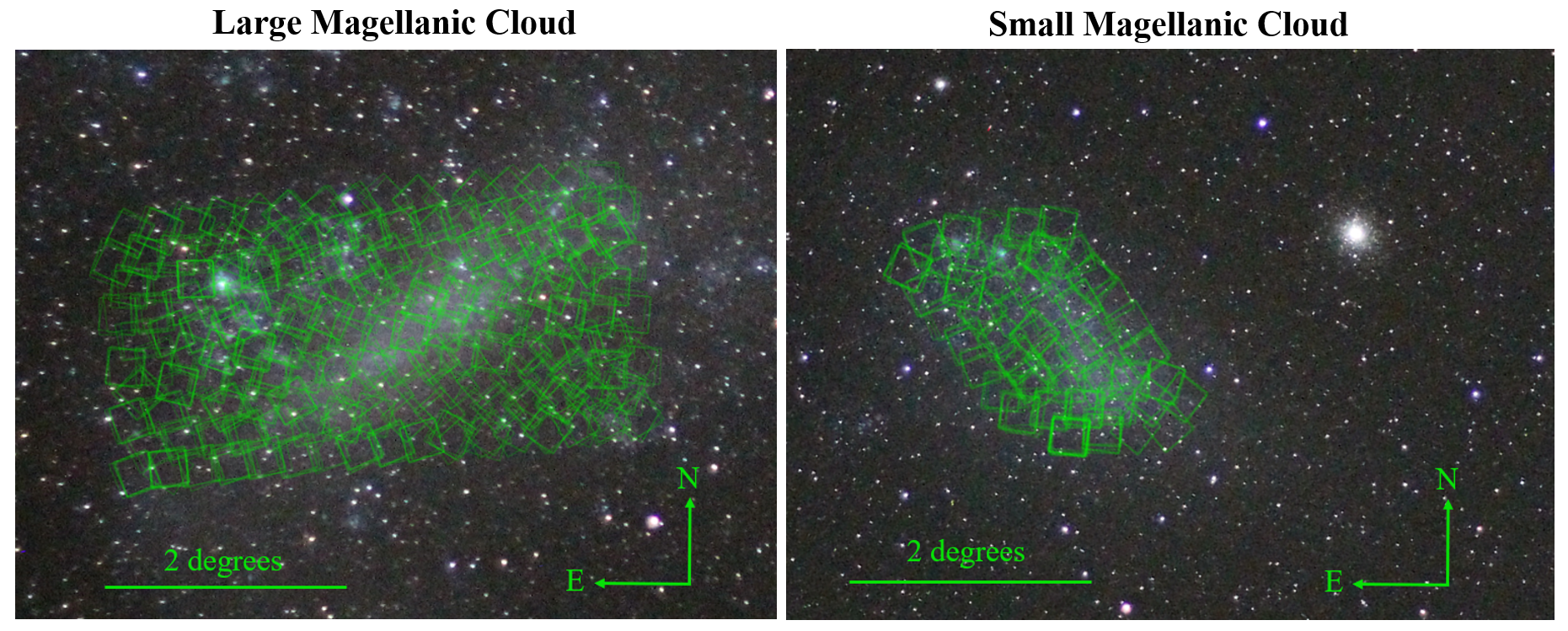}
    \caption{A summary of the spatial coverage of the SUMaC survey, which provided extensive imaging of the Magellanic Clouds in the ultraviolet using the UVOT instrument on \emph{Swift}. 
    The green squares indicate the footprints of individual images associated with the survey, which covered 150 fields in the LMC and 50 fields in the SMC. In total, there was an average of 660 and 219 images obtained in each of the three UVOT UV filters (UVW2, UVM2, and UVM1) in the LMC and SMC, respectively (\S\ref{sec:SUMAC}).  
    Some footprints overlap, in which case photometry for a single source can be averaged over multiple pointings. The background optical image was taken during a SUMS observing run with a Cannon EOS Rebel T5 by Anna O'Grady at the Las Campanas Observatory.}
    \label{fig:coverage}
\end{figure*}

\section{UV Survey Data \& Optical Catalog}
\label{sec:Data}

Identifying a population of stripped  stars via an observed UV excess requires \emph{both} UV and optical photometry across the Magellanic Clouds. Here, we describe the primary data that we utilize in this study.

\subsection{The Swift UVOT Magellanic Clouds Survey: SUMaC}\label{sec:SUMAC}
The \emph{Swift-}Ultraviolet Optical Telescope (UVOT) is a photon-counting ultraviolet and optical (170-600 nm) telescope on the \emph{Swift} satellite with a 17$\arcmin$ $\times$ 17$\arcmin$ field of view and 2.37--2.92$\arcsec$ spatial resolution (defined by the FWHM of the \gls{PSF}; \citealt{Roming2005,Breeveld2010}).  Originally launched to provide prompt follow-up of \emph{Swift} identified gamma-ray bursts, it has since carried out observations for a wealth of science cases---from supernova and tidal disruption events to comets and near-Earth asteroids \citep{Brown2014,Hinkle2021,Bodewits2023,Ofek2024}.  

Between 2010 and 2013 \emph{Swift-}UVOT carried out the \gls{SUMaC}. These observations consist of 150 fields in the LMC and 50 fields in the SMC \citep{Siegel2014, Hagen2017}. Figure~\ref{fig:coverage} shows the survey footprints plotted over optical images of the clouds covering roughly 10.7 and 3.5 square degrees for the LMC and SMC respectively.  These images are taken in the UVW2, UVM2, and UVW1 bands (see Figure~\ref{fig:extinctioncurve} for transmission curves) which have a pixel scale of 1$\arcsec$ per pixel, and correspond to central wavelengths of 1928, 2246, and 2600\r{A} respectively \citep{Breeveld2010}. 

The \gls{SUMaC} dataset has been used for a number of science cases, ranging from the shape of the UV extinction law to measurements of the recent star formation history in the Magellanic Clouds \citep{Hagen2017}. For point sources, it is estimated that there are approximately 250,000 in the SMC and one million in the LMC \gls{SUMaC} images with a 50\% detection limit at approximately 18.7 AB mag \citep{Siegel2014,Hagen2017}. However, no catalog of point sources has yet been released. We therefore construct a pipeline, described in Section~\ref{sec:Photometry}, to assemble our own UV photometric catalog. 

We retrieve the calibrated and filtered level 2 \gls{SUMaC} images using the UK \emph{Swift} Science Data Centre \footnote{\url{https://www.swift.ac.uk/swift_live/index.php}}. This data has already been processed through the \emph{Swift-}UVOT pipeline, which performs bad pixel, flatfield, and boresight corrections, as well as computation of an initial World Coordinate System (WCS) astrometric solution. The official \gls{SUMaC} images have observation IDs 40415 to 40464 (for the SMC) and 45422 to 45586 (for the LMC). In addition, the \gls{SUMaC} team included archival observations to supplement their coverage. In particular, we identify one region in the SMC center without data from the \gls{SUMaC} survey, and we therefore include data from \emph{Swift} observation ID 32214 to cover this area. Each of the \gls{SUMaC} fields was observed by \emph{Swift} on 1 to 5 occasions over the duration of the survey (2010-2013). During each visit, between 2 and 8 individual exposures were obtained in each UV filter. As a result, multiple  images are available for each \gls{SUMaC} field, with exposure times ranging from approximately 30 to 600 seconds. We describe in Section~\ref{sec:Photometry} and \ref{sec:catalog}, below, how we process and combine this data within our final catalog.

\subsection{Magellanic Clouds Photometric Survey: MCPS}
\label{sec:MCPS}

The \gls{MCPS} is an optical photometric survey of both the Large and Small Magellanic Clouds carried out with the Great Circle Camera on the Swope Telescope at Las Campanas Observatory \citep{Zaritsky1996}. The results from this survey have been fundamental to understanding the spatially resolved star formation history of the Magellanic Clouds, as well as providing detailed extinction maps \citep{Harris2004,Harris2009,Zaritsky1999,Zaritsky2002,Zaritsky2004}. 

The LMC portion of the survey consists of over 24 million stars extending over an $8.5\degree$x$7.5\degree$ patch of the sky, while the SMC portion consists of 5 million stars over $4.5\degree$x$4\degree$ of the sky \citep{Zaritsky2002,Zaritsky2004}. Comparatively, the \gls{SUMaC} data described in the previous section covers roughly 17\% and 20\% of the \gls{MCPS} footprint in the LMC and SMC, respectively. However, using the star-formation history maps of \cite{Harris2004} and \cite{Harris2009} (which were derived from the \gls{MCPS} data), we find that the footprint of the SUMaC images cover $\sim$40\% and $\sim$36\% of the recent ($<$100 Myr) star formation in the LMC and SMC, respectively.\footnote{{We note, however, that these star formation rates are reliant on many model assumptions, including that they are derived from single star models. To investigate possible variations, we also examine the LMC SFH maps from the VISTA survey of the Magellanic Clouds survey \citep{mazzi_vmc_2021}. While they still do not account for impacts of binary interaction, these maps are computed assuming just 30\% binaries are found in detached binaries. When comparing to these maps we find that 54\% of the young star formation overlaps with the SUMaC footprint (a 14\% increase over that inferred from the MCPS SFH maps). These differences may be due to a combination of model assumptions, low number statistics, and extinction.
We therefore emphasize that these values should be taken as order of magnitude estimates.}} 
The \gls{MCPS} catalogs contain U, B, V, I photometry to a limiting magnitude of $\sim$21 Vega mag in $V-$band. In the sections below, we utilize both the photometry and astrometry from the \gls{MCPS} survey.

\section{UV Photometry with the Tractor }
\label{sec:Photometry}
Here we present the photometric pipeline used to construct a catalog of UV point sources in the \gls{SUMaC} images of the Magellanic Clouds.
In general, it is not feasible to use the standard \emph{Swift-}UVOT photometry routines on the \gls{SUMaC} images because they rely on aperture photometry within a 5$\arcsec$ radius around a source. In the crowded environment of the Magellanic Clouds, such a region usually encompasses the light from multiple stars. 
We therefore choose to use \theTractor{}\footnote{\url{https://thetractor.readthedocs.io/en/latest/}} image-modeling software \citep{Lang2016,Lang2016B} to perform forced PSF photometry at the location of stars with astrometric positions from MCPS. \Tractor{} is a forward-modeling code that works by optimizing the likelihood of the photometric properties of a set of astronomical sources when provided with an input image and priors on the source positions and brightnesses. By constructing models of the image from the photometric parameters convolved with the image PSF, \theTractor{} is able to more effectively deblend crowded sources than the standard \emph{Swift} aperture photometry routines. The forced-photometry aspect of \theTractor{} is also ideal for our circumstances, as (i) optical catalogs from deep, high-resolution images of the Magellanic Clouds already exist, and (ii) identification of stripped stars via UV excess requires both optical and UV photometry. By calculating UV photometry for sources with known optical magnitudes, we avoid complications that can arise from cross-matching catalogs in different wavelength regimes when source detection was performed independently.

In the subsections below, we provide detailed information on our photometry pipeline including the inputs required by \theTractor{}, the specific settings used, the conversion of source count rates output by \theTractor{} into magnitudes, and an evaluation of \theTractors{} performance computing photometry on \emph{Swift-}UVOT images. Readers interested mainly in the final catalog contents can skip to Section~\ref{sec:catalog}. Those interested in the selection of stripped star candidates from the resulting UV catalog can skip to Section~\ref{sec:Candidates}.

\subsection{Tractor Input}
As input, \theTractor{} requires (i) an astronomical image, (ii) information about the input image, such as a model of the sky background and the filter-dependent PSF and (iii) priors on the source locations and normalization.

\subsubsection{Input Images} \label{sec:inputimages}

We run \theTractor{} independently on each individual \emph{Swift-}UVOT image of a given field, as opposed to averaging them together. While combined images would be deeper, this method provides us with multiple independent measurements for each target star, while also avoiding changes to the \gls{PSF} shape that can result due to small astrometry errors and resampling effects. However, we identify a number of images which are incompatible with our method. In particular, we do not analyze images that either (i) have a large number of pixels with zero counts, as this causes the background estimation procedure described in Section~\ref{sec:background} to fail, or (ii) show clear evidence of tracking issues during the exposure, which results in a highly altered and non-symmetric PSF. For the former, this predominantly affects images with very short exposure times ($\lesssim$ 45 seconds). For the latter, we identify problematic images by visual inspection. In total, across all filters, we reject \NsummactrackingSMC{} SMC and \NsummactrackingLMC{} LMC images ($\sim$\Nsummactrackingpercent{}\% of the total).
We note that because each field was imaged by SUMaC multiple times, this does not impact the overall survey coverage. In total, we run \theTractor{} on  \NsummactotalimagesSMC{} \textit{Swift}-UVOT images in the SMC and \NsummactotalimagesLMC{} in the LMC.

We make several small adjustments to the Level 2 sky images from \emph{Swift} before inputting them into \theTractor{}. First, we divide each image by the exposure time to be in units of count rate, as this will be required for final calibration of the resulting source brightness into magnitudes (Section~\ref{sec:hea}). Second, we recompute the WCS solution on each image using \texttt{astrometry.net} \citep{Lang2010} with a custom index files that include all MCPS sources with U-band magnitudes brighter than 19 mag. This ensures the closest match possible between the astrometry on the input images and the coordinates of the MCPS stars that will be used for forced photometry (Section~\ref{InitGuess}).

\begin{figure}
    \centering
    \includegraphics[width=\columnwidth]{ 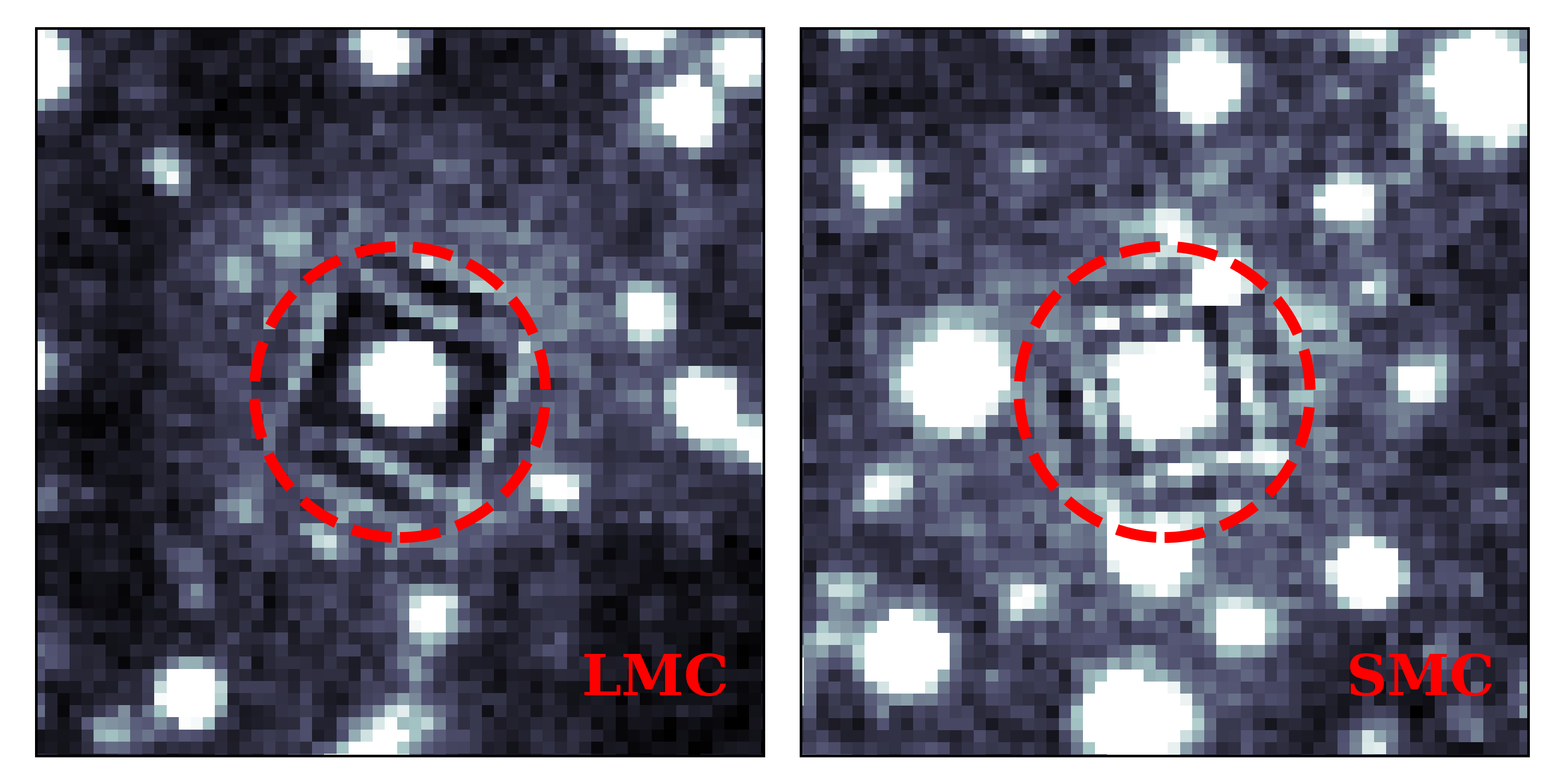}
    \caption{Examples of coincidence-loss distortions that appear around bright sources in \textit{Swift}-UVOT images. These features impact our ability to accurately measure the photometry of sources within the immediate vicinity.
    To account for this, we mask a region with a 12$\arcsec$ radius around the brightest sources in each image (indicated here as the dashed red circle).}
    \label{fig:masking}
\end{figure}

Finally, we mask the regions around very bright stars in the Level 2 images. \emph{Swift-}UVOT is a photon counting device and therefore subject to coincidence loss (which occurs when multiple photons arrive within the same frame; in normal operation, \emph{Swift-}UVOT gathers data at a rate of 90.6 frames\;s$^{-1}$). While coincidence loss can be corrected up to a certain level, for objects with very high count rates a modulo-8 distortion pattern is created on the image \citep[see][for additional details]{Page2014}. Examples are shown in Figure~\ref{fig:masking}. These features are non-correctable, and we therefore choose to mask the area around such objects, to avoid situations such as \theTractor{} moving a nearby source to try to fit a peak in the distortion pattern. 

By visual inspection, we determined that the stars that produce these defects typically have count rates of 65 counts per second or higher (measured within a 5$\arcsec$ radius; by comparison, most stars are under 20 counts per second). To determine which objects are above this threshold and should therefore be masked, we perform aperture photometry at the location of all MCPS sources with the \texttt{Photutils} package of {\tt astropy} \citep{Bradley2022}. However, when doing so, we found that the MCPS catalog is not complete at the brightest magnitudes. We therefore also perform aperture photometry at the location of all stars within the UCAC4 catalog \citep{Zacharias2013}. In total, we identify 1759 and 741 unique sources in the LMC and SMC respectively above our bright threshold. These sources had typical V-band magnitudes of $\sim$16--6 mag. Approximately 1200 of these sources matched within 1$\arcsec$ of a source in the SIMBAD astronomical database \citep{Wenger2000}. Of these, more than 85\% were classified as O/B-type stars (453/570, respectively) and an additional 48 were listed as Wolf Rayet stars. We mask a 12" aperture around each, chosen to encompass the square-like defects visible in Figure~\ref{fig:masking}. The total area masked accounts for less than 1\% of the estimated coverage in each galaxy.

\subsubsection{Image Background and Noise}\label{sec:background}

\TheTractor{} requires both a description of the background and the noise within the input image. We estimate both using a set of routines distributed as part of \texttt{Photutils} \citep{Bradley2022}. When constructing the image background, we first mask an 11$\times$11 pixel box centered on all sources in the image that have at least 5 adjacent pixels with signal-to-noise greater than 5 using the \texttt{make\_source\_mask} routine (for further details on how signal-to-noise is calculated see the \texttt{Photutils} documentation). We then create a model for the background using  the \texttt{Background2D} routine. 
We adopt a box size of 20$\times$20 pixels within which a single background value is computed, apply median filtering using three adjacent boxes to smooth the background map, and apply the \texttt{SExtractor} background estimator \citep{Bertin1996} to create the model.
We also compute a total error image that accounts for both background noise and Poisson noise, following the procedure described in the documentation for \texttt{calc\_total\_error}. The background noise is estimated using the RMS output from \texttt{Background2D}, while the Poisson noise is computed as the square root of the total counts per pixel in the original image. The error image is initially calculated in units of counts, then converted to count rates by dividing by the image exposure time.

\subsubsection{Initial Guesses for Source Positions and Count Rates}\label{InitGuess}

\TheTractor{} requires priors on source positions and count rates for each \textit{Swift}-UVOT image. As described above, we use positions from the MCPS catalog, which is more than $50\%$ complete for sources brighter than 21 Vega mag in the V-band \citep{Zaritsky1997}, and which also served to define the WCS solution for the images (see Section~\ref{sec:inputimages}). However, we choose to exclude sources that would likely be too faint to appear in the UV image as: (i) this can significantly reduce the total computation time and (ii) we found that \theTractor{} will sometimes output negative count rates when trying to fit sources within a region of the image consistent with the background. 

We choose which sources to include based on a combination of their \gls{MCPS} photometry and properties of the UVOT images themselves. First, we require that an input source has a MCPS U-band magnitude of at least 20.5 Vega mag or brighter. We chose this threshold because even very blue stars (e.g., stripped  stars) have $U-UVW1$ colors of less than 1.4 Vega mag, implying fainter stars would fall below the nominal \gls{SUMaC} detection threshold of $\sim$19 Vega mag in the UV. If the MCPS catalog does not provide a U-band magnitude, we apply this threshold to the B-band instead, as all sources in MCPS are required to have a B-band measurement to be included in their catalog. Second, we check that there is evidence for significant flux above the background level within the \gls{SUMaC} images. We perform 5$\arcsec$ aperture photometry using \texttt{Photutils} \citep{Bradley2022} at the position of each source in both the input image and the model background, described in \S\ref{sec:background}. For inclusion in our final input list of stars, we require that the count rate in the image is at least 1.5 times that in the model background. While this threshold ensures that there is evidence for excess flux above the background, it also means that some faint sources for which UV magnitudes could in principle be recovered from the \gls{SUMaC} images may be missing from our catalog (see Appendix~\ref{sec:caveat} for further discussion of this and other caveats of our catalog).

We provide \theTractor{} with the 5$\arcsec$ aperture photometry results, described above, as rough initial estimates for the count rates of input sources. Although these values likely overestimate the true count rates, we test the sensitivity of our final results to these initial guesses. For a small region in a randomly chosen image, we vary these initial count rate estimates by adding between 0.25 and 10 counts/s in steps of 0.25 counts/s, and found that the resulting magnitudes from our pipeline are not significantly altered. For very faint stars ($\gtrsim$19 Vega mag), the percent difference is at most 1.4\%. For brighter stars, the difference is 0.4\% or lower. 

\subsubsection{The PSF}\label{sec:psf}

We provide \theTractor{} with a 2D, radially symmetric PSF model for each \emph{Swift}-UVOT UV filter, which we reconstructed from the curves-of-growth distributed with CALDB. As described in \citet{Breeveld2010}\footnote{See also the \emph{Swift}-UVOT CALDB release note number 104; \url{https://heasarc.gsfc.nasa.gov/docs/heasarc/caldb/swift/docs/uvot/uvot_caldb_psf_02.pdf}}, these curves-of-growth were derived by fitting the brightness profiles of isolated sources in long-exposure images. The PSF is assumed to be radially symmetric: its core is modeled using the cumulative integral of a Moffat profile out to 5$\arcsec$, while the wings are fit with a centered double Gaussian out to 30$\arcsec$. The resulting model is integrated to yield the relative flux as a function of radius. For use with \theTractor{}, we render this 1D model onto a 2D radially symmetric pixel grid, sampling at 0.1$\arcsec$ per pixel and then rebinning to 1$\arcsec$ per pixel to match the resolution of the \gls{SUMaC} images. 
Finally, we normalize the PSF model so that its integral within a 5$\arcsec$ radius is unity, consistent with the requirements of the \emph{Swift}-UVOT photometric calibrations (see Section~\ref{sec:hea}).

Although our PSF model is assumed to be constant, the \textit{Swift}-UVOT PSF is known to vary slightly with source brightness (due to coincidence loss) and spacecraft voltage. To account for these effects, we take several steps: (i) we apply a 5\% systematic uncertainty in the calibration (\S~\ref{sec:hea}); (ii) we mask sources bright enough to induce significant PSF distortions (\S~\ref{sec:inputimages}); and (iii) we perform tests to assess the residuals between our final \Tractor{} models and the input images as well as how close this method agrees with standard aperture photometry routines for isolated stars (\S~\ref{sec:residual} and \ref{sec:isolatedtest}). We also tested an alternative approach in which a PSF model is constructed for each image by fitting isolated stars using the IRAF \texttt{psf} function \citep{Tody1986,Tody1993}. However, we find that these image-specific PSFs do not improve the image reconstructions in a statistically meaningful way and are significantly more computationally intensive.

\begin{figure*}
    \centering
    \includegraphics[trim={0cm 0.cm 0.3cm 0cm},clip,width=\textwidth]{ 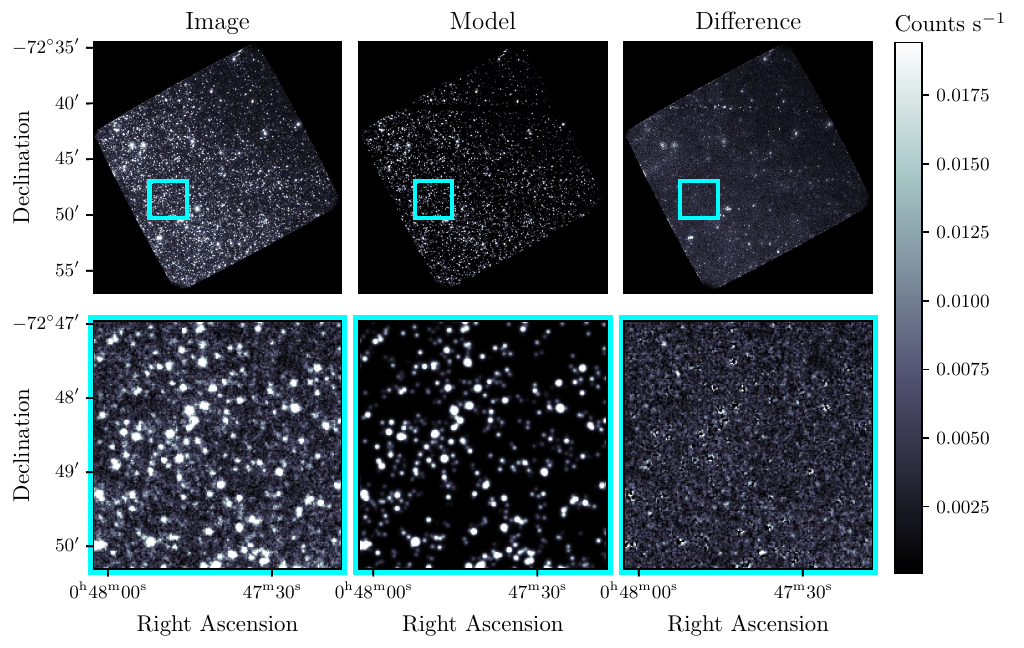}
    \caption{ 
    An example of the performance of \theTractor{} forward modeling pipeline described in Section~\ref{sec:Photometry} on a \emph{Swift}-UVOT image from the SMC. The left column shows the background-subtracted UVOT image in units of counts s$^{-1}$. The middle column shows the corresponding model image from which UV magnitudes can be measured. The right column shows the difference or residual image (i.e., the image minus the model) which highlights the regions where the model deviates from the observed data. The top row shows the full $17\arcmin \times 17\arcmin$ UVOT image, while the bottom row shows a $3.3\arcmin \times 3.3\arcmin$ zoomed-in area (marked by the cyan box in the top row). 
    A horizontal streak visible as a black patch near the top of the full model image and a row of stars in the full difference image shows a region where no sources were modeled. This was due to the absence of optical sources in the MCPS catalog, caused by occasional misalignment in the MCPS scanning strategy.
    }
    \label{fig:model}
\end{figure*}

\subsection{Tractor settings and output}\label{sec:TractorSettings}
Using the inputs described in detail above, we run \theTractor{} on each \gls{SUMaC} image with the following settings: 

\begin{itemize}[leftmargin=1em,itemsep=-0.3em]
    \item \emph{Image Data:} Within \theTractor{} \texttt{Image} object, we set the data to the input images described in \S~\ref{sec:inputimages} after subtracting the background images described in \S~\ref{sec:background}. The units of this data are counts per second.
    \item \emph{Sky:} We use the \texttt{Constant\_Sky} setting to describe the background in the input data, with a value set of the median in the sky-subtracted images after sigma-clipping bright sources. Although our data is already background-subtracted and these input values are very small (typically a few $\times$10$^{-3}$ counts s$^{-1}$), we found this led to better performance than setting \texttt{NullSky}.
    \item \emph{Variance:} We set the \texttt{Invarr} setting to be the inverse square of the error image described in \S~\ref{sec:background}.
    \item \emph{PSF:} We use the \texttt{PixelizedPSF} setting with input described in \S~\ref{sec:psf}. The PSF in the three \emph{Swift-}UVOT UV filters varies slightly.
    \item \emph{Photometric Calibration:} We use the \texttt{NullPhotoCal} setting, because \emph{Swift-UVOT} is a photon counting device and the final photometric calibration varies slightly from typical optical/IR images for which \theTractor{} was originally written. With this setting, the units of the final photometric parameters for each source are in the same units as the input image (counts s$^{-1}$).
    \item \emph{WCS solution:} We use the \texttt{NullWCS} solution, instead choosing to input source positions in pixel space.
    \item \emph{Input Sources:} We model all input objects as point sources (using the \texttt{PointSource} input) with initial guesses for their positions and brightnesses described in \S~\ref{InitGuess}.
    \item \emph{Optimization of Source Positions:} While we originally froze the source positions within the input image, we found that this led to small dipole patterns even for relatively isolated sources. We therefore chose to run \theTractor{} allowing the positions to move, but setting a Gaussian prior on their positions which are centered on their MCPS position and had a width of $\sigma = 0.05\arcsec$ (i.e 1/20th of a pixel). This strongly constrains their positions while allowing for some WCS errors. The impact of this setting will be discussed in \S~\ref{sec:source_movement}.
\end{itemize}

The value output by \theTractor{} corresponds to the scaling factor applied to the input PSF to best fit the image data. Since we normalize the PSF to integrate to unity within a 5$\arcsec$ radius, the output directly represents the source count rate within that aperture.
Using these inputs, \theTractor{} generates a model of the image and iteratively adjusts source positions and brightnesses until a convergence criterion is met ($d(\ln{P}) < 10^{-3}$). The final output includes the best-fit source positions, fluxes, and the modeled image itself. 
Figure~\ref{fig:model} illustrates the performance of \theTractor{} using a full SMC image (image ID: 40418 in UVM2), along with a zoomed-in $200\arcsec \times 200\arcsec$ region. We show the original \gls{SUMaC} image, the best-fit \Tractor{} model, and the residual image, highlighting the differences between them.

\subsection{HEASARC Calibrations}
\label{sec:hea}
We convert the best-fit count rates from \theTractor{} into magnitudes using the standard \textit{Swift}-UVOT calibration tools provided by HEASARC.
Our custom implementation follows the same overall procedure as the HEASARC task \texttt{uvotsource}. This process requires four quantities as input:  (i) the total count rate (source+background) within a 5" circular aperture, (ii) the total count rate error, (iii) the background count rate within a 5" circular aperture, and (iv) the background count rate error. As described above, due to the normalization of our model PSF, the source count rates output by \theTractor{} are already scaled to a radius of 5$\arcsec$. We assume Poisson errors on these values (accounting for the exposure time of each image). For the background count rate and error, we take the results from aperture photometry performed with a 5$\arcsec$ radius at the position of each source on the 2D background images (described in Sections~\ref{sec:background} and \ref{InitGuess}). Finally, for the total count rate, we add these two values together and combine their errors in quadrature. We then follow the following procedure for each source: 

\begin{enumerate}[leftmargin=1em,itemsep=-0.4em]
    \item Correct the total count rate for coincidence loss using \texttt{uvotcoincidence}.
    \item Correct the background count rate for coincidence loss using \texttt{uvotcoincidence}.
    \item Subtract the corrected background count rate from the corrected total count rate to produce a coincidence-loss corrected source count rate.
    \item Apply a recommended additional 5\% systematic error to the corrected count rate to account for variations in the \emph{Swift-}UVOT PSF shape.
    \item Identify individual sources that fall on locations of the \emph{Swift-}UVOT detector with known sensitivity issues by running the task \texttt{uvotlss} while pointing to the input file sssfile5.fits distributed by HEASOFT\footnote{See \url{https://swift.gsfc.nasa.gov/analysis/uvot_digest/sss_check.html}. Note that for more recent releases of HEASOFT (2022+), this step is performed automatically in many UVOT tasks.}. While corrections for these ``small scale sensitivity'' regions are not currently possible, this step flags sources with unreliable photometry, which will be removed before construction of our final catalog, below. 
    \item Apply corrections for large-scale sensitivity variations across the \emph{Swift-}UVOT detector by running \texttt{uvotlss}.
    \item Apply corrections for the degradation of the \emph{swift-}UVOT sensitivity, using the midpoint observation time for each image and the sensitivity correction file distributed within CALDB.\footnote{We specifically use file swusenscorr20041120v005.fits, which is accurate during the time period when the \gls{SUMaC} images were taken (2011--2013). }
    \item Convert these final, corrected, count rates and errors into magnitudes by running \texttt{uvotflux}.
\end{enumerate}

At the end of this process, the HEASARC routines provide calibrated Vega magnitudes and errors for each source. Flags are also automatically produced that indicate any sources that were saturated (generally not applicable in our case due to our procedure of masking bright sources described in Section~\ref{sec:inputimages}) or which fell on a region of the detector with anomalously low sensitivity (as described in step 5, above). We also flag any source located within 5$\arcsec$ of the \emph{Swift-}UVOT detector edge. For such systems, while \theTractor{} could conceivably still output an accurate source count rate (as long as a sufficient fraction of the PSF is on the detector to allow for vertical scaling), our use of aperture photometry on the 2D background image would lead to an underestimation of the background count rate. Hence, our photometry for these edge sources would be less accurate.

\begin{figure*}
    \centering
    \includegraphics[width=0.95\textwidth]{ 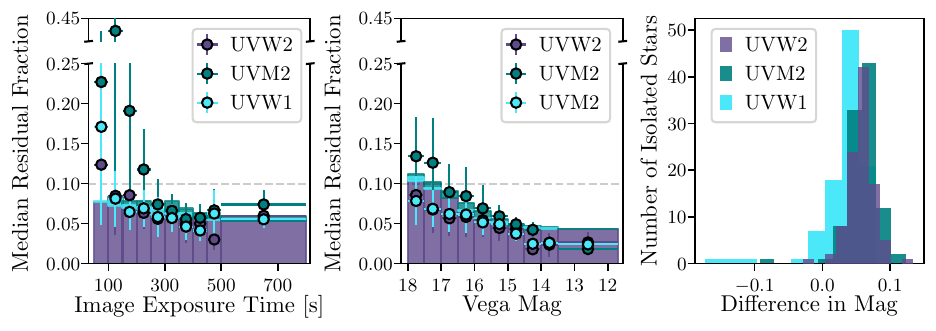}
    \caption{\emph{Left and Center:} Plots of the residual fraction (a metric to quantify the residuals between \theTractor{} model and original image relative to the brightness of a star being measured; defined in Section~\ref{sec:residual}) for a set of isolated stars in the SUMaC images. Circles represent the median residual fraction for stars in that bin, while error bars represent the median average deviation. The left panel shows the median residual fraction as a function of image exposure time, while the center panel shows the median residual fraction as a function of source magnitude. Overall residuals increase for both faint sources and low exposure time images. For reference, we also show the median photometric error, in magnitudes, returned from our photometry pipeline for the same set of sources (shaded regions). \emph{Right:} The difference between magnitudes computed by our \Tractor{} pipeline and the standard  \textit{Swift} routine {\tt uvotsource} for a set of 100 relatively isolated sources in each UV filter. Magnitudes from {\tt uvotsource} (which uses aperture photometry) are slightly brighter than those from \theTractor{}. However, the difference is small ($\sim$0.03--0.05 mag) and within the typical photometric uncertainty of both methods.
    }
    \label{fig:resid}
\end{figure*}

\subsection{Tractor Performance and Validation}

To our knowledge, this is the first time \theTractor{} has been applied to \emph{Swift-}UVOT images. Here, we examine several aspects of our pipeline outputs to assess the robustness of the photometry produced. Specifically, we examine the movement of sources from their initial positions within images, model residuals for isolated sources, and a comparison of our final calibrated magnitudes with the output of the standard routine \texttt{uvotsource}. Finally, we run simulations to estimate how well \theTractor{} performs when the PSF for multiple sources overlap.

\subsubsection{Astrometry and Movement of Sources within Images}\label{sec:source_movement}

As described in Section~\ref{sec:TractorSettings}, while we use the positions from MCPS as initial inputs into \theTractor{}, for our final catalog, we allowed the positions to move slightly. While this both decreased the prevalence of dipole-shaped residuals and lead to smaller residuals overall, it raises the possibility that some sources will move more significantly. In this case, it would be possible that the UV magnitude produced by our pipeline is not actually the counterpart to the input optical MCPS source. We therefore examine the distance that all sources move after running \theTractor{}.
We find that 80\% of sources stay within 0.05" (1/20th of a pixel), 91\% within 0.1", and 99\% within 0.3", and that relative directions were uniformly distributed (i.e., there was no overall systematic shift). These numbers are comparable to the RMS of the new astrometry solutions that we applied to the images (0.25$\arcsec$ on average)\footnote{This is calculated based on the documentation provided by astrometry.net here: \url{https://web.njit.edu/~gary/322/assets/Astrometry.net_uncertainties.pdf}} indicating that positions likely move due to residual WCS uncertainty. We will further discuss the small minority of sources which moved larger distances in Section~\ref{sec:catalog}.

\subsubsection{Model Residuals as a Function of Source Magnitude and Image Exposure Time}\label{sec:residual}

As described in Section~\ref{sec:psf}, we use a single ``idealized'' PSF for each \emph{Swift-UVOT} UV filter based on the curves-of-growth published within CALDB. However, the PSF shape is known to vary somewhat with source magnitude (due to coincidence loss) and the degree to which the PSF is sampled can vary with image exposure time. We therefore examine the residuals between the original \gls{SUMaC} images and our \Tractor{} models---both as a function of source magnitude and image exposure time---to assess the PSF performance.

We first select a set of ``isolated'' stars in each input image, defined as having no other sources in the output catalog within 15$\arcsec$. We then perform aperture photometry on the residual images (defined as the input image minus \theTractor{} model) within a 5$\arcsec$ aperture at the location of each source. We then divide this residual count rate by the total count rate assigned to that specific source (also measured within a radius of 5$\arcsec$), in order to compute what we call the ``residual fraction'' for each source. If the selected stars were truly isolated and the PSF fit equally well for all sources, this number would be expected to be a constant. A benefit of this definition is that if one assumes that all the residual counts should have been assigned to the source in question, then the residual fraction \emph{very roughly} corresponds to the shift in magnitude that would result from adding these counts back in
\footnote{Using the approximation $-2.5\log_{10}(1+X) \approx X$, which is valid to within $\lesssim$10\% for $X<0.5$, one can convert residual fraction to an approximate magnitude correction. This is meant as an interpretive aid only, as it does not account for effects such as coincidence loss (\S~\ref{sec:hea}), or any potential flux arising from undetected stars below our detection threshold.}.

In Figure~\ref{fig:resid} we plot the results from this assessment. In the left panel, we plot the median residual fraction for a set of $\sim$1500 isolated stars (per filter) with magnitudes 
brighter than 16.8 Vega mag as a function of image exposure time. In general, these values range from $\sim$0.05-0.08, but can be seen to diverge to higher values for images with low exposure times. This is likely due to the fact that sources in low exposure time images contain fewer counts and thus may not fully sample the PSF described by the curves of growth. For the UVW1 and UVW2 filters, this divergence appears mainly in $\lesssim$150s images. In contrast, for UVM2, residuals begin to grow for images with exposure times $\lesssim$250s and reach much higher levels than the other two filters.

In the middle panel of Figure~\ref{fig:resid} we again plot the median residual fraction for isolated stars, but as a function of source magnitude. Here, we limit ourselves to sources in images with exposure times of at least 150s. The main visible trend is a slow increase in residual fraction when moving to fainter source magnitudes. For sources brighter than 18 Vega mag (the limit where we have at least 25 isolated sources per filter per bin), these residual fractions remain below 0.1 and 0.15 for the UVW1/UVW2 and UVM2 filters, respectively. For comparison, we also plot the median magnitude errors for these same sources as shaded histograms. They follow a very similar trend, increasing at fainter source magnitudes.

In conclusion, although the \emph{Swift-}UVOT PSF model fits less well for sources with low numbers of counts, we do not find evidence for a strong systematic effect beyond what is included in the reported magnitude errors, except for images with very short exposure times (see also \S ~\ref{sec:isolatedtest} below). In particular, we do not observe any increase in residuals for bright sources in our catalog, as would be expected for a narrowing of the PSF due to coincidence loss. This indicates that most stars in our catalog fall below the count rate, where this is a significant issue.

\begin{figure*}
    \centering

    \includegraphics[trim={0.5cm 0cm 0cm 0cm},clip,width=0.97\textwidth]{ 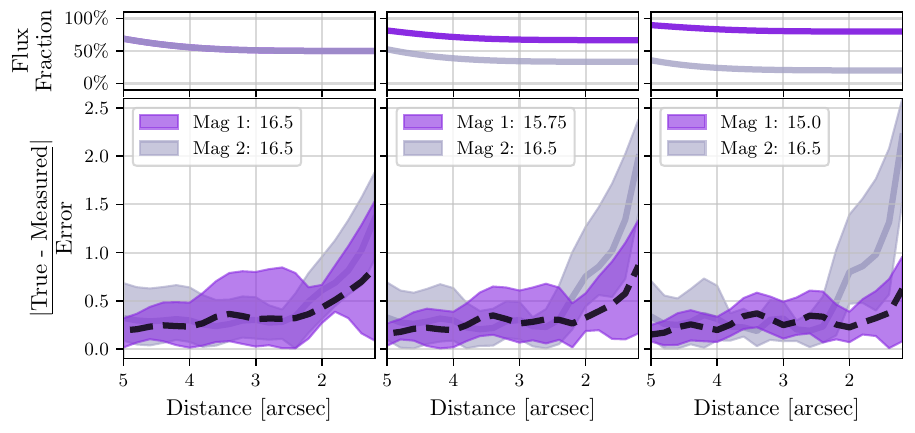}
    \caption{
    Absolute difference between \theTractor{}-measured and true (injected) count-rates, weighted by the reported uncertainty, as a function of decreasing separation between two simulated stars. The three panels correspond to different brightness contrasts between the two injected stars: equal brightness (left), a difference of 0.75 mag (middle), and a larger difference of 1.5 mag (right). The top row shows the ``flux fraction'' for each star, defined as the fraction of flux within a 5$\arcsec$ aperture centered on the star that is attributed to it by \theTractor{}. In general, \theTractor{} is able to accurately recover magnitudes for sources with overlapping PSFs, although some caution is warranted for sources with close neighbors ($\lesssim$1.5$\arcsec$) that are not the dominant sources of light in the region (see Section~\ref{ref:crowding} for further details).}
    \label{fig:cluster}
\end{figure*}

\subsubsection{Comparison of Calibrated Magnitudes from The Tractor with {\tt uvotsource} for Isolated Stars}\label{sec:isolatedtest}

To derive magnitudes from \theTractor{} output, we wrote a custom script to implement most of the HEASARC calibration tasks. Given this, and the fact that some residuals are present between our final \Tractor{} models and the input \gls{SUMaC} images, we wish to test the efficacy of our entire pipeline. To do so, we compare our final \Tractor{} magnitudes with results from running the HEASARC task \texttt{uvotsource} on the ``isolated'' sources within the Magellanic Clouds described above (\S~\ref{sec:residual}).  

{\tt uvotsource} performs aperture photometry on a single source within a UVOT image. As input, it requires two region files: one for the source aperture and one that is used to measure the background count rate. To ensure consistency between the two photometric methods, we take a 5$\arcsec$ source aperture centered at the final \Tractor{} pixel positions. In addition, we test a range of possible background regions, ultimately selecting the region that has a background count rate closest to what was used for that source by our pipeline (which was found by performing aperture photometry on our 2D background images as described in \S~\ref{sec:hea}). 

In the right panel of Figure~\ref{fig:resid} we plot the difference between our final \Tractor{} photometry and \texttt{uvotsource} for a set of 100 isolated stars in each UV filter. In general, our \Tractor{} pipeline produces magnitudes that are slightly fainter than \texttt{uvotsource}. However, these differences are small, with median values of 0.04-0.05 mag. We do not apply any systemic shift to our photometry measured by \theTractor{} to ``correct'' for this shift, as it is possible that these offsets are simply caused by the inclusion of some light due to nearby sources in the aperture photometry performed by \texttt{uvotsource}. In addition, we note that these offsets are all within the 1 sigma errors computed by \theTractor{} for these sources, which range from 0.05--0.1 mag. 

\subsubsection{Accuracy of Flux Recovery by The Tractor for Sources with Overlapping PSFs}\label{ref:crowding}

%\vspace{-0.5cm}
While we establish above that our pipeline performs well for isolated sources, a strong motivation for using \theTractor{} was the ability to disentangle flux from nearby sources. However, while the PSF fitting provides significant improvement over aperture photometry, degeneracies can still exist when distributing light within the image to sources with very small separations. This is particularly true when the PSF is slightly undersampled, as is the case for the \gls{SUMaC} images (which have 1$\arcsec$/pixel and $\sim$2.5$\arcsec$ FWHM PSFs). 
It is therefore essential to identify at what point crowding impacts the ability of \theTractor{} to accurately recover the flux from sources with overlapping PSFs. 

To understand this, we create a series of simulated \emph{Swift-}UVOT images where we inject two stars with known count rates, progressively move them closer together on the sub-pixel level, and then run our \Tractor{} pipeline to test how well the known count rates are recovered. We create our mock stars using our model for the PSF, scaled by the input count rate. We create a model for the noise of a typical \gls{SUMaC} image by finding a relatively blank area in an arbitrary image and randomly sampling from a Gaussian distribution with the same mean and standard deviation. As the noise is random, we generate 5 distinct noise models to add to our images.

In total, we run three series of tests using these simulated images. In the first, we inject two stars with the same input count rate. In the second and third, the count rates are varied to correspond to stars that have magnitudes that differ by 0.75 mag and 1.5 mag, respectively. 
For each test, we start by injecting the two stars 5$\arcsec$ apart. We then move them closer together in steps of 0.25$\arcsec$ until they are only 1.0$\arcsec$ apart. For each separation, we create 5 images using the 5 distinct noise models and run \theTractor{} on each.

The results of these tests are shown in Figure~\ref{fig:cluster}. In each panel, we plot the absolute deviation between the input and \theTractor{}-modeled count rates, normalized by the Poisson error. A value greater than 1.0 on the vertical axis indicates that \theTractor{} result differs from the true value by more than 1$\sigma$. While we plot only the magnitude of the deviation, in practice, the modeled count rate may be either an under- or overestimate of the true value. For each star, the black line shows the average result across five trials at each separation, while the shaded region reflects the variability between trials, driven by differences in the input noise model.

In all three panels, the results for both stars are relatively flat, and \theTractor{} output agrees well with the true injected value down to a separation of at least 2.5$\arcsec$. For separations below this value, \theTractor{} measurements begin to diverge from the true values. However, in all cases, the brighter star stays within $\sim$1 sigma of the true value down to a separation of $\sim$1$\arcsec$ (the pixel scale of the \emph{Swift}-UVOT images.)
In contrast, we see that \theTractor{} measurements of the fainter stars in our second and third trials grow to deviate by $\sim$1.5--2.5 sigma from their true values once they reach separations of only 1$\arcsec$. 

\emph{Some caution is therefore warranted when using the outputs of our pipeline for stars with very close neighbors, if they are not the dominant source of light in the region.} We further quantify this in the top panels of Figure~\ref{fig:cluster} where we plot a quantity that we call the ``flux fraction'' for both stars. We define this quantity to be the fraction of the flux within a 5$\arcsec$ radius centered on a given object that is actually due to the source. If a star were truly isolated, this quantity would be 1. 
The flux fraction decreases when a non-negligible portion of a nearby source’s PSF overlaps with the target position. In the first trial, with two equally bright sources, both stars have flux fractions of $\sim$50\% when separated by less than 1$\arcsec$. 
By contrast, in the third trial, where the two sources differ by 1.5 magnitudes, the fainter star has a flux fraction of only $\sim$20\% at a 1$\arcsec$ separation. 
We emphasize that a low flux fraction does not inherently indicate unreliable photometry. For example, in the third trial, the fainter source already has a flux fraction of only $\sim$21\% at a 3$\arcsec$ separation---yet \theTractor{} still recovers accurate magnitudes for both stars (right panels, Figure~\ref{fig:cluster}). 
However, when a source has both a low flux fraction \emph{and} a very close neighbor, additional caution is warranted, as the reported magnitude may deviate from the true value by several times the quoted uncertainty.

In the final catalog, we provide a number of quantities to help assess whether an individual source has nearby neighbors and whether it is the dominant source within it local region (see \S~\ref{sec:finalcontents} and Appendix~\ref{sec:caveat}).

\section{Final Swift UVOT Catalog of the Magellanic Clouds}\label{sec:catalog}

After running our pipeline on all the input \gls{SUMaC} images, we assemble the newly computed photometry into a final catalog. Here we describe the process we use to combine multiple observations of the same source and the contents of the final catalog. A summary of caveats and limitations relevant for general users is presented in Appendix~\ref{sec:caveat}. 

\subsection{Initial Quality Cuts}\label{sec:qualitycuts}

\TheTractor{} outputs a best-fit count rate for every input source, whether or not it is detected at a significant level. Before we combine multiple observations of the same source from different images in the same filter, we perform a number of initial quality cuts: 

\begin{itemize}
    \setlength\itemsep{-0.1em}
    \item We drop measurements where \theTractor{} inferred a negative count rate or where the final calibrated magnitude error is greater than 0.36 magnitudes (indicating that the source was not detected at $>$3$\sigma$).\footnote{Calculated by applying standard error propagation to $m=-2.5\log_{10}{F}$ and taking $F=3\sigma_F$.}
    \item In \S\ref{sec:hea} we describe both the HEASOFT generated small-scale sensitivity (SSS) flag and an edge flag for sources within 5$\arcsec$ of the edge of the detector. We use these flags to remove measurements located in these problematic areas of the detector.
    \item We remove sources where the final \Tractor{} position is more than 1$\arcsec$ from the initial input coordinates (see  \S~\ref{sec:source_movement}). As this is larger than the typical uncertainties in the WCS solutions, this could indicate a mismatch between the UV source being modeled by \theTractor{} and the original input MCPS source.
    \item In \S\ref{sec:residual} we describe a quantity we call the ``residual fraction''---calculated by dividing the count rate left in the residual image by the count rate assigned to a given source. We remove measurements with residual fractions greater than 0.3 (i.e., an equivalent of more than 30\% of the counts assigned to the source is left in the residual image). This may indicate undetected tracking issues that alter the PSF shape or other problems leading to a systematic uncertainty the in the final magnitude. On average, approximately 25\% of sources are removed from a given image due to this cut, with the percentage increasing as the exposure time decreases. However, since source measurements are based on multiple images, only about 20\% of sources are completely excluded from the catalog for this reason.
\end{itemize}

\subsection{Combining Multiple Measurements of  a Single Source}\label{sec:avg}

After our initial cleaning of the data, each source in our catalog has between 1 and 25 measurements of the source magnitude, with an average of 3.5 per UV filter. Between 24\% and 27\% of the stars observed in a given UV filter had only a single $>$3$\sigma$ detection. However, only 2\% of sources in the catalog had only one observation in all three UV filters.

We combine these source magnitude measurements by taking a weighted mean after outlier rejection. To identify outliers, we use a formalism based on the median absolute deviation (MAD) of each set of observations. In particular, we calculate a modified Z-score for each measurement of a particular source in a given filter as: $0.6745(x_i - \tilde{x})/MAD$, where $x_i$ is an individual measurement and $\tilde{x}$ is the median of a set of measurements. Following \citet{iglewicz1993detect}, we remove measurements with scores of 3.5 or higher as likely outliers. The vast majority, or 99.9\% of our sources, have 2 or fewer outliers, with 86.6\% having none at all.\footnote{We note that only sources with 3 or more measurements will return a possible outlier under this prescription. Indeed, the $\sim$13\% of source/filter combinations that have outliers removed have an average of 4.5--5 measurements.} 

To construct weights for the weighted mean, we take into consideration both the measurement errors on the calibrated magnitude, and the residual fraction of each observation, as the latter can probe systematics specific to individual images. Specifically, we calculate our weights by taking the inverse of the residual fraction and magnitude error added in quadrature. With these, we calculate (i) the weighted average of the source magnitude, (ii) an error on this average magnitude by propagating the individual measurement uncertainties in quadrature, and (iii) the weighted standard deviation of the measurements used to calculate the average.

\subsection{Comparison to UV Photometry Presented in \citetalias{Drout2023}}\label{sec:photom-comp}

As described in Section~\ref{sec:intro}, a preliminary version of the photometry pipeline presented in this manuscript was used in \citetalias{Drout2023} to select targets and verify the presence of a UV excess in their spectroscopic sample of stripped stars. However, several updates have been made since. 

\begin{figure}
    \centering
    \includegraphics[width=\columnwidth]{ 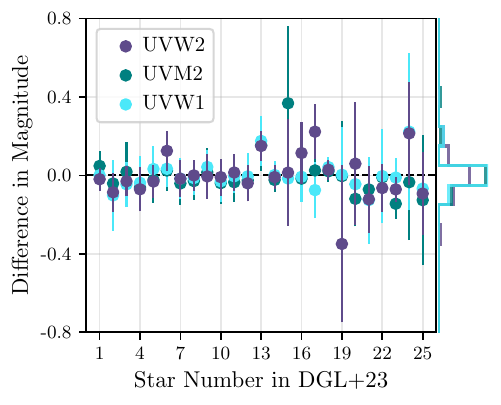}
    \caption{The difference between the photometry in this work and that of \citetalias{Drout2023} (scatter points). Error bars shown represent the 1 sigma uncertainty from each publication added in quadrature. While there are some small differences, they agree within errors. The median difference for all stars is $-0.01$ mag in all three UV filters. See Section~\ref{sec:photom-comp} for further discussion. }
    \label{fig:comparison}
 \end{figure}

In particular, the original target selection in \citetalias{Drout2023} used a version of the UV photometry that was only computed on a single image for each \gls{SUMaC} field and where source positions were not allowed to move when running \theTractor{}. While \citetalias{Drout2023} then recomputed new UV photometry for the final spectroscopic sample of 25 objects where source positions were allowed to vary and multiple observations were averaged together, our final pipeline described here also uses: (i) a different process to reject certain images as poor quality (\S\ref{sec:inputimages}) and (ii) a different weighting scheme and outlier rejection to average together multiple observations (\S\ref{sec:avg}). As a result, the UV photometry in our final catalog varies somewhat from that presented in \citetalias{Drout2023}.

In Figure~\ref{fig:comparison}, we plot the difference between the photometry in our final UV catalog and that of \citetalias{Drout2023}. Points are  color-coded by the band of the photometry and ordered by their ``star number'' in \citetalias{Drout2023}. In general, most of the photometry agrees to within $\lesssim$0.1 mag (with a median offset of $\sim-0.01$ mag). 
While there are some individual cases of stars/filters with larger differences ($\gtrsim$0.2 mag), these generally correspond to stars that had larger standard deviations between observations and thus had correspondingly higher quoted errors (e.g., Star 15, Star 17, Star 19, Star 24). 
In only one case (Star 13) is the photometry presented here and in \citetalias{Drout2023} discrepant at a level higher than 1$\sigma$. 
However, even in this case, the photometry difference is relatively small ($\lesssim$0.15 mag) and does not impact our interpretation of the source as a UV excess system (see Section~\ref{sec:compD23} where we further discuss our photometric classification of all the sources originally presented \citetalias{Drout2023}).

 \begin{figure}
    \centering
    \includegraphics[width=0.47\textwidth]{ 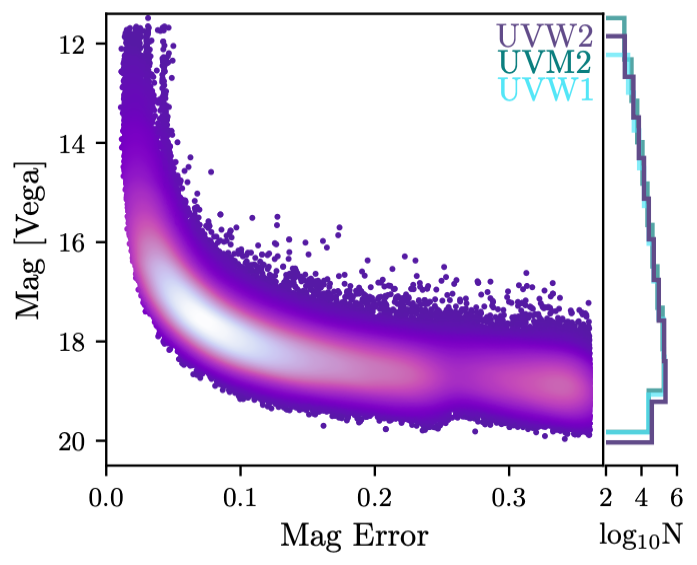}
    \caption{Summary of the source magnitude and associated magnitude errors present in the SUMS UV catalog. The density plot shown was created by randomly sampling the catalog (for both the LMC and SMC and all three UV filters) 100,000 times. Our catalog spans a range of Vega UV magnitudes from roughly 20 to 12 ($\sim$21.7-13.5 in AB magnitudes) with the fainter magnitudes typically having higher errors. Histograms showing the number of sources as a function of magnitude are shown on the right.}
    \label{fig:magerr}
\end{figure}

\subsection{Final Catalog Contents}\label{sec:finalcontents}

The completed SUMS UV photometric catalog contains \NSourcesLMC{} sources in the LMC and \NSourcesSMC{} in the SMC. In Figure~\ref{fig:magerr} we plot UV magnitude versus magnitude error for the catalog, which spans 20th to 12th Vega magnitudes. We make this full catalog, as well as the pipeline used to measure the UVOT magnitudes with \theTractor{}, publicly available for the community. Here we describe the contents (magnitudes, quality flags, etc.) included in the full catalog, a few samples rows of which are presented in Table~\ref{tab:minifullcat}. All photometry in the full catalog is presented in Vega magnitudes (the standard output from the UVOT calibration routines). 

\begin{deluxetable*}{lllllllllll}
     \tabletypesize{\footnotesize}
      \caption{The Stripped-Star Ultraviolet Magellanic Cloud Survey Catalog}
      \label{tab:minifullcat}
    \tablehead{
        \colhead{SUMS\_ID} & \colhead{R.A.} & \colhead{Dec.}& \colhead{UVW2}& \colhead{UVW2\_err}& \colhead{UVM2}& \colhead{$\hdots$}& \colhead{ UVW1\_n2p5}& \colhead{UVW2\_nobs}& \colhead{UVM2\_nobs}& \colhead{ UVW1\_nobs} \\ 
        \colhead{}& 
        \colhead{\scriptsize[$\degree$]} & \colhead{\scriptsize[$\degree$]} & 
        \colhead{\scriptsize[Vega Mag]}& 
        \colhead{\scriptsize[Vega Mag]}& \colhead{\scriptsize[Vega Mag]}& 
       \colhead{}&\colhead{}&\colhead{}&\colhead{}&\colhead{}}
	\startdata
SUMS$\_$cb289 & 17.084715 & -72.264870 & 18.29 & 0.09 & 18.26 & $\hdots$ & 0 & 6 & 3 & 5 \\
SUMS$\_$cbc8c & 17.573010 & -72.166890 & 19.15 & 0.17 & 19.02 & $\hdots$ & 0 & 2 & 1 & 1 \\
\vdots & & & & & & & & & & \vdots\\
SUMS$\_$189b5 & 75.533445 & -70.477690 & 12.62 & 0.03 & 12.55 & $\hdots$ & 0 & 3 & 4 & 3 \\
SUMS$\_$18b73 & 75.583170 & -70.516640 & 13.68 & 0.03 & 13.7 & $\hdots$ & 1 & 3 & 4 & 3 \\
 	\enddata
    \tablenotetext{\dagger}{This table provides an illustrative subset of our UV photometric catalog of the Magellanic Clouds. The complete catalog is accessible in a machine-readable format in the online journal and on VizieR.}
\end{deluxetable*}

\begin{enumerate}[leftmargin=1em]
    \setlength\itemsep{-0.1em}
    \item \textbf{SUMS\_ID:} A hexadecimal code to assist with target identification and cross-matching. 
    \item \textbf{R.A., Dec.:} Coordinates in degrees provided by the MCPS catalog. 
    \item \textbf{UVW2, UVM2, UVW1:} Vega magnitudes in the three \textit{Swift}-UVOT filters. Multiple observations of the same source are measured individually and combined as described in \S \ref{sec:avg}. 
    \item \textbf{\{UVW2,UVM2,UVW1\}\_err:} Statistical errors on the UV magnitudes. For sources with a single observation in a given filter, these are the errors returned by the \textsc{heasarc} routine \texttt{uvotflux} in \S \ref{sec:hea}. If more than one observation of a source occurred, this is the weighted average of these errors, calculated as described in \S \ref{sec:avg}. 
    \item \textbf{U, B, V, I}: The optical photometry
    obtained from the MCPS catalog. 
    \item \textbf{\{U, B, V, I\}\_err}: The associated magnitude errors from the MCPS catalog. 
    \item \textbf{\{UVW2,UVM2,UVW1\}\_std:} Weighted standard deviation of the mean for sources with more than one observation in a given UV filter, described in \S~\ref{sec:avg}
    \item \textbf{\{UVW2,UVM2,UVW1\}\_flux\_frac:} The fraction of  \theTractor{} modeled flux within a 5$\arcsec$ radius around a source that is attributed to the source (first described in \S\ref{ref:crowding}). We calculate this by taking the magnitudes of all sources modeled by \Tractor{} that located within 10$\arcsec$ of the source of interest and determining the fraction of their PSF that overlaps with a 5$\arcsec$ aperture around the source of interest. If \theTractor{} does not model any sources within 10$\arcsec$ of a source, then the flux fraction listed in the catalog is 1. 
    \item \textbf{\{UVW2,UVM2,UVW1\}\_resid\_frac:} The ``residual fraction'' for a given source, calculated as described in \S\ref{sec:residual}. This is the ratio of the counts remaining in the residual (image - model) image divided by the total number of counts assigned to the source, both measured within a 5$\arcsec$ radius.
    \item \textbf{\{UVW2,UVM2,UVW1\}\_dist\_moved:} The average distance in arcseconds \theTractor{} moved a source to best fit the flux as described in \S \ref{sec:source_movement}. 
    \item \textbf{\{UVW2,UVM2,UVW1\}\_dist\_neighbor:} The distance in arcseconds from the source to its nearest neighbor, \emph{that was also detected} in a  \textit{Swift}-UVOT image. We note that because: (i) sources can appear on different parts of the \textit{Swift}-UVOT detector between independent observations, and (ii) a particularly faint neighbor may not be detectable across images with different exposure times, there may be slightly different neighbors detected in different images and/or filters. The number provided is the smallest distance to the nearest neighbor found in any image for a given filter.
    \item \textbf{\{UVW2,UVM2,UVW1\}\_n5:} 
    The maximum number of neighboring sources within 5\arcsec of the source in any image for a given filter. As described above, various factors can lead to a slightly different number of neighbors being detected in any given image. We therefore take the maximum to include all possible neighbors. 
    \item \textbf{\{UVW2,UVM2,UVW1\}\_n2p5:} The maximum number of neighboring sources within 2.5\arcsec of the source, as described above.
    \item \textbf{\{UVW2,UVM2,UVW1\}\_nobs:} The total number of observations that were included in the measurement of the source. 
\end{enumerate}

\begin{figure*}
    \centering
    \includegraphics[trim={0.cm 0.4cm 0.1cm 0.cm},clip,width=0.95\textwidth]{ 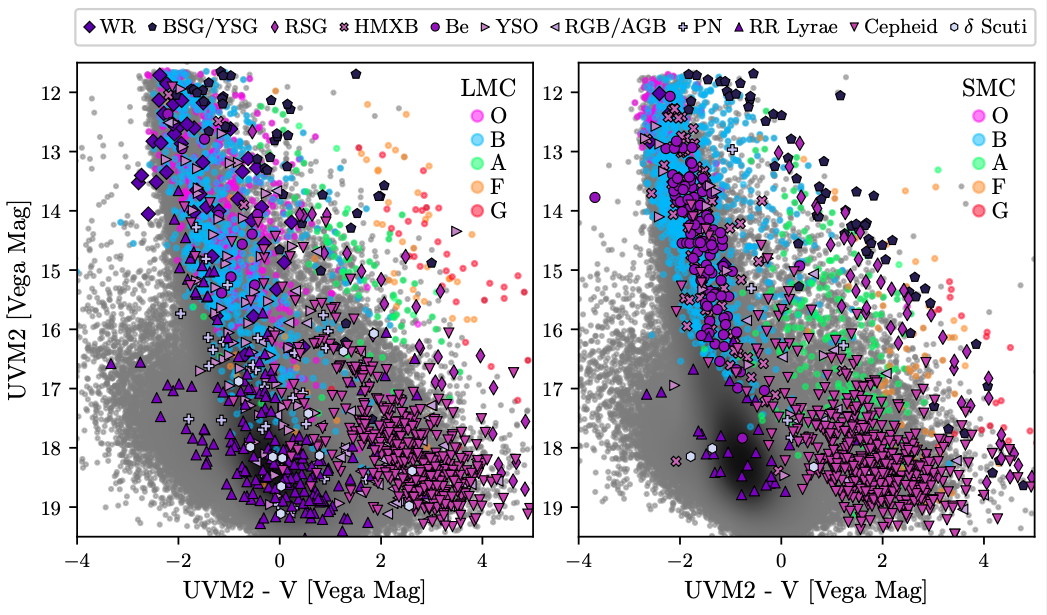}
    \caption{UV-optical CMDs for the full SUMS UV photometric catalog of the LMC (left) and SMC (right). When plotting, we have cross-matched to the SIMBAD database and removed obvious non-members of the Clouds based on a combination of listed object types (from SIMBAD) and parallax/proper motion measured from \emph{Gaia} (see Section~\ref{sec:diverse}). Selected object types, plotted as various purple shaded markers, include Wolf-Rayet (WR), Blue and Yellow Supergiants (BSG/YSG), Red Supergiants (RSG), High Mass X-ray Binaries (HMXB), Be stars (Be), Young Stellar Objects (YSO), Red and Asymptotic Giant Branch stars (RGB/AGB), Planetary Nebulae (PN), Post AGB stars (p-AGB), RR Lyraes, Cepheids, and Delta Scuti Variables. We plot some of the stars without these specific object types by their spectral type instead, shown here as brightly colored scatter points.}
    \label{fig:sums}
\end{figure*}

For a summary of the caveats and limitations to be aware of when using this catalog, please refer to Appendix~\ref{sec:caveat}.

\subsection{Diverse Populations in the Full UV Catalog}\label{sec:diverse}

The complete SUMS catalog contains \NSources{} sources. Of these, we find that only $\sim$33,000 sources ($\sim$4.5\% of the catalog)  match within 1$\arcsec$ of a source with any classification in the SIMBAD astronomical database \citep{Wenger2000}. However, even this small fraction reveals the rich variety of astrophysical populations found in the SUMS catalog. 
In Figure~\ref{fig:sums} we present a UV-optical \gls{CMD} for each galaxy using Vega magnitudes without an extinction correction. 

In this figure, we exclude obvious non-members of the Clouds. 
This includes 7,542 sources with either \emph{Gaia} parallax detected at greater than 4$\sigma$ or \emph{Gaia} proper motions inconsistent with the Clouds (with the latter determined via a method discussed further in \S~\ref{sec:member} and Appendix~\ref{ap:gaia-chi2}) as well as 23 sources classified as quasars or galaxies in SIMBAD. 
However, determining true membership for each source/population, such as using distances derived from the variability of RR Lyrae or Cepheid stars, is beyond the scope of this work. 
Photometry for all sources---regardless of their membership or possible non-membership in the Clouds---are included in the published catalog.

Sources with particular object types in SIMBAD are plotted as various purple scatter points. For many of the populations explicitly labeled, there are between $\sim$50-200 known sources in our catalog with those classifications. Exceptions include RR Lyrae and Cepheid variables with 441 and 2,423 known sources in our catalog respectively. Stars that do not fall into one of the specific categories labeled, but have  spectral types listed in SIMBAD are indicated as colored circles.

As evidenced from Figure~\ref{fig:sums}, the SUMS UV photometric catalog holds great potential for studying a wide range of astrophysical phenomena. 
Interestingly, there is also a notable absence: if we consider the nearly-vertical over density of stars (which coincides with many of the previously classified O$-$ and B$-$type stars) to be a rough identification of stars on the main sequence, then most of the systems in the SUMS catalog with bluer UV-optical colors than main sequence stars of similar brightness have no information in the SIMBAD database. The main exceptions are a handful of WR and RR Lyrae stars, which will be discussed later in Section~\ref{sec:otherblue}.
This suggests an understudied population of stars, possibly overlooked due to optical faintness, yet prominent in the UV.

\section{Selection of Candidate Stripped Star Systems}
\label{sec:Candidates}

Our primary aim when performing UV photometry on the \gls{SUMaC} images was to use the resulting UV catalog to identify candidate 
stripped stars
based on the UV excess they can introduce to the \glspl{SED} of their systems \citepalias[e.g. as demonstrated in][]{Drout2023}. In this section, we use the 
\glspl{SED} of the \NSources{} sources in our final catalog to select a set of candidate systems. Our 
approach is the same as outlined in \citetalias{Drout2023} in that we select systems that appear to be bluewards of the \gls{ZAMS} in multiple UV-optical \glspl{CMD}. We apply a set of physically motivated cuts to try to ensure that selected objects appear bluewards of the \gls{ZAMS} due to their actual SED shape as opposed to data quality issues. In the subsections below we discuss (i) the various stellar and spectral models used to guide our candidate identification (\S~\ref{sec:models}) (ii) our baseline choices for properties such as distance and extinction (\S~\ref{sec:extinction}), (iii) our preliminary identification of stars bluewards of the ZAMS and subsequent quality cuts (\S~\ref{sec:bluewards}--\ref{sec:morequalitycuts}), (iv) assessment of membership of our candidates in the Magellanic Clouds (\S~\ref{sec:member}), (v) candidate rankings and comparison to the stripped star sample of \citetalias{Drout2023} (\S~\ref{sec:rankings}--\ref{sec:compD23}) and (vi) 
the impact of our approach to extinction on the number of candidates we identify (\S~\ref{sec:extinction_on_candidates}). 

Throughout, we emphasize that our goal was to simply select a set of high-quality candidate systems to motivate future follow-up. Use of these systems for quantitative analysis of the rates or prevalence of stripped stars will require both confirmation of the nature of the sources and assessment of the catalog's completeness and various observational biases. These will each be the subject of future work.  

Although our final catalog is presented in Vega magnitudes, we choose to transform to the AB magnitude system when performing our candidate selection, as we found that this allows for a more intuitive assessment of SED shape and quality.\footnote{In the AB system (unlike Vega) the relative magnitudes between two bands are directly proportional to their relative fluxes.} To determine the conversion factors, we use the zero points provided by the Spanish Virtual Observatory filter service. \citep{Rodrigo2020}\footnote{\url{http://svo2.cab.inta-csic.es/theory/fps/}} The conversion factors ranged from $<$0.01 mag (V$-$band) to $\sim$1.7 mag (UVW2) and are listed in  Appendix~\ref{ap:tables} Table~\ref{tab:conversions}. 

\subsection{Stellar Evolution and Spectral Model Benchmarks}\label{sec:models}
Throughout this manuscript, we use a number of different stellar evolution and spectral models to both identify a set of candidate stripped stars and to characterize their properties. Here we summarize these models for reference.

\subsubsection{Models of Stripped Stars}  \label{sec:strippstarmodels}
To assess the expected photometric properties of stripped stars in the LMC and SMC, we perform synthetic photometry on the spectral models from \citetalias{Drout2023} and \citet{Gotberg2018}.
Specifically, we use the models with $Z=0.006$ from \citetalias{Drout2023} for the LMC and with $Z=0.002$ from \cite{Gotberg2018} for the SMC. These models were both generated with the on-LTE radiative transfer code CMFGEN \citep{1990A&A...231..116H, 1998ApJ...496..407H} based on the properties of the evolutionary models produced by \cite{Gotberg2018} with the MESA stellar evolution code \citep{Paxton2011,Paxton2013,Paxton2015, Paxton2018,Paxton2019,Jermyn2023}. {Evolutionary models were computed for an assumed solar metallicity of $Z_{\odot} =0.014$ \citep{2009ARA&A..47..481A}.} Spectral models were computed for stripped stars with masses of 0.37--7.23 M$_{\odot}$ (which were produced from stars with initial masses between 2.0 and 18.2 M$_{\odot}$). The stripped star spectra were computed using properties of the MESA models roughly halfway through the central helium burning, when $X_{\rm He, center}=$0.5. 

While \cite{Gotberg2018} also presented models for the LMC metallicity ($Z=0.006$), the models from \citetalias{Drout2023} were recomputed at lower mass loss rates (to more closely align with the spectral morphology of the observed population).
Ultimately, the difference is relatively minor for broadband photometry: the mean absolute difference in the $Z=0.006$ models with different mass loss rates is only 0.03 mags \citepalias[][]{Drout2023}.
While models with lower mass-loss rates have not yet been computed for SMC metallicity, the wind features present in the original models of \citet{Gotberg2018} are weaker than for the LMC metallicity. Impact on the broadband photometry should therefore be minimal. 

\subsubsection{Models of Binary Stripped Stars with \gls{MS} Companions}\label{sec:compositemodels}

To further describe the expected photometry of observed stripped stars, we place them in binaries with \gls{MS} companions. To accomplish this, we use two different composite grids of stripped stars plus \gls{MS}\footnote{{The MS models used here do not include the effects of binary interaction such as mass accretion. This caveat is further mentioned later in \S\ref{sec:SED_fitting}}.} models.

\textbf{\emph{Grid from \citetalias{Drout2023}:}} First, we use the grid of stripped star plus \gls{MS} star composite models presented in \citetalias{Drout2023}. These were constructed by combining the 0.37--7.23 M$_{\odot}$, $Z=0.006$, core-helium burning stripped star models described in \S~\ref{sec:strippstarmodels} with a new set of CMFGEN models for MS stars. The spectral models for MS stars were computed using the surface conditions of the \citet{Gotberg2018} MESA stellar evolution models as input, but now adopting the conditions during the main-sequence evolution. They were calculated for stars with initial masses of 2.21--18.17 \Msun\ and $Z=0.006$ at three evolutionary phases: 20\%, 60\%, and 90\% through the duration of the MS phase (in terms of time). While \citetalias{Drout2023} provide full spectral models for this grid, in this manuscript we utilize the broadband photometry.

\textbf{\emph{Grid Using MIST MS Stars:}} As noted above, the composite grid of \citetalias{Drout2023} only extends to \gls{MS} companions of $\sim2M_\odot$. However, our photometric candidates may contain objects with even lower mass \gls{MS} companions. We therefore create an additional model grid, using models from the MESA Isochrones and Stellar Tracks \citep[MIST,][]{Dotter2016,Choi2016,Paxton2011,Paxton2013,Paxton2015}\footnote{\label{footnote:mist}\url{https://waps.cfa.harvard.edu/MIST/}} to expand the range of \gls{MS} companions. In particular, we take MIST models with initial rotation of v/v$_{\rm{crit}}$ $=$ 0.4 and metallicity of Fe/H=[-0.37,-0.85] for the LMC and SMC, respectively (corresponding to Z=0.00{6} and Z=0.002 for $Z_{\odot} = 0.0142$). We use stars with masses from 1.0--12.4 M$_\odot$ in steps of 0.2 M$_\odot$, and take their properties when they are 20\% through the duration of core hydrogen burning.
To construct our new composite grid, we take the broadband photometry provided by MIST for these \gls{MS} stars and combine them (in flux space) with the broadband photometry for the LMC and SMC stripped star models described in \S~\ref{sec:strippstarmodels}. To provide broadband photometry for their MESA stellar evolution models in the temperature range of the MS stars used in our grid, the MIST team uses ATLAS model atmospheres \citep{Kurucz1992}.

\subsubsection{Theoretical Model of the Zero-Age Main Sequence}\label{sec:zams} 
In the subsections below, we identify stripped star candidates by selecting systems that appear bluewards of the
\gls{ZAMS} in multiple UV-optical \glspl{CMD}. 
The location of the \gls{ZAMS} is described using the same theoretical models as in \citetalias{Drout2023}(see their Section S1.3.2). In brief, we perform synthetic photometry on a set of Kurucz ATLAS model atmospheres \citep{Kurucz1992} whose temperatures and bolometric luminosities match predictions for the ZAMS in the MESA stellar evolution models computed by \cite{Gotberg2018}. We compute this for stars with masses between 2 and 100 \Msun\ and use the evolutionary models with $Z = 0.006$ to represent the LMC and with $Z=0.002$ to represent the SMC. Overall, the shape of these two ZAMS are similar, but the lower metallicity version is shifted $\sim$0.1 mag to the blue in a variety of UV-optical CMDs. We will further examine the impact of the choice of metallicity on the location of the ZAMS (and hence our candidate selection) in Section~\ref{sec:ms}. 

\subsection{Distances and Reddening Corrections}\label{sec:extinction}

Before comparing our observed magnitudes to the theoretical ZAMS described in the previous section, it is necessary to correct for both distance and reddening. For the former, we adopt 50 kpc for the LMC \citep{Pietrzynski2013} and 60.6 kpc for the SMC \citep{Hilditch2005}.\footnote{Formally, both the LMC and SMC have depth along the line-of-sight. The LMC is broadly consistent with a face on disk \citep{Gaia2021} while the SMC is more complex, potentially even resulting from two star-forming structures superimposed along the line-of-sight \citep{Murray2024} However, in both cases, differences in distance modulus amongst young stars are expected to be a few tenths of a magnitude or less \citep[e.g.][]{Subramanian2009}.}  For the latter, we again follow \citetalias{Drout2023} and do not attempt to correct for extinction on a star-by-star basis at this stage. Instead, we adopt a single ``mean'' extinction value for each galaxy: \Av{}$= 0.38 $ mag for the LMC and \Av{}$= 0.22 $ mag for the SMC (see below for the rational of these values). To determine the bandpass-specific extinction corrections that are associated with these \Av{} values (i.e., A$_{\rm{X}}$/A$_{\rm{V}}$, where X are each of our observed UV-optical filters) we perform synthetic photometry on the set of stripped star models described in \S~\ref{sec:strippstarmodels} both before and after applying the mean extinction values listed above and calculate the magnitude difference. For this analysis, we use the ``average LMC'' and ``SMC bar'' extinction curves from \cite{Gordon2003}. The correction factors for the   \emph{Swift}-UVOT UV and MCPS optical filters, which are broadly applicable to spectra resembling those of hot stars, are listed in Appendix~\ref{ap:tables}, Table~\ref{tab:conversions}.

Motivation for these specific choices of \Av{} value was described in the Supplementary Materials of \citetalias{Drout2023}. However, as these factors have a \emph{significant} impact on our sample of stripped star binary candidates, we both summarize and elaborate on these choices here. The impact of these choices on our final candidate list is then discussed in more detail in Section~\ref{sec:extinction_on_candidates}.

\textbf{\emph{Choice of \Av{}:}} When choosing \Av{}, we first investigated using the mean values from \cite{Zaritsky2002,Zaritsky2004}, who derive extinction values for  millions of stars in the LMC and SMC by fitting the MCPS \glspl{SED} with stellar atmosphere models. However, adopting the mean values from these \Av{} distributions (0.55 and 0.46 mag for the LMC and SMC, respectively) caused the theoretical \gls{ZAMS} to lie in the middle of the large overdensity of stars in Figure~\ref{fig:sums} (which we interpret as the MS). This is likely an indication that our sample of stars (which represents on $\sim$2.5\% of the full MCPS catalog) is not well described by the \emph{mean} values of the MCPS \Av{} distributions. In particular, these \Av{} distributions have significant tails to large \Av{} values that can shift the mean to larger values than the peak or median of the distribution. At the same time, our UV photometric catalog is actually biased against the most crowded regions of the LMC/SMC (which also tend to have higher extinction). Instead, we find that adopting \Av{}$= 0.38 $ and 0.22 mag aligns the blue edge of the large overdensity of stars in the UV-optical \glspl{CMD} with the theoretical \gls{ZAMS} described in \S~\ref{sec:zams}. Notably, these values lie close to the \emph{peaks} of the distributions of \Av{} values for hot stars in \cite{Zaritsky2002,Zaritsky2004}.

\textbf{\emph{Use of a single \Av{} value:}} We also investigated using a spatially varying extinction correction. In particular, \cite{Zaritsky2002,Zaritsky2004} provide 2D extinction maps of both hot and cool stars in the LMC, and SMC (computed as an average over $\sim$ 1$\times$1$\arcmin$ cells). However, adopting the values from the hot star map provides similar results to above: the theoretical \gls{ZAMS} lies in the middle of the large over-density of stars in Figure~\ref{fig:sums}. 
This may again indicate that our sample of UV sources is not well described by the \emph{mean} extinction value within each 2D cell in the MCPS maps. As noted above, these cells are $\sim$ 1$\times$1$\arcmin$ in size and in many cases there is significant standard deviation ($\gtrsim$0.3 mag) between the stars in a given cell. The larger extinction corrections in the UV also amplify even small uncertainties in the optically-derived \Av{} values from MCPS. While \citet{Hagen2017} compute a spatially variable \Av{} maps of the SMC using the \gls{SUMaC} data, coordinates for the pixels in these maps are not yet publicly available, and the analysis has not yet been repeated for the LMC. Overall, we find that improvement over a single mean \Av{} value for each galaxy would likely require a star-by-star fitting analysis, which is beyond the scope of this current work, but will be further discussed below.

\begin{figure}
    \centering
    \includegraphics[width=\columnwidth]{ 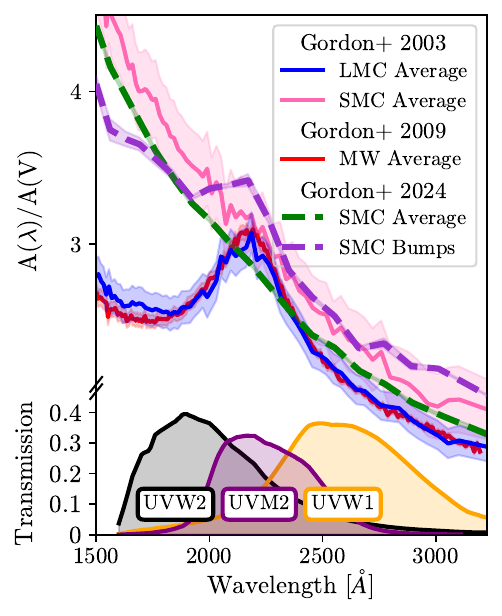}
    \caption{Summary of the shape of various LMC, SMC, and Milky Way extinction curves in the UV \citep{Gordon2003,Gordon2009,Gordon2024} compared to the wavelength coverage of  \emph{Swift}'s UV filters. \emph{Swift}-UVOT's UV filters (black, purple, and orange curves) are thought to be particularly sensitive probes of reddening in the LMC  due to the alignment of the bump in the extinction curves (blue and red lines) resulting from the 2175\AA\ interstellar extinction feature with the UVM2 filter. The SMC average extinction curve (green line) is comparatively smoother, however some regions in the SMC are also affected by this dust bump (dashed purple line). Throughout this paper, we adopt the \cite{Gordon2003} extinction curves, as described in Section~\ref{sec:extinction}.}
    \label{fig:extinctioncurve}
\end{figure}

\textbf{\emph{Choice of extinction curve:} }
We adopt the average LMC and SMC extinction curves from \citet{Gordon2003}. However, especially in the SMC, there is some uncertainty in the appropriate extinction curve along the line-of-sight to individual stars. In particular, \citet{Gordon2024} recently published an updated analysis of the SMC extinction curve, which is based on a larger number of sight lines. These extinction curves are shown, along with the transmission curves of the \emph{Swift-}UVOT UV filters, in Figure~\ref{fig:extinctioncurve}. While the new ``average'' SMC curve from \citet{Gordon2024} is consistent within errors with that of \citet{Gordon2003}, they also found that $\sim$20\% of their target SMC stars showed evidence for the presence of a 2175~\AA\ ``bump'' in their extinction curves. As shown in Figure~\ref{fig:extinctioncurve}, this feature---which is located near the UVM2 filter---is found in the Milky Way and LMC extinction curves. However, it has not been observed for a majority of stars studied in the SMC, to date.  Given that a proper assessment of the 2175~\AA\ bump within the SMC would require a star-by-star analysis, we instead choose to continue using the curves of \cite{Gordon2003} for consistency with the analysis of \citetalias{Drout2023}. Finally, we note that we adopt a single \Av{} value and single extinction curve for each galaxy. We do not separately account for a baseline level of extinction due to the Milky Way along the line-of-sight. While the LMC and Milky Way extinction curves are similar (and the amount of foreground extinction along due to the Milky Way is low: \Av{}$\sim$ 0.1 mag in the analysis of \citealt{Gordon2024}) this is an additional source of uncertainty for stars in the SMC.

Because extinction correction is a major source of uncertainty in identifying stripped star candidates, we examine the underlying assumptions in more detail in the following sections: \S\ref{sec:extinction_on_candidates}, where we evaluate how different extinction choices affect the number of candidates; \S\ref{sec:extinction_on_seds}, where we discuss the influence of extinction on the SEDs of our candidates; and \S\ref{sec:ms}, where we discuss potential contamination from low-reddening main-sequence stars in our sample.

\subsection{Initial Assessment of UV Excess}\label{sec:bluewards}

To perform our initial assessment of which stars within our UV catalog have an apparent UV excess, we only use magnitudes of stars that were detected at a level of at least 5$\sigma$, corresponding to magnitude errors less than 0.217 mag.
For consideration, a given source must have at least 5 measured magnitudes (including a minimum of two in the UV and three in the optical) above this threshold. We additionally restrict ourselves to sources that have extinction corrected UV magnitudes (in all three \emph{Swift} UV filters) in the range of 14 AB mag $<$ m$_{\rm{UV}}$ $<$ 19 AB mag. This range is broadly similar to 
the range of magnitudes predicted for Magellanic Cloud binary systems comprised of stripped stars with masses of $\sim$1--7 M$_\odot$ and MS stars of $\leq$12 M$_\odot$ in the composite models
described in \S~\ref{sec:compositemodels}. However, our adopted range allows for systems that are $\sim$1--1.5 mag brighter in the UV than found in the composite grid, in order to allow for slightly more massive stripped stars \citep{Shenar2020}. Below our adopted threshold, we would be sensitive to primarily lower mass stripped stars such as subdwarfs (and the typical photometric errors in our catalog also increase dramatically). Above this threshold, we would be sensitive primarily to even higher mass systems (which tend to exhibit Wolf-Rayet-like winds). We will further examine the possible inclusion of both of these types of objects in our final candidate list in Section~\ref{sec:otherblue}. 

For each source, we then calculate the difference in color between the extinction corrected absolute magnitude photometry and the location of the theoretical \gls{ZAMS} in nine different UV-optical \glspl{CMD}. Specifically, we consider each combination of (UVW2,UVM2,UVW1) $-$ (B,V,I) versus (UVW2,UVM2,UVW1).\footnote{ Where the UV filter used for the vertical axis is always the same as the one used to construct the color on the horizontal axis.} We neglect the U-band data when constructing \glspl{CMD} because these observations were typically shallower than for the other optical filters in MCPS and because the smaller wavelength difference to the \emph{Swift}-UVOT filters provides less leverage to detect any color excess. To further consider a given system, we require that it lies bluewards of the \gls{ZAMS} in a minimum of one of the nine \glspl{CMD} \emph{by an amount that exceeds the photometric error on the color}. This error is calculated by combining the statistical error output by \theTractor{} in quadrature with the standard deviation found for objects with multiple observations (see \S~\ref{sec:avg}). The median color error for the entire catalog is $\sim0.27$ mag for any of the nine UV-optical filter combinations. This requirement ensures that a given system is inconsistent with being a single ZAMS star at $>1\sigma$ in at least one observed \gls{CMD}, given our extinction assumptions.  Of the \NSources{} stars in our catalogs, we have an initial population of \NQremain{} such stars  (\NQremainLMC{} and \NQremainSMC{} in the LMC/SMC, respectively).

\subsection{Quality Cuts based on SED Shape}\label{sec:SEDquality}

To further refine our selection of stripped star candidates, we next analyze the \glspl{SED} of all sources previously identified as having a potential UV excess.
Our objective is to remove sources that may exhibit an apparent excess due to a data quality issue or that have \gls{SED} shapes that differ dramatically from expectations of binaries containing stripped stars. 
When making the latter assessment, we use the \citetalias{Drout2023} composite stripped star plus \gls{MS} model grid described in \S~\ref{sec:compositemodels} to describe the ``expected'' properties of such systems. Overall, we perform four broad quality cuts for our sample.

\emph{UV-optical source mismatches:} Our first goal is to eliminate sources that falsely appear to have a UV excess due to a mismatch between UV and optical counterparts in our catalog (i.e., cases where the \gls{SUMaC} UV photometry and MCPS optical photometry are incorrectly associated with the same source; see Appendix~\ref{sec:caveat}). We test for this in two ways:

\begin{enumerate}[leftmargin=1em,itemsep=-0.3em]
    \item 
    We first exclude sources for which the UV and optical magnitudes each show very little spread in observed magnitude within their respective groups, while the two groups themselves exhibit significant separation from each other. We consider the optical and/or UV photometry to show little variation if the standard deviation of the magnitudes is less than 0.217 mag (the maximum statistical error we allowed for a photometric point to be considered in our analysis) and we consider the UV and optical magnitudes to be ``significantly separated'' if the means of each group are separated by $>$1 AB mag in excess of the maximum photometric error.
    \item We additionally exclude sources where the most adjacent filters between the UV and optical groups (typically UVW1 and U-band, but if either are missing we utilize UVM2/B-band) are separated by an amount that is more than 0.217 mag larger than the biggest separation found between those filters in the helium plus \gls{MS} star composite model grid. When carrying out this analysis, we allow for the possibility of a  single poorly estimated magnitude by checking the next most adjacent magnitude difference as well (i.e., if the UVW1-U color of a source is too large, but UVW1-B color is within the range allowed by the composite model grid we do not exclude it at this stage).
\end{enumerate}

\emph{Optically-red \glspl{SED}:} When examining the \glspl{SED} in our sample, we found some sources that appear to progressively decrease in flux from the UV to blue optical bands, but then increase in flux again when moving to progressively redder optical bands (resulting in a ``V''-like \gls{SED}). While these may be interesting sources in their own right, they are not consistent with expectations for binary stripped stars with \gls{MS} companions which have absolute magnitudes in the range of our search. We therefore remove any sources that increase in flux between two or more adjacent optical filters (i.e., from U to B, B to V, and/or V to I).

\emph{Poor-quality photometric points:} We next examine whether some sources show apparent UV excess \emph{only} due to one or more photometric points that are ``outliers''---which we define as points that vary from the adjacent bands by a larger amount than would be expected for a stellar continuum.  We note the presence of such outliers could either be caused by data quality/processing errors or by physical effects such as (i) a source with strong emission lines that dominate the flux in a given band or (ii) a variable source where different bands were observed at different times.

To assess the presence of outliers, we compare the magnitude difference between adjacent bands to expectations from the stripped star plus \gls{MS} star model grid. We flag a given color as suspect if it falls outside the range spanned by the model grid by more than 0.217 mag.
By including this additional buffer, we allow for some small variation from model expectations due to known photometric errors. We then consider three cases: 
\begin{enumerate}[leftmargin=1em,itemsep=-0.3em]
    \item If more than three colors were flagged as suspect, we reject the source from our sample due to either having a poor quality \gls{SED} or varying dramatically from expectations for stripped star binaries.
    \item If two adjacent colors were flagged, then we remove the middle photometric point and assess whether the source would still be considered bluewards of the \gls{ZAMS} based on the method in Section~\ref{sec:bluewards}.
    \item If only one color (or two non-adjacent colors) are flagged, then we remove both relevant magnitudes and assess whether the source would still be considered bluewards of the \gls{ZAMS}.
\end{enumerate}
For points 2 and 3, if the source would still have a UV excess without the offending photometric points, then we keep it in the sample. Otherwise, it is removed.

\emph{Faint I-band photometry:} When examining the remaining sources in our sample, we noted multiple occasions where the MCPS I-band photometry appears significantly fainter than would be expected from extrapolating the rest of the optical \gls{SED}. We note that while all of these objects would likely have V-I colors that fall with the overall range of composite model grid (due to the previous criteria for inclusion), this does not necessarily imply that it is consistent with expectations based on the rest of the \gls{SED}. We therefore test whether the remaining objects would still be considered bluewards of the \gls{ZAMS} if we remove the I-band. We remove any objects that only show a UV excess due to this band.

Overall, the cuts described in this subsection remove \NSEDCutLost{} sources from our population, with \NSEDCutRemain{} sources remaining (\NSEDCutRemainLMC{} and \NSEDCutRemainSMC{} in the LMC/SMC, respectively).

\subsection{Additional Quality Cuts}\label{sec:morequalitycuts}

While \NSEDCutRemain{} stars in our catalog both show an apparent excess of UV light (in at least one UV-optical \gls{CMD}) and pass the basic \gls{SED} quality cuts described above, we still find significant variation in both the quality of and our confidence in individual candidates. For example, some of these objects are relatively isolated and show an apparent UV excess in all 9 \glspl{CMD} considered, while others are found in crowded regions and appear bluewards of the \gls{ZAMS} in only one \gls{CMD}. As our goal at this stage is to develop a robust set of candidates to motivate further follow-up (as opposed to perform detailed rates calculations---which will be tackled in later work), here we further restrict our sample based on the quality of the data and strength of evidence for a UV excess. 

First, to be examined further in this manuscript, we require that a source is bluewards of the \glspl{ZAMS} in at least 4 of the 9 UV-optical \glspl{CMD} considered. This selects objects with more robust UV excess. This requirement eliminates \NAdditionalblueincmds{} candidates. 
Second, we impose requirements on the fraction of the UV flux within a 5$\arcsec$ radius around each candidate that must be attributed to the candidate star itself (i.e., the ``flux\_frac'' parameter described in Section~\ref{sec:finalcontents}). This is motivated by our findings in Section ~\ref{ref:crowding}, which show that \theTractors{} UV photometry can be systematically biased when multiple sources are nearby---particularly for the fainter star (see Figure~\ref{fig:cluster}, right panel). Specifically, if another source lies within 2.5$\arcsec$  of the candidate, we require that the candidate contributes $>$25\% of the total UV flux (averaged across the three UVOT filters).

In addition, for candidates where the nearest source detected in the SuMAC images is $>$2.5$\arcsec$ away, we still require that the average UV flux fraction exceeds 10\%. Although Section~\ref{ref:crowding} shows that \theTractor{} photometry of injected sources is generally accurate when the nearest source is $>$2.5$\arcsec$ away, further inspection revealed that many candidate systems exhibit very low UV flux fractions even when the nearest neighbor lies $\sim$2--4$\arcsec$ away (see Appendix~\ref{ap:figures}, Figure~\ref{fig:ffrac_closest}). These cases are typically systems where stripped star candidates with UV magnitudes of $\sim$18–19 AB mag are located within $\lesssim$4$\arcsec$ of a much brighter star ($\sim$14–15 AB mag).

Given that (i) these magnitude differences are more extreme than tested in Section~\ref{ref:crowding}, (ii) the UV magnitudes of these nearby bright sources can be close to the count rate level for the detector when coincidence-loss artifacts begin to appear (see Section~\ref{sec:inputimages}) we have opted to be conservative at this time. However, we emphasize that there may be actual stripped stars, or other blue astrophysical sources, located within these low flux fraction objects. 
Future work (L, Blomberg et al. 2026, in preparation) will perform additional injection tests with \theTractor{} on the SuMAC images to further test its performance in dense/multi object environments. 
This, or higher resolution UV image (e.g., UVIT, HST), may be used to further vet/verify the SED shape of these objects. Overall, after applying the cuts described in this section, we are left with \NAdditionalRemain{} stripped star candidates (\NAdditionalRemainLMC{} and \NAdditionalRemainSMC{} in the LMC and SMC, respectively).

\subsection{Assessing Membership in the Magellanic Clouds}\label{sec:member}
To this point, we have considered all stars within our photometric catalog. However, some objects may be foreground stars, as opposed to true members of the Magellanic Clouds. While radial velocities are often used to gauge membership in the Clouds \citep[e.g][]{Evans2008,Neugent2012,Gonzalez-Fernandez2015}, spectroscopic observations are not yet available for most\footnote{To our knowledge, the only candidates with published spectroscopic information are those in the supplementary material of \citetalias{Drout2023}) and $\lesssim$10 other stars (discussed in Section \ref{sec:otherblue}).} of our candidates. We therefore perform an initial assessment using astrometry (i.e., parallax and proper motion) measurements from \emph{Gaia}---with an aim of removing objects that are very likely foreground stars based on currently available data. 

We cross-match the \NAdditionalRemain{} ranked candidates from \S~\ref{sec:morequalitycuts} with the \emph{Gaia} DR3 catalog \citep{GAIA2023} using a 1$\arcsec$ search radius. In total, there were 
\NNoGaiaMatch{} sources that either (i) did not cross-match with any \emph{Gaia} DR3 sources, (ii) cross-matched with a source that had a \emph{Gaia} G-band magnitude that differed by more than 1.5 magnitudes from both of the MCPS B- and V-band magnitudes indicating potential cross-match errors (iii) cross-matched with a \emph{Gaia} source for which no parallax or proper motion information was reported. We choose to \emph{retain} these objects in our sample, because we do not currently have data required to assess whether they are likely foreground stars.  In particular, we note that many are faint in the optical (m$_{\rm{V}}$ $>$ 19 mag) and lie in dense regions, which can impact the overall sensitivity threshold of \emph{Gaia}.

For the remaining sources with \emph{Gaia} data, we assess whether they have kinematics consistent with expectations for stars located in the Magellanic Clouds. We first remove any sources that have robustly measured parallax values (which we define as sources with parallax greater than four times the parallax error). 
This removes \NGaiaParallaxCut{} sources, all of which have inferred distances of $\lesssim$1.5 kpc. 

We next perform a $\chi^2$ analysis to compare the proper motions of the remaining stars to expectations for stars in the LMC/SMC, following the procedures described in Appendix~\ref{ap:gaia-chi2}. We initially consider a source to be a likely foreground star if its proper motions fall outside the region that contains 99.5\% of a sample of ``likely'' LMC/SMC members (see Appendix~\ref{ap:gaia-chi2} for details). However, to explicitly remove a star from our sample, we require both that the \emph{Gaia} proper motions fall outside this region \emph{and} that they do not have a poor goodness-of-fit statistic (which we define as \texttt{astrometric\_gof\_al} $>$3) in the \emph{Gaia} catalog\footnote{When making this decision, we were motivated by the results of \citetalias{Drout2023}. They found that two objects in their final spectroscopic sample had radial velocities consistent with membership in the Clouds (and spectra that closely resembled other objects in the sample in both spectral morphology and inferred physical properties) but had \emph{Gaia} proper motions that were slightly inconsistent with expectations for members of the Clouds. However, both objects had poor goodness of fit statistics in the \emph{Gaia} database, indicating some issues with the astrometric solution. Similar results have been found for other samples of hot stars in the Magellanic Clouds (e.g., the O-star sample of \citealt{Aadland2018}).}. Based on these criteria, we reject 
\NGaiaPMCut{} additional stars.

In total, we found that 
\NGaiaMembersPercent{} of the candidates with robust \emph{Gaia} data have parallax and proper motion values consistent with expectations for the stars in the Magellanic Clouds. 
This implies that the rate of foreground contamination in this region of the \gls{CMD} is relatively low. It is therefore likely that many of 
\NNoGaiaMatch{} sources without \emph{Gaia} kinematic information and 
\NGaiaGOFOverThree{} sources with poor \emph{Gaia} goodness-of-fit statistics are also true members of Clouds. However, we acknowledge that this subset of our sample may have a higher rate of foreground/background contamination, which will be further discussed in \S~\ref{sec:otherblue-foreback}.
After removing likely foreground stars, our final candidate sample contains \NFinal{} sources (\NFinalLMC{} in the LMC and \NFinalSMC{} in the SMC). 

\subsection{Candidate Rankings}\label{sec:rankings}

We classify our set of \NFinal{} candidate stripped star systems into four groups based on a combination of their UV excess (i.e., how ``blue" they are) and the overall quality of their respective \glspl{SED}. This classification is done along two axes. Specifically, we categorize each target as:

\begin{itemize}[leftmargin=1em]
    \setlength\itemsep{-0.1em}
    \item \emph{`Very Blue' or `Blue':} This distinction is based on the source's average distance from the \gls{ZAMS} across the nine UV-optical \glspl{CMD} considered. Objects that are, on average, more than 0.4 mag bluer than the \gls{ZAMS} are classified as ``Very Blue," while the rest, which lie closer to the \gls{ZAMS}, are designated as``Blue."
    \item \emph{`Excellent' or `Good':} This distinction is made based on a number of data quality metrics. An object is considered `Excellent' if it has (i) an average UV flux fraction of $>$40\%, (ii) at least 6 of the 7 photometric bands present in its \gls{SED} and (iii) no more than one pair of adjacent bands in its \gls{SED} with a color that is inconsistent with the expected range predicted by stripped star plus MS star models (as described in Section~\ref{sec:SEDquality}). If an object does not meet any of these three criteria, it is designated as `Good'.
\end{itemize}

The resulting four categories are: Very Blue-Excellent (VB-E), Very Blue-Good (VB-G), Blue-Excellent (B-E), and Blue-Good (B-G). The number of sources in each category for both the LMC and SMC is listed in Table~\ref{tab:Rankings}. Approximately half of our candidates are classified as ``Very Blue" rather than ``Blue". These sources are the most robust against uncertainties due to extinction (see Section~\ref{sec:extinction_on_candidates}, below). There are three times as many ``Very Blue" candidates in the LMC as in the SMC.

\begin{deluxetable}{l||c|c}	[t]
    \centering
	\tablecaption{Number of Stripped Star Candidates by Rank\label{tab:Rankings}}
	\tablehead{\colhead{\S\ref{sec:rankings} Rankings}\hspace{.5cm} & 
                \colhead{\hspace{.65cm}LMC}\hspace{.65cm} & 
                \colhead{\hspace{.65cm}SMC}\hspace{.65cm} }
	\startdata
        \textbf{Total} & \NFinalLMC & \NFinalSMC \\ 
        \hline 
	\textbf{VB-E:} Very Blue - Excellent & \NVBELMC & \NVBESMC \\
        \textbf{VB-G:} Very Blue - Good & \NVBGLMC & \NVBGSMC \\ 
        \textbf{B-E:} Blue - Excellent  & \NBELMC & \NBESMC \\
        \textbf{B-G:} Blue - Good & \NBGLMC & \NBGSMC \\ 
	\hline
	\enddata
\end{deluxetable}

\begin{figure}
    \centering
    \vspace{-0.3in}
    \includegraphics[width=0.9\columnwidth]{ 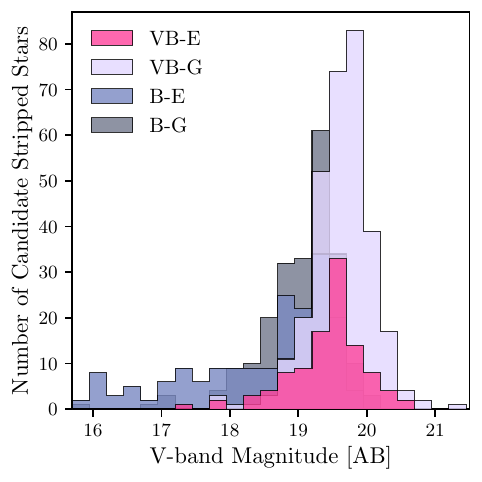}
    \caption{The distribution of optical V-band magnitudes for our candidate stripped star population. Colors correspond to the rank assigned in \S\ref{sec:rankings}. }
    \label{fig:candidatevmag}
\end{figure}

\begin{deluxetable*}{llllllllllll}
     \tabletypesize{\footnotesize}
      \caption{The Stripped Star Candidate Catalog.}\label{tab:candidate_catalog}
    \tablehead{
        \colhead{SUMS\_ID} & \colhead{R.A.} & \colhead{Dec.}& \colhead{Ranking}& \colhead{UVW2}& \colhead{UVW2\_err}& \colhead{$\hdots$}& \colhead{UVW2\_nobs}& \colhead{UVM2\_nobs}& \colhead{ UVW1\_nobs}&\colhead{Gaia\_chi2} \\ 
        \colhead{}& 
        \colhead{\scriptsize[\degree]} & \colhead{\scriptsize[\degree]} & 
        \colhead{}& 
        \colhead{\scriptsize[AB Mag]}& \colhead{\scriptsize[AB Mag]}& 
       \colhead{}&\colhead{}&\colhead{}&\colhead{}&\colhead{}}
	\startdata
        SUMS\_79e24 & 84.238275 & -69.458760 & VB-E & 17.77 & 0.03 & $\dots$ & 8 & 8 & 8 & 1.2 \\
SUMS\_832c6 & 85.523865 & -69.205750 & VB-G & 18.73 & 0.1 & $\dots$ & 1 & 2 & 2 & 5.7 \\
$\vdots$ & & & & & & & & & & $\vdots$\\
SUMS\_8c796 & 10.414935 & -73.377410 & B-E & 18.38 & 0.05 & $\dots$ & 2 & 1 & 3 & 0.1 \\
SUMS\_8dfe7 & 10.831395 & -73.659790 & B-G & 17.44 & 0.03 & $\dots$ & 2 & 2 & 2 & 1.8 \\
 	\enddata
    \tablenotetext{\dagger}{This table provides an illustrative subset of our candidate stripped star catalog. The complete catalog is accessible in a machine-readable format in the online journal and on VizieR.}
\end{deluxetable*}

In Figure~\ref{fig:candidatevmag} we plot histograms of the apparent V-band magnitudes of all candidates, separated by their final ranks. The distributions peak near 19.5 mag, highlighting that many of the candidates are optically faint. This relative faintness may be the reason that many of these systems have been only sparsely studied to date---and why intermediate mass stripped stars were only recently identified.

In Table~\ref{tab:candidate_catalog} we provide a summary of information for our final set of \NFinal{} candidates. Much of the information provided is the same as in Table~\ref{tab:minifullcat}, with three exceptions: (i) UV and optical photometry are now provided in AB mag, following the conversions in Table~\ref{tab:conversions}, (ii) we provide the rank assigned to each candidate and (iii) we provide the $\chi^2$ value computed for each star when comparing their \emph{Gaia} proper motions to a set of likely LMC/SMC members (see \S~\ref{sec:member}).

\subsection{Comparison to Stripped Star Sample in \citetalias{Drout2023}}\label{sec:compD23}

As described in Section~\ref{sec:intro}, \citetalias{Drout2023} present a spectroscopic sample of 25 stripped star candidates, which were initially selected as UV excess stars by a similar process to that described above. In particular, we adopt the same assumptions for distance, reddening, and theoretical \gls{ZAMS} as \citetalias{Drout2023}. However, we now perform an explicit assessment of \gls{SED} quality for all candidates (\S~\ref{sec:SEDquality}) and place stricter requirements on both the strength of the UV excess relative to measurement errors and UV flux fraction of candidate systems (\S~\ref{sec:morequalitycuts}). Here, we assess where the sample of \citetalias{Drout2023} fall in within our current sample of photometric candidates.

Overall, we find that 18 of the 25 system in the sample of \citetalias{Drout2023} are in the final set of candidates presented in Section~\ref{sec:rankings}. This includes all 8 of the ``Class 1'' stars (which show spectroscopic features in the optical consistent with stripped stars), 7 out of the 8 ``Class 2'' stars (which show spectroscopic features indicative of \emph{both} a hot stripped star and B-type star), and 3 out of the 9 ``Class 3'' stars (which have spectra consistent with expectations for B-type MS stars). Of these, all the Class 1 and Class 2 objects in our final candidate list fall in the VB-E category, while the three Class 3 stars (specifically, Stars 18, 21, and 23) fall in the B-E category. This is consistent with the findings of \citetalias{Drout2023} who found that these objects all had high quality \glspl{SED} and that the Class 3 objects generally fell closer to the \gls{ZAMS} than either the Class 1 or Class 2 objects.

We now examine why the remaining seven stars from \citetalias{Drout2023} were excluded from our final candidate list in Section~\ref{sec:rankings}. There are two main reasons. 
First, five systems---Star 15 (Class 2) and Stars 17, 19, 22, and 25 (Class 3)---were excluded because they did not lie significantly blueward of the \gls{ZAMS}, beyond their photometric uncertainties, in at least four UV-optical \glspl{CMD}. 
This was also the case in \citetalias{Drout2023} (see e.g. their Figure 1), where the photometric error bars for these objects overlapped with the \glspl{ZAMS}. However, their selection criteria allowed a source to be considered blueward of the \gls{ZAMS} even if its uncertainties overlapped it.

Second, two Class 3 systems (Stars 20 and 24) were excluded because they contributed only $\sim$5\% and 17\% of the UV flux within a 5$\arcsec$ centered on their locations. Hence, they were excluded based on the quality cuts described in Section~\ref{sec:morequalitycuts}. This was also known to be the case in \citetalias{Drout2023} (see their Table S7, where the UV flux fractions for each source are listed). 
Overall, the photometric cuts applied to the sample presented here are \emph{stricter} than those used in \citetalias{Drout2023}.
While these more conservative criteria increase the likelihood of confirming true stripped stars, they also imply that additional candidates may exist elsewhere in the catalog that were excluded by our selection.

\subsection{Impact of Extinction on Candidates}\label{sec:extinction_on_candidates}

As described in \S\ref{sec:extinction}, to correct our sample of extinction, we apply a single \Av{} value for each galaxy that was chosen to align the large overdensity of stars in Figure~\ref{fig:sums} with the theoretical \gls{ZAMS}. This will preferentially identify stars that are bluer than ``typical'' \gls{MS} stars in the Clouds. While this will be true for many systems with an intrinsic UV excess, we emphasize that: (i) some of our candidates may be \gls{MS} stars that occur in areas of particularly low reddening within the Clouds and (ii) we may be missing additional stripped star binaries that come from regions of higher extinction. For the latter case, we note that when \cite{Gotberg2023} performed spectrophotometric fitting on the Class 1 systems from \citetalias{Drout2023}, they found a range of extinction values--ranging from  0.21 mag $<$ \Av{} $<$ 0.63 mag. These were all very blue sources, which showed an apparent UV excess despite our underprediction of their extinction in some cases. However, other systems with different flux ratios between the stripped star and \gls{MS} companion may be excluded from our sample if they fall in regions of higher extinction.
To further assess the impact of our extinction choices on our population of candidate stripped star binaries, we run a number of tests outlined in the following subsections.

\subsubsection{Number of Candidates Retained for a range of Assumed \Av{}}\label{sec:candextinction}

First, to understand what fraction of our selected sources would remain in our sample even if the true extinction along the line of sight were lower, we run the same analysis described in \S~\ref{sec:bluewards}--\ref{sec:member} but with lower \Av{} values. The results are shown in Table~\ref{tab:av}, where the number of candidates progressively drops with lower \Av{}, as expected. However, $\sim$13\% of our candidates (105 stars) would show UV excess even if there was \emph{zero} extinction along the line-of-sight. These objects are dominated ($>$90\%) by sources in the `Very-Blue' categories described in Section~\ref{sec:rankings}. In addition, we emphasize that zero extinction along the line-of-sight is unlikely, as there is expected to be a contribution of \Av{}$\sim$0.1 mag from the Milky Way \citep[e.g.][]{Edenhofer2024}.

\begin{deluxetable}{c||c|c}
	\label{tab:Extinction}
    \centering
	\tablecaption{Impact of Extinction on Number of Candidates \label{tab:av}}
	\tablehead{\colhead{\hspace{.75cm}$A_v$ [mags]}\hspace{.75cm} & 
    \colhead{\hspace{.75cm}LMC}\hspace{.75cm} & 
    \colhead{\hspace{.75cm}SMC}\hspace{.75cm} } 
	\startdata
	0 &  \NAVZEROLMC{} & \NAVZEROSMC{} \\
    0.1 &  \NAVPONELMC{} & \NAVPONESMC{} \\ 
    0.22  & \NAVPTWENTYTWOLMC{} & \textbf{\NFinalSMC{}} \\
    0.3 &   \NAVPTHREELMC{} & -  \\ 
    0.38 &  \textbf{\NFinalLMC{}} & - \\ 
	\hline
	\enddata
\end{deluxetable}

\begin{figure}
    \centering
    \includegraphics[trim={0.in, 0.0in, 0.in, 0.in},clip,width=0.47\textwidth]{ 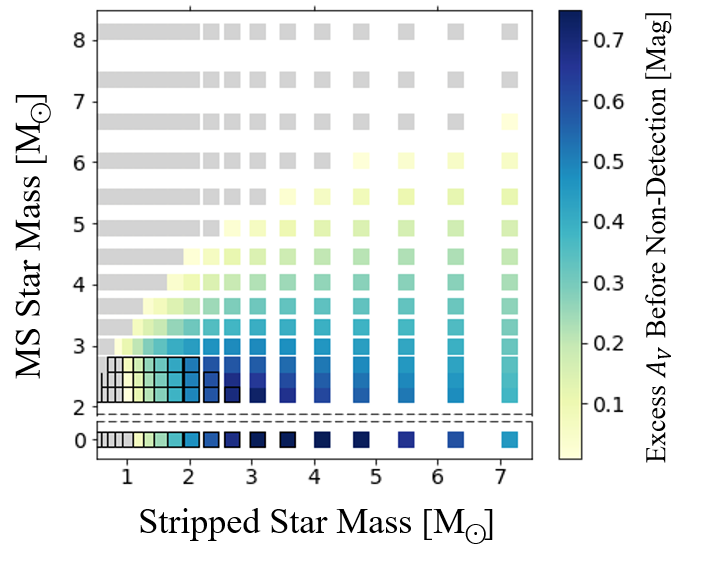}
    \caption{The grid of stripped star plus \gls{MS} star binaries from \citetalias{Drout2023}. Points are color coded by the amount of excess extinction---beyond what we correct for in our analysis---that could be present along the line-of-sight before they would no longer be identified by our selection process. Systems shown in grey are not expected to show a detectable UV excess even with accurate extinction correction. Systems with under-corrected extinction could be missing from our sample either because (i) they no longer appear bluewards of the \gls{ZAMS} or (ii) they fall below our brightness threshold of 19 AB mag in the UV. Systems that are missed because they are too faint are outlined in black. }
    \label{fig:test-Av}
\end{figure}

\subsubsection{Expectations for the Presence of a UV Excess as a Function of Extinction and Stripped Star Binary Type}\label{sec:missing}

Second, to broadly understand what types of binary systems we may be missing by underestimating the extinction, we examine the stripped star plus \gls{MS} star composite grid from \citetalias{Drout2023} described in \S~\ref{sec:compositemodels}. For each system, we examine how much extra extinction along the line of sight (beyond what we correct for in our baseline procedures) would be necessary for the system to either no longer exhibit a UV excess or to fall below our brightness threshold of 19 mag AB (see \S~\ref{sec:bluewards}). The results are shown in Figure~\ref{fig:test-Av}, where systems in grey are not expected to show a detectable UV excess even with accurate extinction correction. For this example we (i) consider only the UVM2--V versus UVM2 \gls{CMD} as opposed to the full set of nine \glspl{CMD}, (ii) assume an average color uncertainty of 0.15 mag when assessing if a system is bluewards of the \gls{ZAMS} by more than it associated uncertainty (this is the median UVM2-V error in our candidate sample), and (iii) adopt an LMC extinction curve. Slight differences may be expected for the SMC, due to the lack of a 2175 \AA\ bump in its extinction curve, which falls near the UVM2 band. 

From Figure~\ref{fig:test-Av}, we find that stripped stars with masses above $\sim$2.5 M$_{\odot}$ that do not have a luminous companion (i.e., single stripped stars disrupted from their binaries or with compact object companions) would need to have extinction values \Av{} more than 0.65 mag higher than our current assumptions (\Av{}$> 1$ mag for the LMC) to be missing from our sample. Such extinction values are on the high end of the \Av{} distribution reported by \citet{Zaritsky2004}. This suggests that our candidate sample likely includes many of these systems, provided they have reliable UV photometry. However, we note that some may still be missed due to missing UV data, for example in crowded regions.

In contrast, both low mass stripped stars ($\lesssim$2 M$_\odot$) and composite systems where the \gls{MS} star dominated the flux in the optical can be lost from our sample if we even moderately underestimate their extinction (in the former case because they become too faint). Stripped stars at higher masses ($\gtrsim$ 7 M$_{\odot}$) can also become more influenced by extinction as these approach closer to the \gls{ZAMS} where hotter \gls{MS} stars push the \gls{ZAMS} bluewards. While this is a moderate effect, we emphasize that stripped stars around $\sim$3-4 M$_{\odot}$ seem to be in a particular sweet spot for detection and indeed \citet{Gotberg2023} found that most of the Class 1 stars from (stripped star dominated systems) had spectroscopic and evolutionary masses overlapping this regime. 

\subsubsection{Extinction required for a MS star to Masquerade as a UV Excess Source}\label{sec:MSMasquerade}

\begin{figure}
    \centering
    \includegraphics[scale=0.8]{ 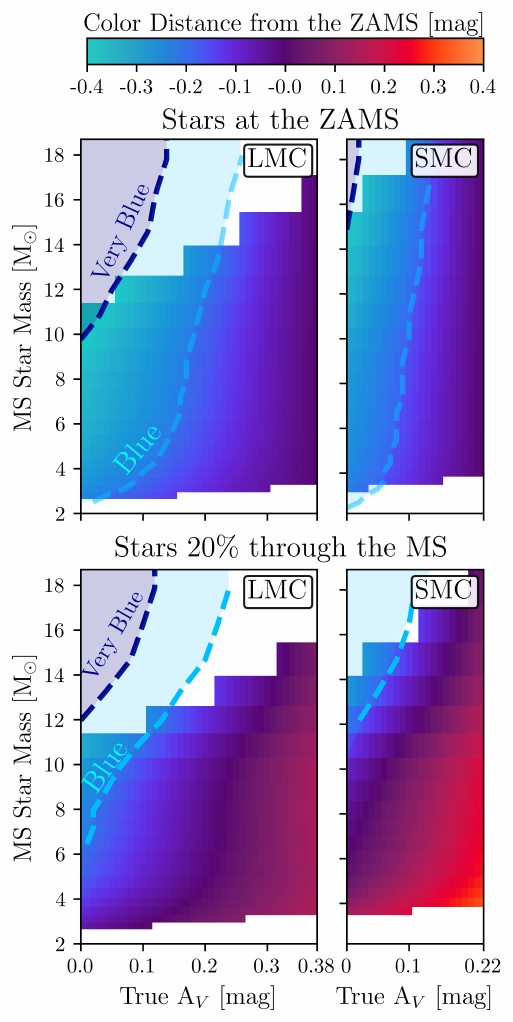}    \caption{
    Schematic demonstrating which MS stars may appear in our candidate sample if they lie in regions of particularly low reddening. In each panel, the MS star mass is shown on the y-axis and the assumed line-of-site extinction on the x-axis. Points are color-coded by the distance (in color) between the star and the ZAMS in a UVM2-V CMD, if it were \emph{over-corrected} for extinction using the nominal values we adopt in Section~\ref{sec:extinction} (negative values are bluewards of the ZAMS, positive values are redward). LMC stars are shown in the left and SMC stars on the right. The top panels show results for stars located at the ZAMS and the bottom panels for stars stars 20\% through their MS lifetimes. {In each panel, the purple dashed line indicates stars that would lie 0.4 mag bluer than the ZAMS (our threshold for designating a source as ``Very Blue'') while the blue dashed lines indicate stars that would bluewards of the ZAMS by an amount equal to the median photometric uncertainty in each galaxy (0.2 mag and 0.16 mag in the LMC and SMC, respectively). Thus, the shaded purple and shaded blue regions highlight the combinations of MS star mass and line-of-site extinction that could potentially contaminate our ``Very Blue'' and ``Blue'' samples, respectively.}
    Stars that would fall outside the brightness threshold of our candidate sample are not plotted in this figure. Overall, we find that while some young MS stars could potentially contaminate our ``Blue'' sample, they are unlikely to be located in the ``Very Blue'' region.}
    \label{fig:MScontaminant}
\end{figure}

Third, we examine how low the line-of-sight extinction would need to be for a MS star to masquerade as a UV excess source in our sample. As in Section~\ref{sec:missing}, we consider just the UVM2-V versus UVM2 \gls{CMD} for this test. The results are shown in Figure~\ref{fig:MScontaminant}, where we plot MS stellar mass versus the assumed true line-of-sight extinction towards a given star. Each point is then color-coded by the distance from the \gls{ZAMS} that a given object would appear \emph{if it were over-corrected for the nominal extinction values we adopted in Section~\ref{sec:extinction}} (i.e., $A_V = 0.38$ mag and 0.22 mag for the LMC and SMC, respectively). Points that are not color-coded correspond to systems that would fall outside the brightness threshold used in our sample selection (i.e., systems with either UVM2 $<$ 14 mag or UVM2 $>$ 19 mag).

We carry out this assessment for two fiducial sets of MS stars with different ages: (i) the stars that were used to construct our model for the \gls{ZAMS} and (ii) the set of MS stars that are 20\% through their MS lifetimes that were used to construct the stripped star plus MS star composite grid from \citetalias{Drout2023}. The former are shown in the top panels for the LMC and SMC, respectively, and (by definition) will lie bluewards of the \gls{ZAMS} for even a small over-correction of the intrinsic reddening. The latter are shown in the bottom row and, since they are located somewhat redwards of the \gls{ZAMS} will only appear as UV excess sources for lower intrinsic reddening values.

In Figure~\ref{fig:MScontaminant} we additionally plot lines to highlight when given objects would lie both 0.15 mag and 0.4 mag bluewards of the \gls{ZAMS}. The former represents the median UVM2-V error in our sample, and we take this as a proxy for when we would select an object as being robustly bluewards of the \gls{ZAMS}. The latter is the distance bluewards of the \gls{ZAMS} that we use to separate the `Very Blue' and `Blue' candidates in Section~\ref{sec:rankings}. From this, we see that our sample could potentially contain very young MS stars with initial masses between roughly 2 and 20$\Msun$ that have intrinsic extinction of \Av{} $<$0.2 mag in the LMC and $<$0.1 mag in the SMC. However, once a star is $\sim$20\% through its MS lifetime, the subset of systems that could potentially contaminate our sample decreases and primarily consists of stars with initial masses between $\sim$6-12 M$_\odot$ and \Av{}$<$0.1 mag in the LMC). 
In summary, Figure~\ref{fig:MScontaminant} shows that (i) 
the `Very Blue' category is  \emph{unlikely to be polluted by \gls{MS} stars with low extinction,} but must instead contain hotter sources and (ii) very few \gls{MS} stars $>20\%$ through hydrogen burning are  
likely to enter either sample.

\begin{figure*}[ht]
    \centering
    \includegraphics[width=\textwidth]{ 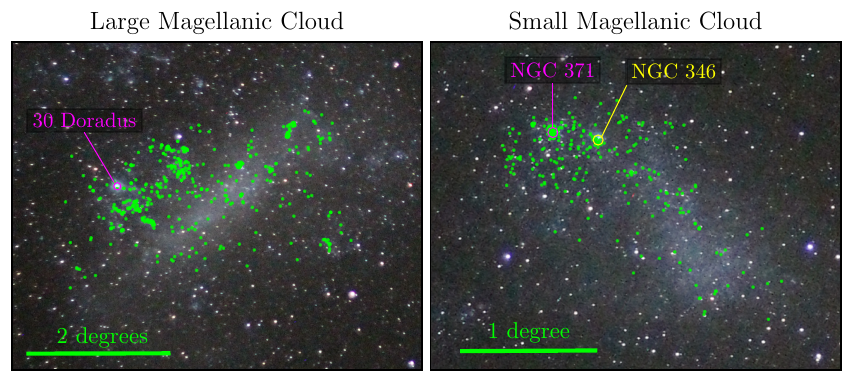}
    \caption{The spatial distribution of our final sample stripped star candidates described in \S~\ref{sec:rankings} (green points) within the Magellanic Clouds. A few specific open clusters are labeled here for reference. The background optical image is the same as Figure~\ref{fig:coverage}. }
    \label{fig:candidatecoverage}
\end{figure*}

\subsubsection{Comparison of Adopted Extinction Values to those in the 2D maps of \cite{Zaritsky2002,Zaritsky2004}}\label{sec:2ddustmap}
Finally, we examine the 2D extinction maps of \citep{Zaritsky2002,Zaritsky2004} to further assess whether we are likely under or over-estimating the extinction on average. We caution, however, that extinction (\Av{}) values are highly uncertain for any individual star in their survey, given that they are derived from SED fits of four photometric points to models of single \gls{MS} and evolved stars. As a result, these maps are divided into 60$\arcsec$ $\times$ 60$\arcsec$ regions, and they present the average of the estimated \Av{} values for all stars in each region. 
When we consider the ``hot star'' map for each galaxy (constructed based on stars with estimated temperatures in the range of 12,000--45,000 K) and restrict ourselves to cells within the 2D maps that overlap with the \gls{SUMaC} footprint, we find that only $\sim$25\% and 10\% of the regions for the LMC and SMC respectively have lower average \Av{} values than we assume, suggesting that our correction likely errs on the side of being conservative. 

\subsubsection{Summary of Impact of Extinction on Candidates}
From the above analysis, it is clear that our extinction assumptions have a large impact on our resulting candidate list. 
Ultimately, it will be necessary to perform a combination of star-by-star extinction modeling and spectroscopic follow-up to both improve upon the candidate list provided here and confirm the nature of individual sources. While this analysis is outside the scope of this manuscript, in \S~\ref{sec:extinction_on_seds} we do provide some discussion and demonstration of extinction-sensitive features that may be present in some of our \glspl{SED} that would facilitate this process in the future. We also emphasize again that, as demonstrated in \S~\ref{sec:MSMasquerade}, \gls{MS} stars at any extinction are unlikely to contaminate our `Very Blue' sample.

\section{Properties of Candidate Systems}\label{sec:properties}

In the previous section, we compiled a sample of \NFinal{} stripped star candidates. A small subset of these have already been confirmed in \citetalias{Drout2023} demonstrating the effectiveness of our approach. Here, we discuss some overall properties of these candidates as a population.

\begin{figure*}[hbtp] 
    \centering
    \includegraphics{ 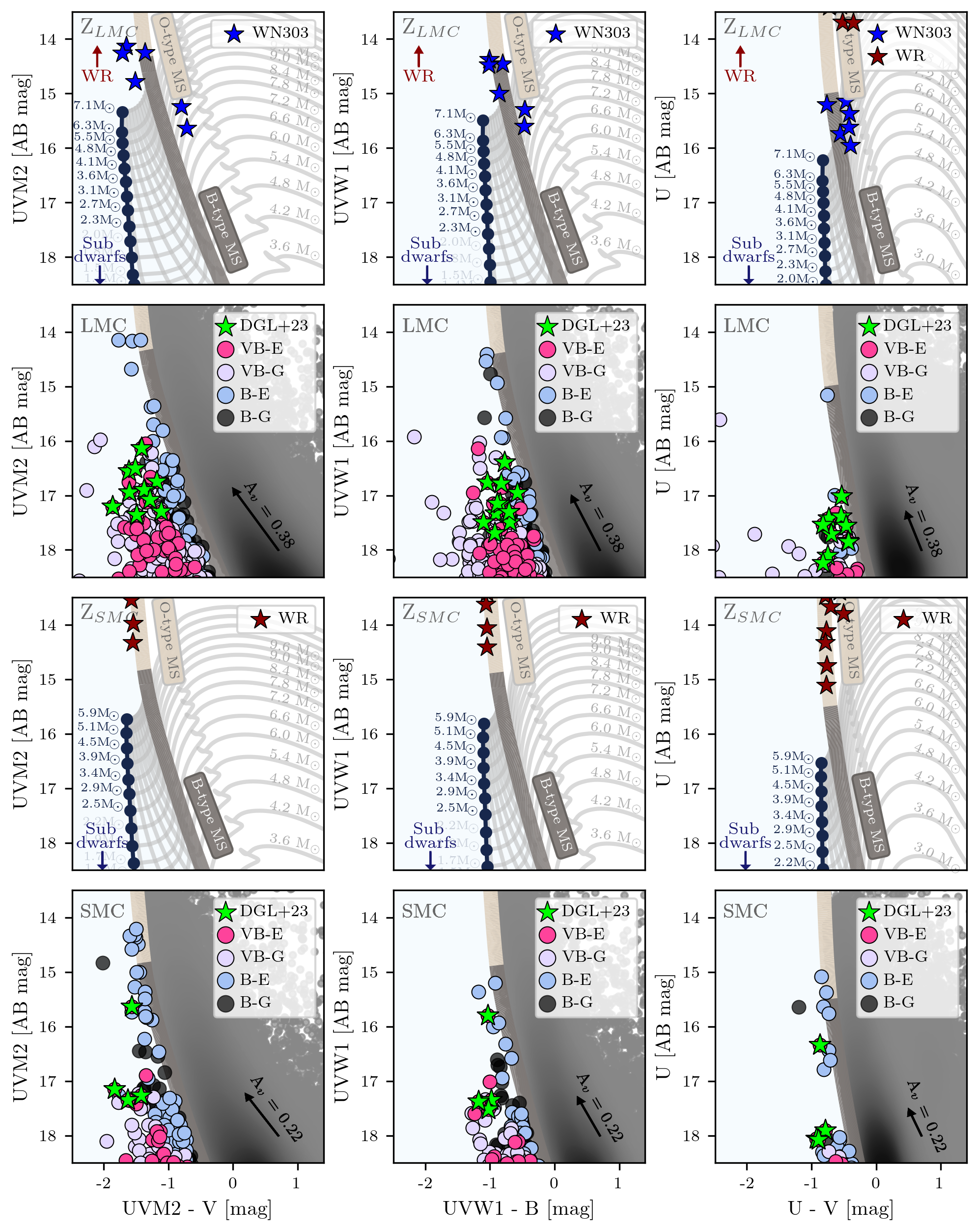}
    \caption{\emph{Top row:} Theoretical \glspl{CMD} showcasing evolutionary models at the metallicity of the LMC. Stripped star models are seen here as the dark blue dots connected with a vertical line and \gls{MS} star models are gray lines that are located to the right of the thick beige and gray \gls{ZAMS} line. Between them is a composite grid of these models, simulating the photometry of a range of binary systems. The locations of WR stars (including a set of weaker-wind objects known as WN3/O3 stars; see \S~\ref{sec:WR}) and subdwarfs are plotted for reference \citep{Breysacher1999,Massey2001,Neugent2017}. The amount of separation (in color) between the stripped star and ZAMS models decrease when considering photometry bands that are either closer together or located in the optical instead of the UV. \emph{Middle and bottom rows:} observational \glspl{CMD} for the LMC and SMC, respectively. Stripped star candidates appear bluewards of the \gls{ZAMS} and are colored by the ranking system applied in \S\ref{sec:rankings}. Black arrows indicate the direction of the reddening vector in each CMD. The length of the arrows are scaled to the baseline \Av{} value adopted for each galaxy (0.38 mag for the LMC and 0.22 mag for the SMC). 
    }
    \label{fig:cmd}
\end{figure*}

\subsection{Location within the Magellanic Clouds}

The location of various populations of stars within the Magellanic Clouds has been used as a proxy for their age and evolutionary history---both by comparing directly to maps of star formation history \citep[e.g.][]{Badenes2015,Sarbadhicary2021,Ogrady2023} and by assessing their relative proximity to other classes of stars \citep[e.g.][]{Smith2015}. While performing a similar analysis on our candidate systems will require a detailed assessment of our sensitivity in different regions of the Clouds due to the impact of clustering on the \emph{Swift}-UVOT images, we can still very broadly assess their global environments.   

In Figure~\ref{fig:candidatecoverage} we plot the positions of our final candidates to assess their global distribution across the Magellanic Clouds. The bulk of the SMC candidates lie towards the northern star forming regions of the galaxy \citep{Rubele2018}, in the direction of two massive star clusters NGC 346, and NGC 371. 
Within the LMC, candidate sources are generally sparse along the bar, with a few exceptions. Instead, a higher density is observed northeast of the bar, possibly due to the relative lack of active star-forming regions within the bar itself \citep{Harris2009}.

We note that we have very few candidates (5) within 30 Doradus, the most active star forming region in the Local Group with the highest density of massive stars \citep{WalbornBlades1997}. This is likely due to crowding effects. In particular, numerous areas within 30 Doradus are above the count rate level where coincidence loss artifacts dominate the \emph{Swift-}UVOT images (and are hence masked in our analysis, as outlined in \S \ref{sec:inputimages}). Indeed, when we compare to the HTTP survey, which conducted HST photometry of 30 Doradus \citep{Sabbi2016}, we find that less than a percent of their sources have UV magnitudes in the final SUMS catalog presented in Table~\ref{tab:minifullcat}.
Identification of stripped star candidates in this and other dense regions of the Clouds requires photometry from a higher resolution instrument such as HST. Though we also note that for very young regions such as 30 Dor ($\sim$1-10 Myr; \citealt{Schneider2018}) they could be too young for the formation of intermediate mass stripped stars \citep[cf.][]{Gotberg2019}.

{
We can also \emph{broadly} assess whether the candidates are located in environments consistent with the expected lifetimes of stripped stars (~10–100 Myr), by comparing their positions to the star formation history (SFH) maps derived from MCPS data for the LMC \citep{Harris2009} and SMC \citep{Harris2004}. For each candidate, we adopt the SFH corresponding to the \emph{nearest} spatial cell in these maps. Each cell is typically 12\arcmin{}x12\arcmin{}, as defined in the Harris \& Zaritsky studies, and thus represents the typical spatial resolution of this comparison.

First and foremost, in every case, the nearest cell shows at least some star formation activity within the 10–100 Myr range. Moreover, 43.4\% (35.9\%) of the candidates in the LMC (SMC) fall in cells where star formation from 10–100 Myr ago dominates the recent history. The remainder are found in regions where a larger fraction of the star formation occurred either very recently ($<$10 Myr; 24.5\% and 34.2\%)  or long ago ($>$100 Myr; 32.2\% and 29.9\%) in the LMC and SMC, respectively.

Spatially, we observe that the candidates found within regions dominated by older star forming regions are more widely dispersed than those in younger regions, potentially indicating that some are the surviving members of older stellar populations or have been displaced from their birth sites via binary supernova kicks or dynamical interactions. A subset of candidates are also clustered near regions dominated by very recent star formation. While this may indicate the presence of some younger objects within our catalog list, in several cases these clusters lie near the edges of the 12\arcmin{} regions within the SFH maps, where adjacent areas are dominated either by 10-100 Myr or $>$100 Myr activity. This may suggest that these candidates occupy zones with more complex or overlapping star formation histories, such as transitioning or mixed-age regions.
}

\subsection{Location on the CMD}\label{sec:cmd_location}

The stripped star candidates in our sample are identified based on theoretical predictions that they should contribute a UV-excess to their binary systems that moves them bluewards of the \gls{ZAMS}. In Figure~\ref{fig:cmd} we compare these theoretical expectations to our final candidate population. 
We plot a series of three \glspl{CMD} built using different UV and/or optical colors, which span a range of separation in wavelength (from UVM2-V on the left to U-V on the right). In the middle and bottom rows, we plot the extinction corrected apparent magnitudes for the candidates identified in \S\ref{sec:rankings} for the LMC and SMC, respectively. Candidates are color-coded by their rank, and in each panel we plot only the candidates that are bluewards of the \gls{ZAMS} \emph{in that particular \gls{CMD}}. Candidates span the full range of magnitudes considered by our selection process (14 mag $<$ m$_{\rm{UV}}$ $<$ 19 mag) and their colors relatively uniformly fill the space between the \gls{ZAMS} and $\sim$1--1.5 mag bluer, depending on the filter combinations used. Notably, the number of candidates which show colors bluewards of the \gls{ZAMS} dramatically drops if only optical filters are used (e.g., in the right panels which use U-V on the horizontal axis) compared to the \glspl{CMD} that utilize UV filters. 

To provide context for these colors and magnitudes, in the top row of panels, we plot the expected location for a number of theoretical models. In particular, we plot:
\begin{itemize}[leftmargin=1em]
 \setlength\itemsep{-0.1em}
    \item A set of single star evolution models (grey lines to the right of the \gls{ZAMS}) with masses between 1.0 and 12.0 M$_{\odot}$, in steps of 0.6 M$_{\odot}$. Models are from MIST with an LMC-like metallicity, as described in \S~\ref{sec:compositemodels}.
    \item A set of stripped stars half-way through central helium burning (dark blue lines) with $Z=0.006$, described in \S~\ref{sec:strippstarmodels}.
    Models shown have masses between 
    $\sim1\mathrm{-}7\; \mathrm{M}_\odot$
     which are intermediate between a majority of Wolf-Rayet stars and subdwarfs.
    \item A set of composite models (a grey ``web'') built by combining some of the MIST \gls{MS} and stripped star models, as described in \S~\ref{sec:compositemodels}.
\end{itemize}

From this, it is clear that our photometric sample has similar UV magnitudes as predicted for $\sim$3--12M$_\odot$ B-type \gls{MS} stars, but bluer colors. Instead, they broadly occupy the region of parameter space predicted for binaries containing $\sim$1--12~M$_{\odot}$ \gls{MS} stars with $\sim$1--7~M$_\odot$ stripped stars. Moving from left to right, the distance between the stripped star models and the \gls{ZAMS} decreases. As a result, fewer combinations of stripped stars plus \gls{MS} companions are expected to show a detectable blue excess, once photometric error is taken into account, consistent with our observed candidates.

While there are some outliers, it is possible to identify a ``color cutoff'' in most of the middle and lower panels, bluewards of which many fewer candidates are found. This color closely aligns with the location of the stripped star models in the top panel,\footnote{We note that this is by design, as we placed constraints on the colors of our candidates relative to expectations for these models (see \S~\ref{sec:SEDquality}. The fact that the observed cutoff is at slightly bluer colors than the models likely reflects the buffer of $\sim$0.2 mag we allowed when selecting candidates.} and corresponds to the expectation when the observed color probes the Rayleigh-Jeans tail of a blackbody distribution and is hence no longer sensitive to temperature. As discussed in \citetalias{Drout2023}, candidates at these bluest colors are expected to have \glspl{SED} dominated by the stripped star. 
Conversely, the closer the object is to the ZAMS, the more the flux contribution from the \gls{MS} companion increases. However, we note that for the system to show UV excess, the stripped star must have a significantly larger bolometric luminosity than its \gls{MS} companion \citepalias{Drout2023}

\begin{figure*}[ht!]
    \centering
    \includegraphics[trim={0, 0, 0, 0},clip,width=0.9\textwidth]{ 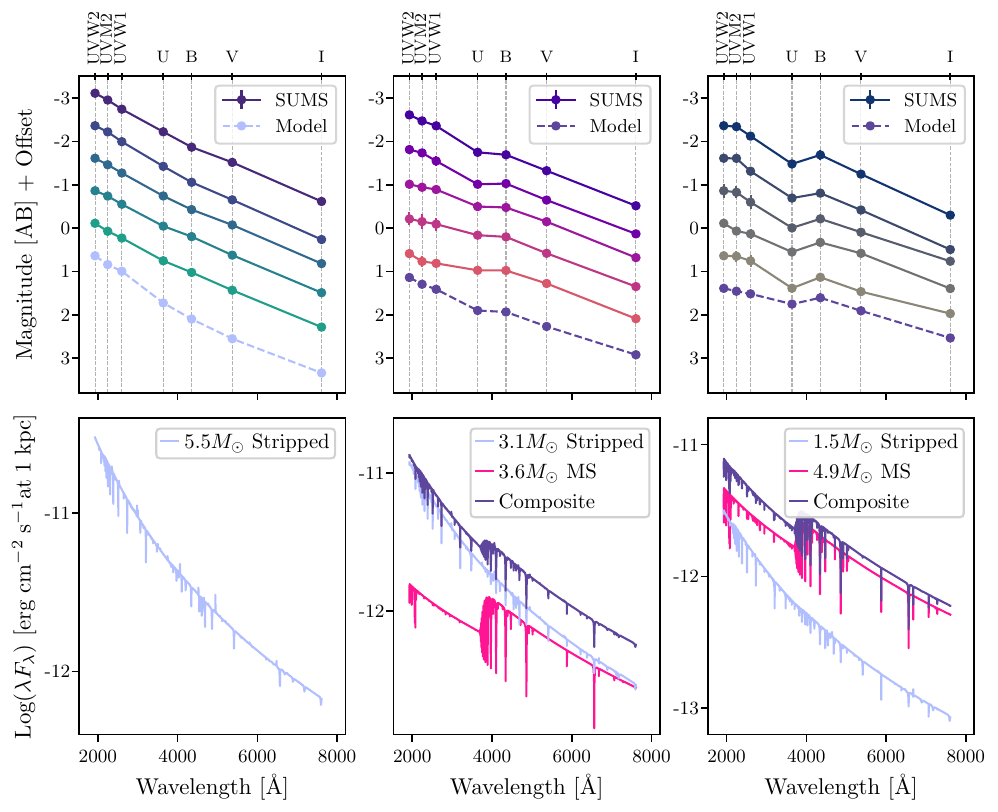}
    \caption{\emph{Top Panels:} {Examples of the three main \gls{SED} types found in our candidate sample, distinguished by their behavior between the U-band and B-band (located at $\sim$3700~\AA\ and $\sim$4400~\AA, respectively): increasing (left), approximately constant (middle), or decreasing (right) flux from U to B. While all \glspl{SED} broadly show increasing flux toward bluer wavelengths overall, this behavior in the U–B region helps separate the classes. This trend aligns with theoretical expectations for binaries containing stripped stars that contribute varying amounts to the system’s total flux. The colored dashed line in each panel shows the expected \glspl{SED} for systems where the stripped star contributes 100\% (left), 55\% (middle), and 16\% (right) of the optical flux.}
    \emph{Lower panels:} Full UV-optical spectra for the model systems that were shown in the top panels. In the middle and right panels we show spectra for both the stripped star and MS companion as well as the total flux from the system.}
    \label{fig:sedtypes}
\end{figure*}

As described in \S~\ref{sec:rankings}, approximately half of our candidates lie within $\sim$0.4 mag of the \gls{ZAMS}, while the remainder exhibit bluer colors corresponding to our `Blue' and `Very Blue' rankings respectively. While these categories may offer insight into the relative numbers of stripped stars with different companion types, we caution that spectroscopic confirmation of each system is necessary as contamination from typical \gls{MS} stars is expected to be more significant near the ZAMS, as discussed in \S\ref{sec:extinction_on_candidates}. Black arrows in the middle and lower panels of Figure~\ref{fig:cmd} illustrate the magnitude and direction that our candidates move given our assumed extinction correction. Given this vector direction, any \gls{MS} contaminants introduced by over-corrected reddening 
are expected to be B-type stars. We further address possible contamination sources present in our candidate sample in \S~\ref{sec:otherblue}.

\subsection{Main SED Shapes in Candidate Sample}

While we select our candidates based on their locations in multiple \glspl{CMD}, we can also examine the broad-band \glspl{SED} from our UV to optical photometry. Upon visual inspection of the candidates with the highest data quality (i.e., those in the `Excellent' categories described in \S~\ref{sec:rankings}) we identify three broad \gls{SED} shapes. These are differentiated mainly by the overall morphology between the U and B bands, which span the Balmer break (they have reference wavelengths of $\sim$3634 \AA\ and $\sim$4405 \AA\, respectively). In particular, we identify systems with \glspl{SED} that (i) increase smoothly in flux between all bands when moving to bluer wavelengths (ii) generally increase in flux when moving to bluer wavelengths but are 
roughly equivalent between the U and B bands and (iii) generally increase in flux to the blue, but show a decrease in flux between the U and B-bands. Examples of all three \glspl{SED} types are shown in color in the top panels of Figure~\ref{fig:sedtypes}.

These three distinct SED shapes are broadly consistent with expectations for a population of stripped star binaries where the stripped star and a \gls{MS} companion contribute varying amounts to the overall flux.\footnote{We note that this is not unexpected, as we use the composite stripped star plus MS model grid from \citetalias{Drout2023} as a baseline when performing cuts on \gls{SED} quality in \S~\ref{sec:SEDquality}. 
We note that a set of stars identified by their apparent UV excess in that section show an increasing flux at redder filters, specifically between B and I band, and were therefore removed from our sample. Otherwise, these systems would have represented an additional SED shape present in our candidate population.} 
This is demonstrated in Figure~\ref{fig:sedtypes}, where we also show one of the theoretical models from \citetalias{Drout2023} in each of the top panels, connected by a dashed line. From left to right, the theoretical models shown are: (i) a stripped star with 5.5 M$_{\odot}$, (ii) a composite system with a 3.1 M$_{\odot}$ stripped star and a 3.6 M$_{\odot}$ \gls{MS} star, and (iii) a composite system with a 1.5 M$_{\odot}$ stripped star and a 4.9 M$_{\odot}$ \gls{MS} star. These three examples, where the stripped star contributes 100\%, 55\%, and 16\% of the flux in the V-band, respectively, show the same \gls{SED} morphology as the observed candidates in the top row.

\begin{figure*}
    \centering
    \includegraphics[trim={0, 0, 0, 0},clip,width=0.95\textwidth]{ 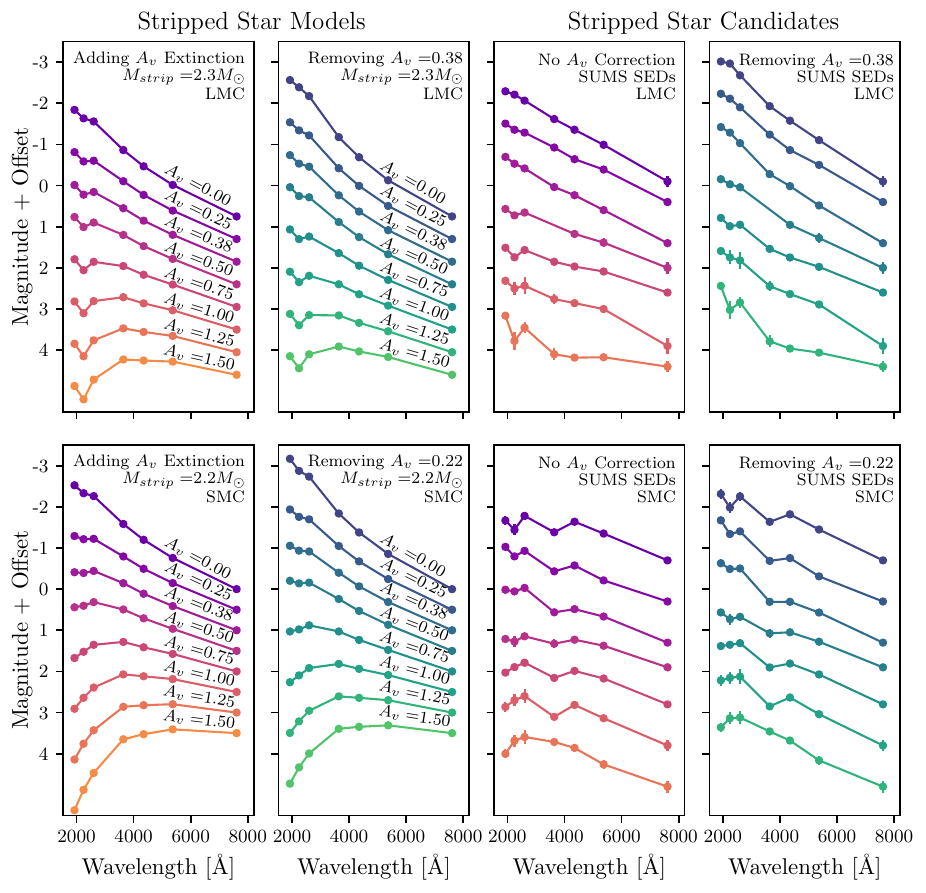}
    \caption{A demonstration of the impact of extinction on the shape of the UV-optical SED. \emph{First column:} A roughly 2$M_\odot$ stripped star model with a range of extinctions applied at the metallicity of the LMC (top) and the SMC (bottom). \emph{Second column:} We correct each stripped star model in the former column with the baseline amount of reddening assumed for our catalog. When under-corrected for reddening, stars in the LMC can show a `V-like' morphology between the three \emph{Swift} UV bands, due to the presence of the 2175 \AA\ bump in the LMC extinction curve (see Figure~\ref{fig:extinctioncurve}). \emph{Third column:} Selected SEDs from our candidate stripped star sample that demonstrate different shapes that may be due to spatially variable levels of extinction. \emph{Fourth column:} the observed SEDs from the previous column, corrected for extinction.}
    \label{fig:sedextinction}
\end{figure*}

For reference, the lower panels display the full UV-to-optical spectra of the models, highlighting the relative flux contributions from each stellar component. While not strictly one-to-one, the three \gls{SED} morphologies identified here broadly correspond with the \emph{spectroscopic} classes defined by \citetalias{Drout2023}: Class 1 systems, where the optical spectra are dominated by the stripped star; Class 2 systems, which show features from both the stripped star and a \gls{MS} companion; and Class 3 systems, whose optical spectra resemble a B-type \gls{MS} star. In \citetalias{Drout2023}, these classes were associated with stripped star contributions of $\gtrsim$80\%, $\sim$20–80\%, and $\lesssim$20\% of the total optical flux, respectively.

\subsection{Impact of Extinction on SED Shapes}\label{sec:extinction_on_seds}

As described in detail  (\S\ref{sec:extinction} and \S\ref{sec:extinction_on_candidates}), one of the largest uncertainties when identifying candidate stripped star binaries based on the presence of a UV excess is the amount of extinction assumed. While we adopted a single \Av{} value for each galaxy when performing our initial candidate selection, the impact of extinction on the \glspl{SED} presented in our catalog could, in principle, be used to better constrain the level of reddening towards individual stars. To illustrate this, in the left panels of Figure~\ref{fig:sedextinction} we apply increasing \Av{} extinction values to a $\sim$2\Msun{} stripped star model. In the LMC (top left panel) this produces a deepening ``V-shape" in the three \emph{Swift} UV magnitudes. However, in the SMC (lower left panel) these magnitudes flatten out and turn downwards. This difference occurs due to the presence or absence of the 2175 \AA\ bump in the extinction curve for each galaxy, which overlaps the middle UV filter (UVM2), as shown in Figure~\ref{fig:extinctioncurve} (see also the discussion in \citealt{Hagen2017}). 

In the second panel of the top row in Figure~\ref{fig:sedextinction}, we show the impact of correcting the same set of LMC models for the baseline extinction value of \Av{} $=$ 0.38 mag used in our candidate selection. After applying this correction, stars with initially low extinction values exhibit relatively smooth \glspl{SED}. In contrast, stars with higher initial extinction—those that are therefore \emph{under}-corrected in our baseline analysis—retain a characteristic ``V-shape'' in their \glspl{SED}, even after correction.

This effect offers a potential diagnostic for identifying stars with higher extinction than we adopt---either among our current candidate sample or among systems that overlap with the \gls{ZAMS} in our analysis, but that may exhibit an intrinsic UV excess if located in regions of greater reddening. The third and fourth panels in the top row of Figure~\ref{fig:sedextinction} illustrate this further: the third panel shows candidate LMC stars without any extinction correction, while the fourth shows the same stars corrected for \Av{} $=$ 0.38 mag.

Overall, the \gls{SED} shapes of these candidates, particularly in the UV, resemble those of the models in the first two panels. Candidates that continue to show a prominent ``V'' in the UV portion of their \glspl{SED}, even after correction, may lie in regions of higher extinction. While the observed flux between the U and B-band does not always match the stripped star models, this discrepancy may be attributed to the presence of a companion, as discussed in the previous section.

The bottom row of Figure~\ref{fig:sedextinction} presents an analogous set of plots for the SMC. The second panel shows the theoretical models corrected for our baseline extinction value of \Av{} $=$ 0.22 mag. The third panel displays observed SMC candidates with no extinction correction, and the fourth shows those same stars after applying the baseline correction. As seen in the second panel, the lack of a 2175~\AA\ bump in the SMC extinction curve makes it more difficult to identify hot stars in regions of elevated extinction without spectroscopy. Without a predicted UV ``V'' feature, such stars may more easily resemble cooler stars.

Nevertheless, the third and fourth panels reveal that some SMC candidates do exhibit a ``V-shaped'' \gls{SED}, even after correction. This could reflect additional extinction along the line of sight—either due to unaccounted foreground Milky Way dust (which we do not explicitly model; see  \S~\ref{sec:extinction}) or due to localized variations in the SMC extinction curve itself. In particular, these stars may trace sightlines where the 2175~\AA\ bump is present.

Our analysis adopts the classical SMC extinction curves from \citet{Gordon2003}, who detect the bump only in the SMC wing—outside the \gls{SUMaC} footprint. However, as noted in \S~\ref{sec:extinction}, the prevalence and distribution of the 2175~\AA\ bump in the SMC remain uncertain. The \citet{Gordon2003} curve is based on only five stars; this number increased to nine in \citet{MaizApellaniz2012}, and most recently to twenty-two in \citet{Gordon2024}. Of these, four show evidence of the 2175~\AA\ feature, with only one located in the wing. The small sample size has historically limited the robustness of the SMC extinction law, largely due to a lack of stars with reliable spectral types.

It is possible that new spectroscopic surveys such as \citet{Shenar2024} coupled with photometry from our catalog could be used to further improve our understanding of the SMC extinction curve. It would also be interesting to investigate whether the SMC stars with the V-shape \gls{SED} in our sample correspond to regions in the SMC with higher abundances of polycyclic aromatic hydrocarbon molecules (PAHs) (e.g., in the map of \citealt{Chastenet2019}) as these are a popular hypothesis for the origin of the 2175 \AA\ extinction bump \citep{Joblin1992}. 
In fact, such an analysis was recently carried out in the southwest bar of the SMC by \citet{Petia2025}, who modeled \glspl{SED} incorporating near-UV HST photometry of approximately 500,000 sightlines from the SMIDGE survey (GO-13659; \citealt{Petia2017}) to examine their dust properties. From their sample, they identify nearly 200 candidate sightlines exhibiting the 2175~\AA\ extinction bump. However, they found no correlation between these candidates and regions enriched in PAHs. While the HST photometry used in their study is deeper and more precise, their analysis covers roughly only 0.5\% of the SMC area included in our survey. Expanding their approach to a broader region—such as that covered here—could provide valuable insight into the spatial variation of SMC dust properties and any potential links to PAH abundance.

In conclusion, it is possible that the morphology of the photometric SED can be used to construct a more precise extinction estimate on a star-by-star basis. This could be achieved either from photometry alone (especially in the LMC and SMC regions with a 2175~\AA\ bump) or coupled with a best-fit spectral model derived from spectroscopic fitting \citep[e.g., as in][]{Gotberg2023}. Such an approach could refine the list of candidate stripped star binaries and improve our understanding of dust properties in the SMC.

\subsection{SED Fitting of the Candidate Sample}\label{sec:SED_fitting}

Follow-up spectroscopy can be used to both confirm the stripped nature of the candidates and provide more reliable mass estimates. To offer a preliminary mass estimate, we fit the \glspl{SED} of our systems to a grid of stripped star plus \gls{MS} companion models. In doing so, we  gain insight into the candidate mass distribution and the types of systems accessible via the UV-excess method. 
Here, we perform $\chi^2$ fitting between our observed sample and a grid of models where we vary the (i) \gls{MS} mass (ii) stripped star mass, and (iii) line-of-sight extinction. 

From Figure~\ref{fig:cmd} it is clear that our sample likely includes objects with \gls{MS} companions below 2 \Msun, which was the lower limit of what was included in the binary grid of \citetalias{Drout2023} (see \S~\ref{sec:compositemodels}).
Therefore, for this assessment, we use the expanded grid of composite binary models described in \S~\ref{sec:compositemodels} which includes MIST MS stars. In addition, we also include cases of single stripped helium and \gls{MS} stars. We then apply a range of extinction values. We consider $A_{\rm{V}}$ $=$ 0.01 to 1.0 mag, in steps of 0.02 mag.  The upper end of this range is defined by the point at which we would no longer expect most stripped star binaries to have a detectable UV excess (see description in \S~\ref{sec:extinction_on_candidates}).

\begin{figure*}
    \centering
    \includegraphics{ 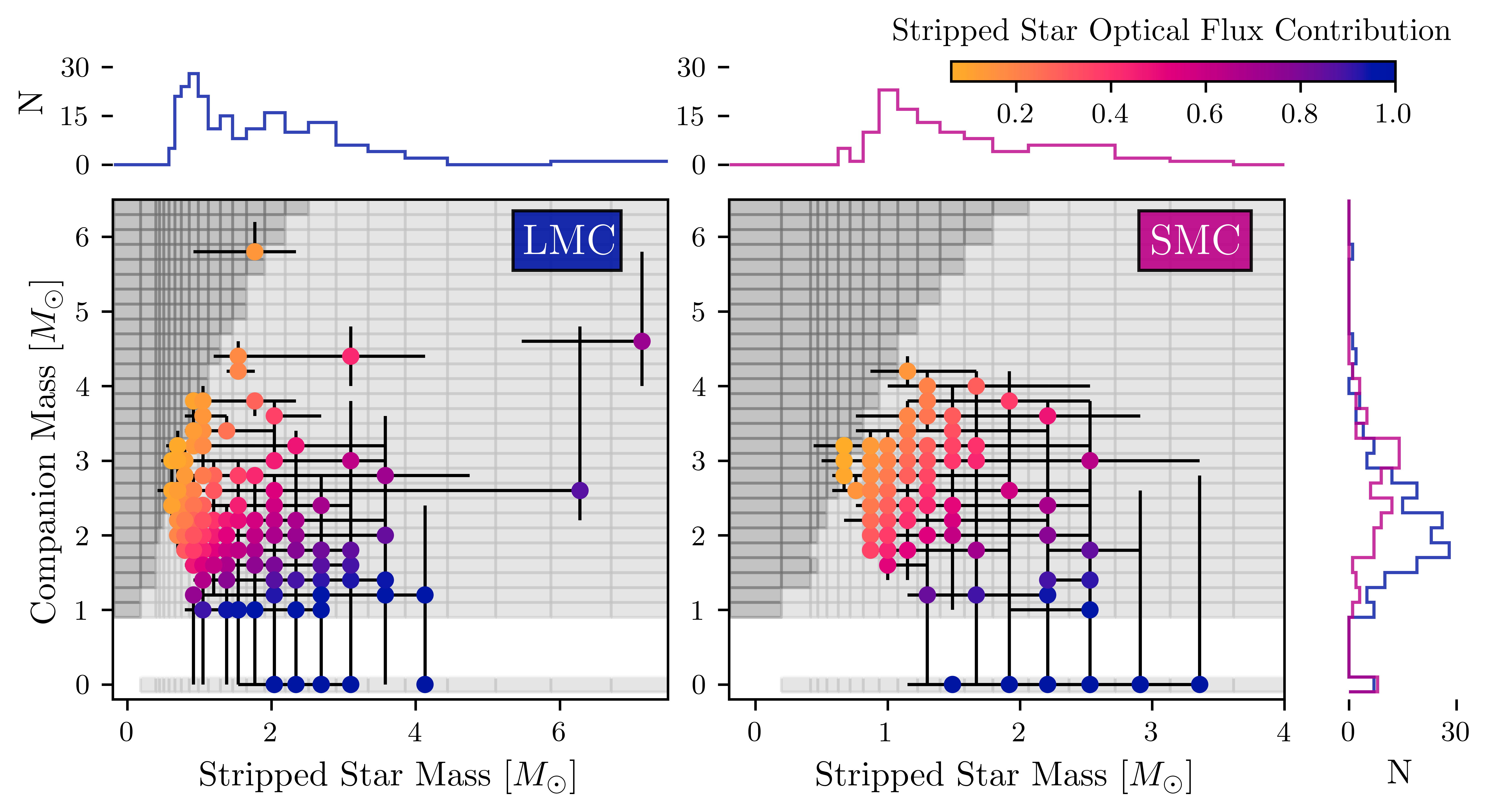}
    \caption{Results from performing SED fitting between our candidate sample and a grid of stripped star plus MS star binaries. LMC stars stare shown on the left and SMC stars on the right. Best-fit stripped star candidates cover a wide range of stripped star and companion \gls{MS} star masses. The grey boxes indicate our composite grid of stripped star and MS star models. Regions that are not predicted to have a UV-excess in the UVM2-V color combination are a darker shade of grey. Our candidates are shown as scatter points, colored by their prediction contribution to the system's optical light. We note that points for multiple sources overlap within the plot due to the spacing of the model grid. Histograms on the top and right indicate the number of stars found for a given mass. See Section~\ref{sec:SED_fitting} for further discussion.}
    \label{fig:sedfitting}
\end{figure*}

We note that the brightness of our candidate sample also extends beyond our composite grid, which only contains combinations with stripped stars up to 7.23 \Msun{} (see \S~\ref{sec:models}). To partially mitigate this, we chose to only fit objects in our candidate sample with UVM2 magnitudes which are fainter than 16 AB mag prior to extinction correction. This excludes about \NSEDBrightCutPercent{}, or \NSEDBrightCut{} of our  
candidates. However, we caution that the results of our fitting are still constrained by our input grid. For example, if a binary with a relatively high mass stripped star were located in a region of higher reddening, but its intrinsic properties are outside the extent of our grid (i.e., similar to the preferred properties for Star 1 in \citealt{Gotberg2023}) the $\chi^2$ fitting performed here may instead return a lower mass stripped star in a region of lower reddening. In addition, our input grid only contains stripped stars and MS stars at a specific point in their evolution (see \S~\ref{sec:models}). It therefore may not fully capture the morphology of systems containing younger/older stars or the potential impact of mass transfer onto the MS companion, such as envelope inflation \citep{lau_expansion_2024} or the development of emission line features \citep{rivinius_classical_2024}. 

With these caveats in mind, we compute the $\chi^2$ statistic between each observed candidate and all models in our grid. For each candidate, we record the stripped star mass, companion mass, and \Av{} value corresponding to the best-fit (minimum $\chi^2$) model, along with the 1$\sigma$ confidence intervals for each parameter. We find that  \NSEDChiCutPerent{} or \NSEDChiCut{} candidates yield poor fits—defined here as being inconsistent with all models at a significance level of $p > 0.0005$ (i.e., $>5\sigma$). Additionally, some candidates are only weakly constrained in one or both of the component masses.

This may be due to a number of reasons including: poor data in one or more photometric bands, a mismatch between the evolutionary states of the component stars and what is included in our grid, or a different origin for some objects in our candidate sample (which will be discussed further in \S~\ref{sec:otherblue}). In Figure~\ref{fig:sedfitting} we plot results for \NSEDPlotted{} candidates where the minimum $\chi^2$ implies a reasonable fit and which also meet one of two sets of additional criteria:
\begin{itemize}[leftmargin=1em]
    \setlength\itemsep{-0.1em}
    \item We plot \NSEDFracErrTotal{} candidates (\NSEDFracErrLMC{} in the LMC and \NSEDFracErrSMC{} in the SMC) where the masses of both the stripped star and MS companion are constrained within a factor of $\sim$2. We enforce this by requiring that the ``fractional error'' on the mass of both stars is less than one, where the ``fractional error'' is defined as the total width of masses allowed within the 1$\sigma$ level divided by the mass of the best fit model. 
    \item We plot \NSEDIsoTotal{} additional objects (\NSEDIsoLMC{} in the LMC and \NSEDIsoSMC{} in the SMC, corresponding to \NSEDIsoPercent{} of the well-fit candidates) that are consistent with zero contribution from a \gls{MS} companion within the 1$\sigma$ level and the maximum MS mass allowed with 1$\sigma$ is $<$3 M$_\odot$. 
\end{itemize}

While objects in the second category do not meet the fractional error requirement on the MS mass that we enforce for objects in the first category, we note that the presence of a $\sim$1--2 M$_\odot$ MS star has very little effect on the photometry of binaries containing higher mass stripped stars (as shown by the colorbar in Figure~\ref{fig:sedfitting} these objects have optical flux contributions from the stripped star of over 90\%). These objects may be similar to the Class 1 objects from \citetalias{Drout2023}, which appear to be dominated by a stripped star and show no clear signatures of a luminous companion in their optical spectra. In fact, for $\sim$42\% of the stars in this category, the mass of a luminous MS companion is constrained to be $<$2 \Msun\ (within 1$\sigma$).
Although mass estimates from SED fitting are rough, if these candidates are confirmed to be stripped star systems without signatures of a luminous companion it would be relatively straight forward to obtain their surface properties via spectral fitting as was done in \citet{Gotberg2023} (whereas systems with visible signatures of MS companions would require disentangling.) As the overall properties of stripped stars still suffer from low number statistics these measurements would be useful in constraining uncertain parameters in binary physics. Also, stripped star dominated systems may be hosts to compact object companions---likely a necessary evolutionary phase for some hydrogen-poor supernovae \citep{Yoon2017,Tauris2015} and gravitational wave mergers \citep{Tauris2017}. 

On the other hand, we note that 72 of the well-fit objects in our sample (16\%) are consistent (within 1$\sigma$) with models containing only a MS star. However, $\sim$90\% of these objects also had solutions within 1$\sigma$ where the stripped star mass ranged up to $3\Msun$. Nevertheless, as some of these objects may represent cases of contamination in our sample by \gls{MS} stars in lower reddening regions of the Clouds (as will be discussed further in Section~\ref{sec:ms}) we have chosen not to add these systems to Figure~\ref{fig:sedfitting}. While these could be systems hosting lower mass stripped stars, (possibly subdwarfs), we are less confident that a stripped star is present based on these SED fits.

Overall, the stripped star masses returned by the $\chi^2$ fitting span $\sim$1--7$\Msun$ 
(the latter of which is the edge of our grid). While we emphasize that the completeness of our sample has not yet been characterized, we find larger numbers of candidates at lower masses, consistent with expectations from the initial mass function. The best-fit models within our grid typically have MS stars spanning $\lesssim$1--6 M$_\odot$. Given that stripped stars with masses between $\sim$1--7 M$_\odot$ are expected to have initial masses in the range of $\sim$5--18 M$_{\odot}$ \citep{Gotberg2018}, the mass estimates found here highlight that the entire hydrogen-rich envelope that was stripped from the primary star likely did not accrete onto the companion. This is consistent with the finding from \citetalias{Drout2023} that the UV excess method is primarily sensitive to stripped star binaries that have undergone non-conservative mass transfer \citep{soberman_stability_1997}. This contrasts with some other samples of post-interaction binaries, such as the Be star plus subdwarf binaries, where conservative mass transfer is favored \citep[e.g.][]{Wang2022,Klement2022}.

\subsection{Order of Magnitude Comparison of the Number of Observed Candidates to Theoretical Predictions}\label{sec:theoretical_predictions}

The number of stripped stars in a stellar population can be predicted using binary population synthesis. This has for example been done for coeval populations \citep{Gotberg2019, Gotberg2020B, Wang2024} and populations with continuous and constant star-formation \citep{Gotberg2019, Shao2021, Yungelson2024, Beryl2024}. 
In this subsection, we provide an order-of-magnitude comparison of the number of objects in our candidate sample with such theoretical predictions. 

To do this, we adopt the number predictions from \cite{Beryl2024}, since they present estimates for the Magellanic Clouds. They also use a binary population synthesis code that interpolates detailed binary evolution models and therefore more accurately models the masses of the stripped stars than for example codes that adopt the standard assumption that stripped stars can be modeled as pure stripped stars \citep[e.g.,][]{Hurley2002}. \cite{Beryl2024} predict that the LMC and SMC should each contain $N_1\sim 6,000$ and $N_1\sim 1,600$ stripped stars with mass $M_{\rm strip} > 1 M_\odot$. The subscript 1 indicates that these are numbers of stripped stars created through the first binary interaction stage. Because the predictions of \cite{Beryl2024} do not include the contribution from stripped stars produced at a later stage, through interaction with a compact companion, we adopt a simplified assumption to account for these. We follow the results from \citet{Zapartas2017} and simply assume that stripped stars produced by interaction with a compact companion is $N_2 = 0.07 N_1$. We thus obtain $N_2 \sim 420$ and $N_2 \sim 112$ for the LMC and SMC, respectively.

The UV excess identification technique that we described (see Sections \ref{sec:Candidates} and \ref{sec:properties}) and the data that we use (see Sections \ref{sec:Data} and \ref{sec:Photometry}) limit the number of stripped stars that our detection technique is sensitive to. This is because of four main reasons: (1) not all stripped star systems are expected to display UV excess because they have a bright companion \citep{Gotberg2018}, (2) some stripped star systems could have higher extinction than what we correct for and therefore do not appear to show UV excess (Section~\ref{sec:extinction_on_candidates}), (3) we removed some parts of the \emph{Swift} images because of crowding (Section~\ref{ref:crowding} and Section~\ref{sec:qualitycuts}), and (4) the \emph{Swift} images do not cover the entirety of the Magellanic Clouds (Section~\ref{sec:SUMAC} and Figure~\ref{fig:coverage}). To account for these limitations, we introduce the following formula to estimate what number of stripped star systems we are sensitive to, $N_{\rm observable}$: 

\begin{equation}\label{eq:Nobs}
\begin{split}    
N_{\rm observable} = &f_{\rm coverage} \times f_{\rm no\_crowding} \times \\ &(f_{\rm UV excess, 1} \times f_{\rm extinction, 1} \times N_1 + \\ &f_{\rm UV excess, 2} \times f_{\rm extinction, 2} \times N_2),
\end{split}
\end{equation}
where $f_{\rm UV excess, 1}$ and $f_{\rm UV excess, 2}$ represent the fraction of stars in the $N_1$ and $N_2$ groups that should show UV excess, $f_{\rm extinction, 1}$ and $f_{\rm extinction, 2}$ describe what fraction of the stars in the $N_1$ and $N_2$ groups that should show UV excess have been sufficiently corrected for extinction, $f_{\rm no\_crowding}$ is the fraction of systems that do not suffer from excess crowding, and $f_{\rm coverage}$ is the fraction of the star-forming region that is covered by the \emph{Swift} images. 

\begin{table}
\centering

\caption{Assumptions and theoretical predictions for the expected numbers of stripped stars in the Magellanic Clouds (see \S~\ref{sec:theoretical_predictions}).}

\label{tab:theory}
\begin{tabular}{lcc}
\toprule\midrule
& LMC & SMC \\
\midrule
$N_1$ & 6,000 & 1,600 \\
$N_2$ & 420 & 112 \\ 
$f_{\rm UV excess, 1}$ & \multicolumn{2}{c}{0.2} \\
$f_{\rm UV excess, 2}$ & \multicolumn{2}{c}{1} \\
$f_{\rm extinction, 1}$ & \multicolumn{2}{c}{0.7} \\
$f_{\rm extinction, 2}$ & \multicolumn{2}{c}{0.9} \\
$f_{\rm no\_crowding}$ & \multicolumn{2}{c}{0.85} \\
$f_{\rm coverage}$ & 0.40 & 0.36 \\
$N_{\rm observable}$ & 414 & 99 \\
\bottomrule
\end{tabular}
\end{table}

The various fractions described above are not trivial to determine accurately. Future work (Blomberg et al., in prep.) is performing a full forward-model of our selection process, and here we instead provide order-of-magnitude estimates. In particular, we assume the following: 

\begin{itemize}[leftmargin=*]
\setlength\itemsep{-0.1em}
    \item $f_{\rm UV excess, 1}$ and $f_{\rm UV excess, 2}$: we assume that 20\% of stripped stars with main-sequence or post-main sequence companions exhibit detectable UV excess, while 100\% of stripped stars with compact companions show UV excess. The former is based on the estimates provided in \citetalias{Drout2023} who consider the limiting case where only common envelope evolution leads to the non-conservative mass transfer necessary to produce a detectable UV excess.
    
    \item $f_{\rm extinction, 1}$ and $f_{\rm extinction, 2}$: we assume that 30\% of stripped stars with stellar companions (that have intrinsic UV excess) and 10\% of stripped stars with compact object companions are lost due to under-correcting extinction. To estimate these, we take the fraction of the cells in the 2D extinction maps of \cite{Zaritsky2002,Zaritsky2004} that overlap with the SUMaC coverage and have average extinction values that are 0.3 mag and 0.5 mag higher than what we adopted in our analysis (thresholds are based on the analysis presented in \S~\ref{sec:missing}).
    
    \item $f_{\rm no\_crowding}$: we assume that approximately 15\% of systems are lost due to crowding. To estimate this, we examine the ``blue'' sources in MCPS (defined as sources with B$-$V $<$ 0.5 mag) that fall in a similar magnitude range as our candidates. We find that $\sim$15\% of these stars in both the LMC and SMC have another star (with m$_{V}$ $<$ 20.5 mag) located within 2$\arcsec$.
    
    \item $f_{\rm coverage}$: we assume that 40\% and 35\% of the star-forming regions of the LMC and SMC are captured by the \emph{Swift} UV images, respectively. This corresponds to the fraction of the recent ($\lesssim$100 Myr) star-formation in the 2D star-formation history maps of \citet{Harris2004,Harris2009} that is covered by the SUMaC survey footprint (see \S~\ref{sec:MCPS}). 
\end{itemize}

Plugging in these numbers into Eq.~(\ref{eq:Nobs}), we find that we should be sensitive to $\sim$414 and 99 stripped stars in the LMC and SMC, respectively. We list the assumptions and predicted observable numbers in Table~\ref{tab:theory}.

This estimate for the number of stripped stars with $M_{\rm strip}>1M_\odot$ that should be detectable by our method are slightly low compared to the number of candidates presented in Section~\ref{sec:rankings}. However, the numbers of predicted observable systems and candidates are in the same order of magnitude. {If we instead work backward, assuming our candidate numbers to be ground truth and otherwise keeping the same assumptions, we would estimate that the number of stripped stars present in the clouds are a factor of 1.6 and 3.8 higher than the predictions from \cite{Beryl2024} for the LMC and SMC, respectively.} Given (i) the uncertainty in many of the parameters estimated in Table~\ref{tab:theory}, (ii) the uncertainty in the predicted number of stripped star systems due to various binary evolution and mass transfer prescriptions (see e.g., \citealt{Yungelson2024, Beryl2024}) {(iii) the fraction of young star formation ($f_{\rm coverage}$) that we assume is based on SFH maps that use only single star stellar evolution models and do not account for binaries and (iv)} 
the fact that the nature of many of our candidates systems has yet to be confirmed (see \S~\ref{sec:otherblue} for a discussion of other possible origins) we consider these numbers to be in broad agreement. 

{
We can also get a very broad sense of whether the number of candidates we identify is consistent with expectations by comparing to the number of \glspl{RSG} present in the Clouds. In particular, \glspl{RSG} are expected to correspond to stars which come from a similar initial mass range as intermediate mass stripped stars ($\sim$8-25\Msun), are in a similar evolutionary phase (post-MS,    predominately core-helium burning), but have not had their hydrogen envelope removed.\footnote{{We differentiate between an \gls{RSG} and, e.g., an AGB star if the star is  expected to be massive enough ($\gtrsim 8$\Msun{}) to reach later nuclear burning \citep[e.g.,][]{ekstrom_grids_2012}.}} While the outcomes of binary interactions are diverse, it is expected that roughly $\sim$1/3 of massive stars should undergo envelope stripping and $\sim$1/3 should evolve as ``effectively single'' stars \citep{Sana2012}. Thus, given that that observed RSG population may also include mass accretors and/or stars that underwent mergers earlier in their evolution \citep{Eldridge2018,Zapartas2019}, we expect the number of RSGs to be comparable to or greater than the number of stripped stars. We can therefore examine the number of \glspl{RSG} present in the Clouds that overlap with our survey as an upper limit to the number of stripped stars we might expect. 

Although the selection function of each population is likely not one-to-one,\footnote{{In particular, our candidates likely contain some lower mass systems, as a $\sim$1\Msun{} stripped star is thought to originate from an $\sim$5\Msun{} MS star \citep{Gotberg2018}.}} we find 1,713 \glspl{RSG} overlap our footprint in the LMC \citep{neugent_red_2020,yang_evolved_2024} and 1,499 in the SMC \citep{yang_evolved_2020,yang_evolved_2023} which is a factor of 3.3 and 5 above the number of candidates in our sample. Given that only a fraction of stripped stars are expected to show UV excess, we therefore also consider these numbers to be in broad agreement within the uncertainties present in our assumptions.
Future work further characterizing the detection efficiency of our survey (Blomberg et al., \emph{in prep.}) as well as follow-up spectroscopy will be necessary to test these predictions.}

\section{Other Possible Origins for Photometric Candidate Sample}\label{sec:otherblue}
In the previous section, we demonstrated that the broad photometric properties of our candidate systems are consistent with expectations for systems containing  intermediate mass stripped stars. However, it is possible that our catalog also contains other classes of objects. Here, we highlight other possible origins for the objects in our candidate sample. We begin with a discussion of any previous classifications for these systems listed in SIMBAD (\S~\ref{sec:candidatesimbad}) followed by a more detailed discussion of other classes of stars both in the Magellanic Clouds (\S~\ref{sec:otherblue-MC}) and foreground/background objects (\S~\ref{sec:otherblue-foreback}) that may also pass all the selection criteria outlined in Section~\ref{sec:Candidates}. Ultimately, spectroscopic follow-up will be required to confirm the nature of our candidate systems.

\subsection{Cross-match with SIMBAD}\label{sec:candidatesimbad}

Crossmatching our sample of \NFinal{} stripped star candidates with SIMBAD reveals only \Nsimbad{} matches within 1$\arcsec$ of our sources, listed in Table \ref{tab:simbad}. Of these, only \Nsimbadspectype{} have listed spectral types, suggesting that the majority of these candidate sources have not been studied in detail. In fact, over 60\% of the matched candidates are simply classified as `stars' with no further information. 
Among the \Nsimbadspectype{} classified by spectral type, 4 are identified as \gls{WR} stars, and 2 as B-type stars. Other stellar classifications within our sample include planetary nebulae, candidate Be stars, and a candidate \gls{WD}. We will further discuss these as possible alternative explanations for some of the candidate stripped stars in our photometric sample below.

We note that in Figure~\ref{fig:sums}, which presents UV–optical \glspl{CMD} for the full SUMS catalog, a population of stars listed as RR Lyrae in SIMBAD showed a UV excess. While these stars are not expected to reach temperatures sufficient to produce an intrinsic UV excess (see e.g. \citealt{For2011}), they are \emph{highly} variable in the UV, with amplitudes ranging between 2 and 8 magnitudes \citep{Downes2004,Wheatley2005,Wheatley2008,Wheatley2012,Siegel2015}. We therefore attribute this UV-excess to the non-simultaneity of the SUMaC UV and MCPS optical observations. The mismatch in observation time can lead to misleading colors in the CMD, causing some variable stars like RR Lyrae to appear anomalously blue. Such sources are subsequently excluded from our candidate sample by the SED quality cuts described in Section~\ref{sec:SEDquality}.

It is nevertheless noteworthy that these objects are bright enough in the UV to be detected in our survey. Known populations of Galactic RR Lyrae stars with published UV magnitudes and distances (e.g., \citealt{Siegel2015}) would be too faint to be observed at the distance of the Magellanic Clouds. Although Gaia-based membership assessments were considered in Section~\ref{sec:diverse}, some of these stars may still be foreground contaminants. However, their intrinsic variability offers an independent distance constraint, and a preliminary crossmatch with the catalog of \citet{Bellinger2020} indicates that at least a subset is truly consistent with Magellanic Cloud distances. These objects merit further investigation in future studies.

\begin{deluxetable}{l||c|c}
	\centering
	\tablecaption{SIMBAD Classifications\label{tab:simbad}}
	\tablehead{Classification & LMC & SMC}
	\startdata
	Total Stripped Star Candidates (\S~\ref{sec:rankings}) & \NFinalLMC & \NFinalSMC \\ 
	SIMBAD Crossmatched (\S~\ref{sec:candidatesimbad}) & 44 & 15 \\\hline
    \multicolumn{3}{c}{Main Object Type}\\\hline
    Star & 29 & 9  \\
    Planetary Nebula (PN) & 4 & 0  \\
    Wolf Rayet (WR) &  3 & 1 \\ 
    Be Candidate (Be?) & 0 & 2 \\
    White Dwarf Candidate (WD?) & 1 & 0 \\ \hline
    Eclipsing Binary (EB) & 6 & 3 \\ 
    Variable (V) & 1 & 0 \\\hline
    \multicolumn{3}{c}{Spectral Type}\\\hline
	WN-Type \big(WN3+O3V; WN3ha\big) & 3 & 1\\
    B-Type \big(B0.2V; B0-5(V)\big) & 0 & 2 \\
    \hline
	\hline
	\enddata
   
\end{deluxetable}

\subsection{Discussion of Specific Classes of Objects within the Magellanic Clouds}\label{sec:otherblue-MC}

\begin{figure*}
    \centering
    \includegraphics[width=0.9\textwidth]{ 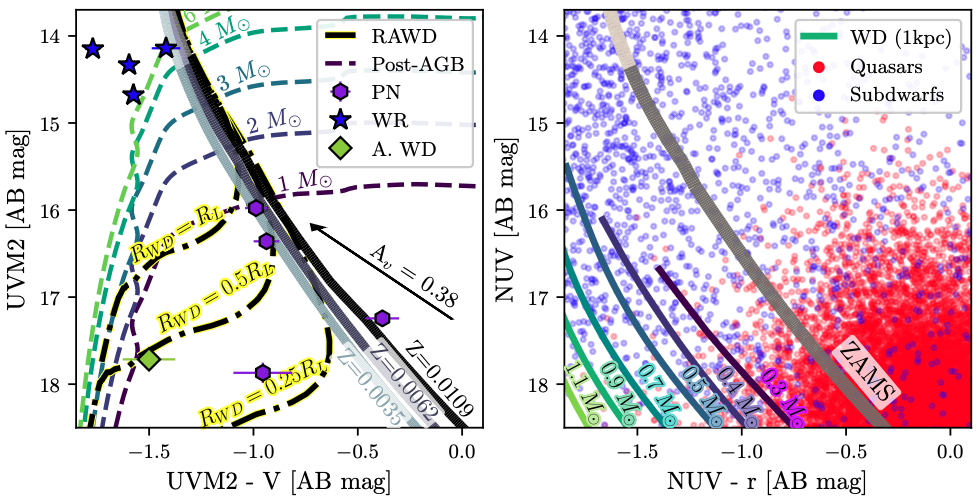}
    \caption{UV-optical color-magnitude diagrams that illustrate alternative objects that could exhibit UV excess within our candidate sample. In the left panel, we highlight systems likely belonging to the Magellanic Clouds, including theoretical models for rapidly accreting white dwarfs (RAWD,\;\S~\ref{sec:accretingwd}) 
    and post-AGB systems (\S~\ref{sec:postagbandpn})
    Additionally, we show observed sources classified in SIMBAD as planetary nebulae (PN,\; \S~\ref{sec:postagbandpn}), Wolf-Rayet stars (WR,\;\S~\ref{sec:WR}) or an accreting white dwarf (A.\ WD,\;\S~\ref{sec:accretingwd}). To demonstrate how our assumptions about extinction and metallicity within the Clouds affect the estimated UV excess, we (i) plot the reddening vector for the LMC as an arrow, with a length that corresponds to our baseline extinction correction described in \S~\ref{sec:extinction} and (ii) we  present three \gls{ZAMS} models that broadly span the LMC's metallicity range (\S\ref{sec:ms}).
    In the right panel, we show examples of non-member (foreground/background) systems that may overlap in color-magnitude space with our candidate sample. The axes show GALEX NUV magnitude versus GALEX NUV – SDSS r color, chosen for because they probe similar wavelengths as the filter combination used in the left panel. In blue, we show a set of Milky Way subdwarfs (described in \S\ref{sec:foreground}) and in red we show a set of background quasars (described in \S\ref{sec:qbackground}). We emphasize that the samples plotted are drawn from all-sky catalogs and do not overlap with the Magellanic Clouds. They are shown for representative purposes only, to demonstrate the range of apparent magnitudes and colors these objects can exhibit. Finally, cooling tracks for white dwarfs at a distance of 1 kpc are shown; beyond 4.2 kpc, these tracks fall outside our parameter space.}
    \label{fig:otherblue}
\end{figure*}

Here, we discuss in more detail other classes of objects \emph{within the Magellanic Clouds} that may have similar colors and magnitudes to our candidate sample. We visualize many of the types of objects discussed below in the left panel of Figure~\ref{fig:otherblue}, where we plot an UVM2-V versus UVM2 color-magnitude diagram in apparent magnitudes. Specific objects/models are described in the following subsections.

\subsubsection{Main Sequence Stars}\label{sec:ms}

The stripped star candidates in our sample are selected for their UV excess, however, some \gls{MS} stars could enter the sample if either (i) they are less reddened than we assume (as noted in \S~\ref{sec:MSMasquerade}) or (ii) they have some stellar property (e.g., metallicity, rotation) that is not well described by the theoretical \gls{ZAMS} model that we assume. 
Candidates in our sample that lie close to the \gls{ZAMS}, such as our `Blue' sample, are more susceptible to this type of contamination.
Here, we further discuss the potential impact of each of these effects on the sources in our current sample of candidates.

\emph{MS Stars with Low Reddening:} The impact of extinction on our set of candidates was discussed extensively in Section~\ref{sec:extinction}. There, we found that roughly 13\% of our candidates would show UV excess even if there was zero extinction along the line of sight,  (a fraction that increases to 28\% if we assume a floor of \Av{}$= 0.1$ mag due to the Milky Way, see \S\ref{sec:candextinction}), while the rest of the sample could \emph{potentially} be \gls{MS} stars with low reddening. In addition, in Section \ref{sec:2ddustmap} we found that $\sim$37\% (8\%) of our LMC (SMC) candidates are found in regions of the Clouds with lower average extinction values than our baseline assumptions in the 2D extinction maps of \citet{Zaritsky2002,Zaritsky2004}. 
Ultimately, extinction remains one of the largest uncertainties in our candidate selection and spectroscopic follow-up will be required to confirm the nature of individual systems. In particular, \citetalias{Drout2023} were able to rule out \gls{MS} stars with low reddening for the Class 1 and Class 2 systems in their spectroscopic sample due to the presence of He~\textsc{ii} absorption, along with the fact that O-type \gls{MS} stars with He~\textsc{ii} should be more luminous. In contrast, systems with higher flux contributions from the \gls{MS} companion may resemble normal B-type stars in the optical (e.g., as was observed for the Class 3 systems in \citetalias{Drout2023}). These systems would require a combination of radial velocity monitoring and/or UV spectroscopy for confirmation.

\emph{Variation in \gls{ZAMS} Location:} Our theoretical \gls{ZAMS} described in \S~\ref{sec:zams} are based on MESA stellar evolution models with zero initial rotation and the metallicities of each respective galaxy (Z$=$0.006 for the LMC, Z$=$0.002 for the SMC). Both of these factors can in principle impact the inferred location of the \gls{MS}, and variation is observed for individual stars in both galaxies \citep[e.g.][]{Choudhury2021,Bouret2021}. 
However, higher stellar rotation rates than we assume would shift the ZAMS to cooler temperatures \citep{Brott2011}. Therefore, unless the rotation is extreme, we expect these stars to appear redward of our adopted ZAMS. However, at sufficiently high rotation rates, efficient rotational mixing could cause main-sequence stars to enter the UV-excess regime \citep{Szecsi2015}. Such objects would themselves be exceptional discoveries, since they would provide the first direct evidence of rotational mixing in massive stars. Likewise, the assumed metallicity of our \gls{ZAMS} model should not significantly increase the number of \gls{MS} contaminants. To illustrate this, in Figure~\ref{fig:otherblue} we plot three \gls{ZAMS} MIST\footref{footnote:mist} models which encompass not only our above stated baseline value but also an upper and lower extreme for the LMC from \citealt{Choudhury2021}. We note that although the lower metallicity models are bluer, in general the difference is small. Even given this wide range of metallicities, the difference in the UVM2-V color is 0.05 mag on average and 0.09 mag at maximum.
These variations are smaller than the overall photometric uncertainty in our catalog (the median combined UVM2-V error for our =candidate sample is 0.2 and 0.16 mag in the LMC and SMC, respectively). 
Thus, while it is hypothetically possible that a particularly lower metallicity \gls{MS} star could pass our cuts without showing an intrinsic UV excess, overall we find the magnitude of this effect is small compared to other uncertainties.\footnote{This is especially true because we chose our extinction correction to align the theoretical \gls{ZAMS} with the large over density of stars in the \gls{CMD}. Thus, it would have to be a star with a significantly lower metallicity than the bulk of stars in the LMC/SMC, as opposed to an overall systematic offset in the metallicity adopted.}

{Finally, Oe/Be spectral types refer to rapidly rotating OB stars which show emission features in the Balmer lines due to the formation of a decretion disk \citep[e.g.][]{conti_spectroscopic_1974,rivinius_classical_2024}. Although emission lines can skew photometry, known Oe/Be stars generally do not show a UV excess (see Figure \ref{fig:cmd}) and would be easily identifiable spectroscopically due to these emission features.}

\subsubsection{Wolf Rayet Stars}\label{sec:WR}
Wolf Rayet (WR) stars are generally understood to be high mass stripped stars (M$\gtrsim$9 M$_\odot$; \citealt{Crowther2007}) which can either be self-stripped due to their strong winds, stripped through binary interaction, or a combination of the two \citep[e.g.][]{Shenar2019,Shenar2020}. Morphologically, WR stars are identified via prominent emission lines in their optical spectra that are broadened by their strong stellar winds (for a review, see \citealt{Crowther2007}). Similar to the intermediate mass stripped stars that are the target of this study, most WR stars are hot (T$_{\rm{eff}}$ $\sim$ 50-120 kK; \citealt{Hainich2014,Hainich2015}), however, their luminosities are comparatively brighter ($\log$[L/L$_{\odot}$] $=$ 5.3--6.5 versus 3.5--5.0; \citealt{Gotberg2018}).

Given their high temperatures, which in many cases exceed that of even high-mass \gls{MS} stars, one may expect WR stars to also exhibit a UV excess. However, after cross-matching with SIMBAD, we find that only 3 of the 154 known WR stars in the LMC, and 1 of the 12 known WR stars in the SMC are in our candidate sample. Upon investigations, we find other known WR stars are excluded due to one of three reasons:
\begin{itemize}[leftmargin=1em,itemsep=-0.3em]
    \item \emph{Lack of UV Data:} Cross-matching with our full UV catalog, we find that only 54 of the LMC WR stars and 7 of the SMC WR stars have UV magnitudes available. Those remaining either fall outside the \gls{SUMaC} footprint or were located in regions that are near a masked source (\S\ref{sec:inputimages}) and/or too crowded to obtain reliable UV magnitudes from the \emph{Swift} images.
    \item \emph{High Luminosities:} When selecting our target candidates, we explicitly removed sources with UV magnitudes brighter than 14 AB mag (corresponding to an absolute magnitude of M$_{\rm{UV}}$ $=$ -4.5/-4.9 mag in the LMC/SMC) to exclude very high luminosity systems.  
    \item \emph{Overlap with \gls{ZAMS}:} Interestingly, as shown in Figure~\ref{fig:cmd}, most of the WR stars present in our UV catalog have colors and magnitudes that overlap with stars at the location of the \gls{MS}. We found this was due to a combination of: (i) systems with cooler temperatures (T$_{\rm{eff}}$ $\sim$30 kK; \citealt{Hainich2014}), (ii) systems with high local extinction (A$_{\rm{V}} \gtrsim 0.8$ mag; \citealt{Hainich2014}), which can be partially due to their winds/nebular environments, and (iii) the presence of strong emission features, which can skew their UV-optical photometric colors.
\end{itemize}

Notably, the three LMC WR stars found in our candidate sample (LMC199-1, LMC079-1, and LMC172-1) are members of the new class of WN3/O3 star \citep{Neugent2012a,Massey2014,Massey2015,Massey2017,Neugent2017}. These stars show both N~\textsc{v} and He~\textsc{ii} $\lambda$4686 emission typical of WN-type WR stars and other He~\textsc{ii} absorption typical of O-type stars. However, while the spectrum appears to be a composite of a \gls{WR} and O-type star, they are too faint to include an O-type \gls{MS} star, signatures of a O-type \gls{MS} companion were not present in the UV spectra \citep{Neugent2017} and the WN3/O3 with orbital solutions suggest that for a wide range of orbital inclinations, they are unlikely to be in binaries with a massive companion \citep{Massey2023}. Instead, their differing spectral morphology is due to their having weaker winds compared to classical \gls{WR}, with mass loss rates of (\.{M} $\sim 10^{-6}$ M$_{\odot}$ yr$^{-1}$; \citealt{Neugent2017}).
The one SMC WR star in our sample (SMC AB 12) is a 
hot WN3ha star with low surface hydrogen ($T_*\sim112$kK and $X_H\sim0.2$; \citealt{Hainich2015}) and which, like most of the WR stars in the SMC, has a similar mixed emission/absorption spectral morphology as the WN3/O3 stars \citep{Massey2003,Foellmi2004}. Overall, these four stars (which are plotted for reference in Figure~\ref{fig:otherblue}) are both among the visually faintest WR stars in the Clouds (M$_{\rm{V}}$ $>$ -3.5 mag), and have some of the highest temperatures (T$_{\mathrm{eff}}$: 95-112kK; \citealt{Hainich2015,Neugent2017}), placing them within the photometric parameter space of our candidates.

More broadly, however, we do not expect to find a significant number (if any) of additional WR stars amongst our candidate sample. Using the online PoWR model grid\footnote{\url{https://www.astro.physik.uni-potsdam.de/PoWR/powrgrid1.php}} \citep{Powr1,Powr2,Powr3} we find that models corresponding to some of the lowest luminosity WR stars in \citet{Hainich2014} and \citet{Shenar2019} could have far UV magnitudes (similar to the UVW2 filter) extending down to 15.2 mag AB. 
However, \citet{Neugent2018} demonstrate that existing emission line surveys of the Magellanic Clouds should be sensitive to even weak-lined WR stars several magnitudes fainter than any identified to date---arguing that the known WR sample is complete. This is consistent with the findings of \citetalias{Drout2023} and \citet{Gotberg2023} that intermediate mass stripped stars in the Clouds show absorption line spectra. In particular, \citet{Gotberg2023} find only very weak infilling in He~\textsc{ii} $\lambda$4686 in their highest luminosity stripped star (Star 1 with log[L/L$_{\odot}$] $\sim$ 5, only slightly below the lowest luminosity WN stars). They estimate a mass loss rate of \.{M}$\sim$10$^{-6}$ M$_{\odot}$ yr$^{-1}$ for this object and place limits of {M}$\lesssim$10$^{-9}$ M$_{\odot}$ yr$^{-1}$ for a set of other systems with luminosities log[L/L$_{\odot}$] $\lesssim$ 4. While mass loss rates for intermediate mass stripped stars are still uncertain, this is broadly consistent with predictions that they can drop dramatically over a small luminosity range as their Eddington factors fall below a critical threshold \citep{Sander2020}.

\subsubsection{Accreting White Dwarfs}\label{sec:accretingwd}

White dwarfs accreting material from binary companions appear as a diverse range of astrophysical systems \citep{Warner1995}. The physics governing accretion under varying conditions has important implications for understanding several types of transient phenomena, from classical novae \citep{Gallagher1978} to Type Ia supernovae \citep{Maoz2014}. While most cataclysmic variables (CVs) are faint,\footnote{For example, a sample of CVs identified by the Sloan Digital Sky Survey \citep{Inight2023} shows absolute optical magnitudes of $\sim$5–12 mag, corresponding to apparent magnitudes of $>$23.5 mag in the LMC.} and thus would generally not appear in our sample (unless they are foreground objects; see Section~\ref{sec:otherblue-foreback} for further discussion), the position of an accreting white dwarf on the HR diagram depends strongly on its mass accretion rate.

In particular, while WDs with low mass accretion rates will undergo nuclear burning in unstable ``flashes'', fusion can proceed stably on the surface of WDs that accrete above a critical threshold \citep[$\sim$10$^{-7}$ M$_{\odot}$ yr$^{-1}$ for hydrogen-rich material and $\sim$10$^{-6}$ M$_{\odot}$ yr$^{-1}$ for helium-rich material,][]{Nomoto2007,  Wolf2013,Kato2014, Kawai1988, Wong2019}. 
With predicted values of log$[\rm{L}/\rm{L}_{\odot}] \sim 3-5$, these stably accreting WDs have similar luminosities to the stripped stars that are a target of our search, although with hotter temperatures of T$_{\rm{eff}} \sim$ 10$^5$--10$^6$ K \citep{Kawai1988,Nomoto2007,Wolf2013,Woods2016}. As a result, a majority of their flux is emitted in the soft X-rays, and they are often associated with super-soft X-rays sources (SSS) in the Milky Way and Magellanic Clouds \citep{vandenHeuvel1992,DiStefano2010}. However, with even higher accretion rates, the WDs are expected to eventually reach a (mass dependent) critical luminosity. Beyond this point, the accretion rate will exceed the stable nuclear burning rate, which will cause the WD radius to inflate and/or an optically thick wind to be driven from the surface of the WD \citep{Nomoto1979,Kato1994,Kahabka1997}. In either case, the photosphere will move outward, decreasing in temperature, and the peak of the SED can shift into the UV regime. It was with this as motivation that \citet{Lepo2013} performed a search for these ``rapidly accreting WDs'', or RAWDs, in the central bar of the SMC using a very similar technique to our search for stripped stars: i.e., searching for UV bright sources bluewards of the ZAMS and following up on them with optical spectroscopy.

Despite identifying a substantial number of UV blue sources and obtaining spectra, \cite{Lepo2013} did not identify sources which had the expected He~\textsc{ii} emission associated with accreting white dwarfs, or that were underluminous for their given temperature. However, this may be due to their candidate identification\footnote{For instance, $\sim$50\% of the observed systems in \cite{Lepo2013} are stated to be likely mismatches between UV and optical photometry. We vet our candidates for such mismatches in Section \ref{sec:qualitycuts}.} or the comparatively sparse coverage of their catalog. We therefore perform two tests to assess whether accreting white dwarfs in the Magellanic Clouds could still occupy a similar region of the UV–optical color–magnitude diagram as our sample. 

First, we examine the case of a WD accreting hydrogen or helium at rates required for stable fusion on the surface using the steady state models of \cite{Nomoto2007} and \cite{Kawai1988}. These models were calculated for a range of WD masses (0.5 M$_\odot$ $\lesssim$ M$_{\rm{WD}}$ $\lesssim$ 1.38 M$_\odot$) and accretion rates. To approximate the observables of these systems, we perform synthetic UVM2 and V-band photometry on blackbodies of the same temperature and luminosity as the models. We find that, despite their high luminosities, high-mass WDs undergoing stable burning on their surfaces have hot enough temperatures that their UV magnitudes would remain below our sensitivity limit in the Magellanic Clouds.\footnote{For example, the 1.38 M$_{\odot}$ hydrogen burning WD from \citet{Nomoto2007} would always remain fainter than 22 mag AB in the UVM2 filter at a distance of 50 kpc, despite having a luminosity of $log(L/L_{\odot}) = 4.7$. Instead, a majority of its flux is emitted in the soft X-ray band.} However, the lower mass WD models (M$_{\rm{WD}} \lesssim$ 0.7 M$_\odot$) remain at sufficiently cool temperatures ($log(\rm{T}_{\rm{eff}}/\rm{K}) \lesssim 5$) when they accrete at the upper end of the stable regime such that their emission remains in the UV \citep[e.g.,][]{chen_population_2015}. This leads to predicted UVM2 magnitudes above our threshold of 19 AB mag. As a result, we find that---especially in the fainter portion of our catalog---we could be sensitive to lower mass WDs burning hydrogen or helium stably on their surfaces.
 
Second, we repeat the procedure described in \citet{Lepo2013} to estimate the observed properties of a RAWD that has undergone radius expansion. As a fiducial test case, we consider a WD with a mass of 0.6 M$_{\odot}$ with luminosity of $\log(L/L_{\odot}) = 3.95$ (corresponding to the critical luminosity before inflation and the onset of optically thick winds in the models of \citealt{Nomoto2007}). We place this WD in a binary with a set of \gls{MS} companion stars with masses between 0.16--18.17 M$_\odot$, choosing the binary separation such that the \gls{MS} star fills its Roche Lobe. 
We then consider three scenarios for the expansion of the white dwarf photosphere, chosen to encompass a representative range of possible configurations: (i) complete Roche lobe filling, (ii) 50\% Roche lobe filling, and (iii) 25\% Roche lobe filling. 

Finally, we model the systems as the combination of a blackbody with the same WD radius and luminosity described above and a \gls{MS} stellar atmosphere model. For \gls{MS} companions with masses above 2 M$_\odot$ we use the \gls{MS} spectral models described in \citetalias{Drout2023} based on the MESA evolutionary models of \citep{Gotberg2018}. For \gls{MS} companions with masses below 2 M$_{\odot}$ we use the \cite{Pickles1998} stellar models, adopting masses, radii, and absolute magnitudes for the corresponding spectral types from the ``Modern Mean Dwarf Stellar Color and Effective Temperature Sequence'' distributed by E. Mamajek\footnote{\url{https://www.pas.rochester.edu/~emamajek/EEM_dwarf_UBVIJHK_colors_Teff.txt}} \citep{Pecaut2013}.

The results from performing synthetic UVM2 and V-band photometry on these models are shown as the yellow highlighted lines in the left panel of Figure~\ref{fig:otherblue}. While the exact location of a system is very sensitive to the amount of radius inflation, as well as the WD mass and luminosity, these models fall in the same region of parameter space as our stripped star candidates---indicating that we are also sensitive to this type of system. 
In fact, when cross-matching with SIMBAD (\S~\ref{sec:candidatesimbad}) we found that one of our candidates had previously been identified as the most likely optical counterpart of the post-nova super-soft X-ray source 1RXSJ050526.3-684628 \citep{Vasilopoulos2020}. We plot the location of this source, thought to be a $\sim0.7\Msun$ CO WD accreting at $10^{-9}\Msun\;\textrm{yr}^{-1}$, as a green diamond in the left panel of Figure~\ref{fig:otherblue}. While it falls close to the RAWD models, we note that the \emph{Swift-}UVOT observations of this source were taken during the peak of the X-ray light curve. If the system were a RAWD, and had moved into the optically thick wind regime, an anti-correlation between the UV and X-rays would be expected. Alternatively, the UV emission could be coming from the irradiated disk and companion \citep{Orio2010}, an  interpretation also favored by \citet{Vasilopoulos2020} due to a correlation between the X-ray light curve and the OGLE light curve of the optical source.  

This further emphasizes that there may be multiple means for WDs accreting near or above the rate for stable nuclear fusion to have photometric properties similar to stripped stars. Spectroscopically, though, these systems are  expected to show emission lines due to a combination of ongoing accretion/disks or optically thick winds \citep{Steiner1998,Hamann2003,Lepo2013}. This contrasts with stripped stars, which are not expected to have strong enough winds at the luminosities and metallicities probed by our survey to show strong emission features \citep{Gotberg2023}.

\subsubsection{Post-AGB Stars and Planetary Nebulae}\label{sec:postagbandpn}

The post-\gls{AGB} phase is the final stage in the life of a star with mass $\lesssim 8 M_\odot$ prior to \gls{WD} formation \citep{Garcia-Berro.E.1994.sAGBFormation}. During the \gls{AGB} phase, pulsation-driven mass loss sheds the hydrogen-rich envelope of the star until the core contracts and heats up \citep[e.g.][]{Bertolami2016}. Once the central star is hot enough to ionize the surrounding circumstellar material, a \gls{PN} is formed \citep{Weidmann2011}. 

During the final phase of post-\gls{AGB} evolution, when then central star is hottest, these systems can have similar luminosities and temperatures as our stripped star candidates. This is illustrated in Figure~\ref{fig:otherblue} where we show a set of post-\gls{AGB} models from MIST.\footref{footnote:mist} However, the duration of this phase is brief: using the MIST models for stars with a range of initial masses, we find that they are in the same UV-optical photometric region as our candidates for between 20 and 8500 years (note that the lifetime of the entire post-\gls{AGB} phase is only $\sim$$10^4-10^5$ years; \citealt{VanWinckel2003}). By comparing these lifetimes to those for the \gls{AGB} phase itself, \citetalias{Drout2023} used the number of known \gls{AGB} stars in the Clouds to estimate that there should be $\lesssim$1 post-\gls{AGB} star in our candidate samples for the LMC/SMC.

When cross-matching our candidate sample with SIMBAD, we do not find any objects explicitly classified as post-\gls{AGB} stars. However, we do identify four sources classified as \gls{PN}, all of which are located in the LMC as shown in Table \ref{tab:simbad}. These sources, plotted as purple hexagons in the left panel of Figure~\ref{fig:otherblue}, were identified by their nebulae: all four are found in optical emission line surveys (stars 215, 894, 1237, 1771 in \citealt{RP2006}), as well as in various infrared-excess studies \citep{Hora2008,KWV2015}. 

The \gls{PN} in Figure~\ref{fig:otherblue} all show redder colors than the MIST single-star post-\gls{AGB} models at similar brightnesses. This could be due to either external effects---such as interstellar reddening or the impact of the nebulae itself on the photometry of the system---or indicate that the nature of the central star(s) of these \gls{PN} differ from those in the MIST models plotted (e.g. in mass and/or binarity). In particular, while the nebulae do emit in the UV, \citet{Gomez2023} find that they tend to show redder colors than their central stars. In addition, many \gls{PN} have very strong emission lines which, similar to the WR stars described above, can impact the broadband photometry of the system. For example, [OIII] $\lambda$5007---which is one of the strong lines used to identify \gls{PN} in narrowband surveys---is located within the Johnson V-band filter used in MCPS. Additional light in the V-band could make objects appear redder in Figure~\ref{fig:otherblue}. Indeed, the one \gls{PN} that appears redwards of the \gls{ZAMS} in Figure~\ref{fig:otherblue} has an unusually high V-band magnitude compared to the rest of its UV-optical \gls{SED}.

In terms of the nature of the central star(s), we note that \citet{Gomez2023} found that binary central stars to \gls{PN} tend to show bluer colors/larger UV excess. Inclusion of this  population could therefore increase the number of PN/post-AGB stars expected within our candidates beyond the $\sim$1--2 estimated by \citetalias{Drout2023}. In particular, we note that \gls{PN} can also originate from the ejection of a common envelope following binary interaction \citep{jones2017binary}. 
It is even possible that some binary central stars of PN could 
{contain helium burning} stripped stars. 
If additional binaries are found within this population it would be notable, as there are only 10 confirmed binary central stars of \gls{PN} in the Magellanic Clouds \citep{Hajduk2014,Gladkowski2024}.

\subsection{Discussion of Possible Foreground and Background Contaminants}\label{sec:otherblue-foreback}

In Section~\ref{sec:member} we excluded sources from our list of stripped star candidates if they had \emph{Gaia} kinematics that indicated they were very likely to be foreground objects. Objects were excluded if either (i) they had a \emph{Gaia} parallax measured at more than 4$\sigma$ or (ii) they had proper motion values that were inconsistent with a sample of probably LMC/SMC stars \emph{and} had an \texttt{astrometric\_gof\_al} flag that indicated the astrometric fit was of high quality. However, it is still possible that some foreground or background objects remain in our final sample. In particular, \emph{Gaia} parallax and proper motion information was not available for \NNoGaiaMatchPercent{} of our sample, and we did not a priori reject these objects. In addition, objects located in the Galactic halo and beyond would not be expected to have significant parallax measurements and may show proper motions similar to those of objects in the Magellanic Clouds by chance. 

Here, we discuss possible types of foreground and background objects that may have blue colors and apparent magnitudes similar to our candidates sample (\S~\ref{sec:foreground} and \ref{sec:qbackground}). We then quantify the expected level of contamination using a control field offset from the Magellanic Clouds (\S~\ref{sec:control_field}). We visualize many of the classes of objects discussed below in the right panel of Figure~\ref{fig:otherblue} where we plot an (apparent magnitude) GALEX NUV versus NUV-r color-magnitude diagram. These bands were chosen as they have broadly similar effective wavelengths to the UV and optical bands used for our candidate selection, but are more widely available for cross matching with catalogs of known objects. The ZAMS plotted in the right panel of Figure~\ref{fig:otherblue} is constructed from MIST models at LMC metallicity using GALEX NUV and SDSS r-band photometry distributed with the models. We emphasize that the specific comparison objects shown in the right panel of Figure~\ref{fig:otherblue} come from all-sky catalogs and are therefore not in the direction of the LMC/SMC. They are included to illustrate the typical colors and magnitudes exhibited by these classes of objects. 

\subsubsection{Foreground White Dwarf and Subdwarf Stars}\label{sec:foreground}
Two classes of stars that are relatively hot ($\gtrsim$10,000 K) are white dwarf (WD) and hot subdwarf (sd) stars. When located in regions of the sky not heavily impacted by reddening these stars could, in principle, appear bluewards of the B-type \gls{MS} and contaminate our sample from within the Milky Way.

In the right panel of Figure~\ref{fig:otherblue}, we show the WD cooling tracks from \citet{Bedard2020},\footnote{ \url{https://www.astro.umontreal.ca/~bergeron/CoolingModels/}} scaled to a distance of 1 kpc. These tracks indicate that at relatively small distances, young WDs near the beginning of their cooling sequences can exhibit apparent magnitudes and colors comparable to those in our sample. However, by $\lesssim$4 kpc, the tracks predict that all WDs would fall below the UV magnitude limits probed by our catalog.
In order to place them on the CMD in Figure~\ref{fig:otherblue}, we approximated the white dwarfs as blackbodies (with the temperature and radii defined by the cooling tracks) and compute synthetic GALEX NUV and SDSS r-band photometry. 
While \emph{Gaia} precision depends both on source magnitude and location, we find that the median parallax error for objects in our candidate catalog is 0.26 mas. Thus, with some exceptions, we would expect most foreground WDs within our sample to have parallaxes detected at $\gtrsim$4$\sigma$ if \emph{Gaia} astrometry is available. 

Hot subdwarfs are low mass stars that have been stripped of their hydrogen-rich envelopes via interaction with a binary companion \citep[for a review, see][]{Heber2016}. As the name suggests, these objects can be hot, with sdO and sdB referring to subdwarfs with spectral types O and B, respectively.  We first note that the distinction between hot subdwarfs and intermediate mass stripped stars is not sharp, and many of these systems may be formed by the same mechanism \citep[e.g.][]{Gotberg2018}. \citetalias{Drout2023} broadly defined ``intermediate mass'' stripped stars as those with masses of $\sim$2--8 M$_\odot$ (roughly corresponding to the mass range that may subsequently explode and produce neutron stars \citealt{Tauris2015}). However, from Figure~\ref{fig:cmd} it is clear that our photometric sample would be sensitive to some stripped stars with masses between 1 and 2 \Msun{} even in the Magellanic Clouds. 
Such low-mass helium stars have been considered subdwarfs, although they are more massive than the average subdwarf \citep[cf.][]{mourard_spectral_2015,mereghetti_ultramassive_2009}.

That being said, most known subdwarfs in the Milky Way are even lower-mass systems ($\sim$0.5 M$_\odot$, e.g.\ \citealt{Fontaine2012,Schaffenroth2022,Lei2023}).  These objects typically have absolute magnitudes between 2--6 mag in the optical \citep[e.g.][]{Eisenstein2006, Geier2020}, and could contaminate our sample if they are located in the halo at distances of $\sim$2-20 kpc.
This is demonstrated in the right panel of Figure~\ref{fig:otherblue}, where we plot the catalog of hot subdwarf stars from \citet{Geier2020}, which was identified using data from \emph{Gaia} DR2. Significant overlap is observed with the region of color-magnitude space where our sample is found. In particular, we note that while \citetalias{Drout2023} disfavored ``typical'' foreground sdO and sdB stars as a possible origin for their spectroscopic sample, this argument was based in part on temperature: their Class 1 objects have temperatures of $\sim$60-90~kK (\citetalias{Drout2023}, \citealt{Gotberg2023}), while most subdwarfs have effective temperatures below these values \citep[][]{Drilling2013}.
This is because the helium main sequence moves to cooler temperatures at lower luminosities. However, as we do not yet have temperature estimates for most of our photometric sample, it is possible it includes examples of such systems.
We quantify the possible level of this contamination in Section~\ref{sec:control_field} below, noting that spectroscopically, it is possible to distinguish foreground subdwarfs from intermediate-mass stripped stars, as demonstrated by the foreground subdwarf star 26 that was analyzed in \citet{Gotberg2023}.

\subsubsection{Background Quasars}\label{sec:qbackground}

Quasars, or quasi-stellar objects, are luminous active galactic nuclei. They can appear as point sources in optical images and their sky density is high enough that examples have been found along the line-of-sight to Local Group Galaxies \citep[e.g.][]{Massey2019}. While they can be easily distinguished spectroscopically due their strong emission lines \citep{Peterson1997}, quasars can exhibit a wide range of observed colors and magnitudes depending on both their type and redshift. To examine whether quasars could potentially appear bluewards of the ZAMS if located behind the Magellanic Clouds, we cross match the SDSS Quasar catalog \citep{Lyke2020} with GALEX \citep{Bianchi2017}. After correcting for a fiducial extinction of \Av{}$=0.38$ mag (mimicking the corrections applied to objects in the direction of the LMC), we plot the results in the right panel of Figure~\ref{fig:otherblue}. While a majority of quasars are found at redder colors and fainter magnitudes than our candidates, there is overlap.

Due to their cosmological distances, quasars are not expected to have detectable parallaxes or proper motions \citep[e.g.][]{Souchay2022}. Indeed, quasars have been used both to define the \emph{Gaia} celestial reference frame \citep{Klioner2022} and measure the \emph{Gaia} parallax zero-point offset \citep[e.g.][]{Groenewegen2021}. However, given that the expected proper motion for stars in the Magellanic clouds is low ($\lesssim$2 mas yr$^{-1}$) and systematic errors can be present (e.g. if some of the quasars are moderately resolved), we test whether it would be possible for a confirmed quasar to have passed the kinematic cuts applied to our observed sample. After cross matching with \emph{Gaia} DR3 \citep{GAIA2023} we find that, as expected, none have significant ($>4\sigma$) parallax values. However, we find that $\sim$10\% of the quasars bluewards of the ZAMS in Figure~\ref{fig:otherblue} have reported proper motion values that would have resulted is a low $\chi^2$ based on the analysis described in \S~\ref{sec:member} (implying consistency with expectations for stars in the Magellanic Clouds). We therefore conclude that it would, in principle, be possible for a background quasar to pass both or color/magnitude and kinematic cuts by changes. We quantify the expected level of this contamination in the next subsection.

\subsubsection{Estimated of Rate of Foreground and Background Objects}\label{sec:control_field}
To estimate the rate of both foreground and background objects that may be contained in our sample, we examine a control field located at a similar galactic latitude to, but offset from, the Magellanic Clouds. Specifically, we query \emph{Gaia} DR3 \citep{GAIA2023} in a circle with a radius of 10$\degree$ centered at the coordinate: (R.A., Dec.) $=$ (03:41:50.6, –55:06:25.0) (J2000). We restrict ourselves to sources with apparent \emph{Gaia} G-band magnitudes brighter than 21 mag, consistent with the brightness range of our candidate sample. We then obtain UV and optical photometry similar to that in our catalog by cross matching the sample to both the revised catalog of GALEX ultraviolet sources \citep{Bianchi2017} and Skymapper DR4 \citep{Onken2024}. This results in a sample of $\sim$50,000 objects with data from all three surveys. Of the three surveys, the GALEX footprint was the most limited. We estimate that it covered $\sim$80\% of the original field queried, resulting in a total control field area of $\sim$ 251 deg$^2$.

We first consider how many objects located in this control field have colors and magnitudes consistent with those in our sample. 
We apply our fiducial LMC extinction correction (A$_{\rm{V}} = 0.38$ mag) and then
identify which objects lie bluewards of the ZAMS in a NUV versus NUV-r color magnitude diagram. We use the ZAMS from the MIST models shown in Figure~\ref{fig:otherblue} at the distance of the LMC (50 kpc). 
In total, there are 644 sources in the control field that would lie bluewards of the ZAMS and have (extinction corrected) NUV apparent magnitudes brighter than 19.0 AB mag. Thus, before considering kinematics, the ratio of the area covered by the control field compared to the \emph{Swift} coverage of the Magellanic Clouds (3.6 and 10.7 deg$^{2}$, see Section~\ref{sec:SUMAC}) implies that we should only expect $\sim$9 and $\sim$27 foreground or background sources amongst our candidates in the SMC and LMC respectively.
This number drops to $\sim$1--2 objects, however, when we apply the kinematics constraints used on our candidate catalog in Section~\ref{sec:member}. In detail, the 644 blue sources in the control field, 312 have parallaxes measured at above 4$\sigma$ and hence would have been removed from our sample. Of the remaining 332 sources, 47 have proper motions that are consistent with LMC members and 81 have proper motions that are consistent with SMC members. The ratio of the area covered by the control field to that of the Magellanic Clouds is again applied leaving 
$\sim$1--2 foreground or background objects within each of the LMC and SMC that are both (i) bluewards of the ZAMS and (ii) pass the kinematics cuts applied in Section~\ref{sec:member}. Thus, while some individual sources in our candidate catalog may be foreground or background sources (especially those which lack \emph{Gaia} kinematic information) we expect overall contamination to be minimal.

\section{Summary and Conclusions}\label{sec:summary}

In this manuscript, we present the Stripped-Star Ultraviolet Magellanic Cloud Survey (SUMS) catalog---a new UV photometric catalog of over seven hundred thousand sources in the Magellanic Clouds. We use this catalog to identify a population of several hundred candidate intermediate mass stripped stars based on the UV light they contribute to the SEDs of their host systems. Here we summarize our main results.

\emph{UV Photometric Catalog:} 
In building our new UV catalog, we perform point-spread function photometry on over 2,400 archival \emph{Swift}-UVOT images of the Magellanic Clouds, primarily sourced from the \gls{SUMaC} survey \citep{Siegel2014,Hagen2017}. We develop a custom pipeline based on the \theTractor\ software \citep{Lang2016}, which forward-models the UV images using the astrometric positions of known sources from the optical \gls{MCPS} catalogs \citep{Zaritsky2002,Zaritsky2004}. We validate the pipeline through a series of tests, showing that it reproduces standard \emph{Swift} aperture photometry for a set of isolated stars. Additionally, simulations demonstrate that \theTractor\ reliably disentangles light from nearby sources in most cases, except when neighboring stars are extremely close ($\lesssim1.5\arcsec$). We combine detections of individual sources across multiple images and require a detection significance of $>3\sigma$ for inclusion. 

The final SUMS catalog contains new UV photometry for \NSources{} sources in three UV filters (UVW2, UVM2, and UVW1 covering a wavelength range of $\sim$1900-3000 \AA) as well as the optical photometry available from the MCPS catalogs. The UV magnitudes presented in the catalog range from $\sim$12 to 20 (Vega). 
A crossmatch with the SIMBAD astronomical database  indicates that fewer than 5\% of sources have been previously studied in detail, yet the catalog includes a diverse set of objects that offer rich opportunities for exploration.
To support such an effort, we make the SUMS UV catalog, candidate sample, and associated pipeline
publicly available on Github,\footnote{\url{https://github.com/AstroLudwig/SUMS_UVPhotometricCatalog}} with version 1.0 permanently archived on Zenodo (\!\dataset[DOI: 10.5281/zenodo.17551743]{https://doi.org/10.5281/zenodo.17551743}; \citealt{sums_github}). Appendix~\ref{sec:caveat} outlines important caveats and limitations to be aware of for best-use of this dataset. 
 
\emph{Stripped Star Candidate Identification:}
Our primary scientific motivation for performing new UV photometry is to search for massive stars that have been stripped of their hydrogen-rich envelopes by binary companions. These stripped stars are expected to be hot, and a subset of them should exhibit a detectable UV excess relative to \gls{MS} stars of similar luminosities \citep{Gotberg2018}. To identify candidate systems, we first select sources that lie blueward of the \gls{ZAMS} in multiple UV–optical \glspl{CMD}, after correcting for distance and extinction. We then apply a series of quality cuts to ensure that the \glspl{SED} of selected objects are broadly consistent with expectations for stripped star binaries and to exclude sources that appear anomalously blue due to poor photometry or other systematic issues. Finally, we assess their membership in the Magellanic Clouds using Gaia proper motions and parallaxes.

In total, we identify \NFinalLMC{} candidates in the Large Magellanic Cloud and \NFinalSMC{} in the Small Magellanic Cloud. We further subdivide these candidates into four categories based on the strength of the detected UV excess and the overall data quality. Overall, the dominant uncertainty in this selection process is extinction correction. We examine its impact on our candidate sample both in terms of potentially missing true stripped star systems (due to under-correction) and misclassifying MS stars as candidates (due to over-correction).

A preliminary version of this candidate selection method was presented in \citetalias{Drout2023}, where we spectroscopically confirmed a subset of candidates. That sample represents the first identified population of observed intermediate mass stripped stars and demonstrates the overall efficacy of our approach. Of the 25 systems reported in \citetalias{Drout2023}, 18 are included in the final candidate list presented here.
These 18 stars contain all but one of the stars from the original sample that showed clear evidence of high temperature through the presence of HeII spectral line absorption. The remaining seven are excluded from the final sample due to the adoption of more conservative selection criteria in this work, including stricter thresholds on UV excess significance and the exclusion of sources with potentially contaminating neighbors in the UV imaging.

\emph{Candidate Properties:} 
Our candidate systems have optical magnitudes similar to B-type \gls{MS} stars in the Magellanic Clouds (16 mag $<$ m$_{\rm V}$ $<$ 21 mag), but exhibit UV magnitudes up to 2 magnitudes brighter. Roughly half of our candidates lie within 0.4 mag of the ZAMS in UV–optical CMDs, while the remainder show a more pronounced UV excess. Notably, over 70\% have V-band magnitudes fainter than 19 mag, which may have precluded their earlier discovery. 

The candidates display a range of SED shapes consistent with expectations for stripped star + MS star binaries, in which the flux contribution from the stripped star varies according to what companion it is paired with.

By comparing the observed SEDs to composite models of stripped and MS stars, we estimate the stellar mass range probed by our selection. The sample is broadly consistent with systems containing stripped stars of $\sim$1–7~$\Msun$ paired with MS companions of $\lesssim$5~$\Msun$. Remarkably, over 50 sources are consistent (within 1$\sigma$) with models that assume no detectable contribution from a luminous companion, suggesting some may be stripped stars with faint or compact companions.

Furthermore, after applying order-of-magnitude corrections for undetected systems---accounting for the absence of a UV excess, extinction, and crowding---we find that the number of candidate systems is broadly consistent with predictions from binary population synthesis models.

\emph{Other Possible Origins:} 
Finally, we consider other types of objects that may occupy similar regions of UV–optical parameter space and could be misidentified as stripped star candidates. While some contamination may arise from foreground or background sources not physically associated with the Magellanic Clouds, or from rare objects rapidly evolving through this space (e.g., post-Asymptotic Giant Branch stars), we show that such cases are expected to be uncommon. Alternatively, some objects may include other binary products, such as hot subdwarfs, Wolf–Rayet stars, accreting white dwarfs, or central stars of planetary nebulae. These systems are scientifically valuable in their own right and warrant future investigation.

The most probable source of contamination, however, is early-type \gls{MS} stars located in regions of very low reddening. We demonstrate that this should only impact the $\lesssim$50\% of our candidate sample that lies closer to the ZAMS and that 13\% of our candidates show UV excess regardless of extinction. To minimize such contamination, we adopt a conservative baseline for extinction correction and impose stringent significance thresholds on the UV excess required for inclusion. A more refined treatment of line-of-sight extinction for individual objects may therefore uncover additional stripped stars currently excluded from our sample.

{The newly identified candidate systems presented here represent a transformative shift in our knowledge of potential hot intermediate mass stripped stars--well beyond the few dozen objects we presented in \citetalias{Drout2023} and \citet{Gotberg2023}. 
The required spectroscopic follow-up is already underway, and we encourage the community to join in observing these targets. This work significantly advances our understanding of stripped stars, suggesting that they may form a larger population and, consequently, exert a substantial impact on the evolution of stellar populations and galaxies. Finally, our results lay a clear foundation for future studies to unravel the complexities of binary interactions and to establish the broader role of stripped stars across the Magellanic Clouds and beyond.}

\section*{Acknowledgements} \label{sec:ack}
We thank Anna O'Grady, Christopher D. Matzner, Joshua S. Speagle, Carles Badenes, Marten van Kerkwijk, Selma de Mink, Hugues Sana, Kareem El-Badry, and Lisa Blomberg for helpful conversations.
{We would also like to thank the anonymous referee for their thoughtful and constructive comments, which helped to improve the clarity and depth of this work.}

BL has received financial support from
the Flemish Government under the long-term structural Methusalem funding program by means of the
project SOUL: Stellar evolution in full glory, grant
METH/24/012 at KU Leuven.
M.R.D. acknowledges support from the NSERC through grant RGPIN-2019-06186, the Canada Research Chairs Program, and the Dunlap Institute at the University of Toronto.  A.L.
acknowledges support from the NSERC and is
funded through a NSERC Canada Graduate Scholarship—Doctoral. A.L. is also supported by the Data Sciences Institute at the University of Toronto through grant number DSI- DSFY3R1P02.

\facilities{We acknowledge the use of public data from the NASA Neil Gehrels Swift Observatory Mission data archive.}

\software{\texttt{astropy} \citep{astropy:2013, astropy:2018, astropy:2022}, \texttt{astroquery} \citep{astroquery:2019}, \texttt{matplotlib} \citep{Hunter:2007}, \texttt{numpy} \citep{harris2020array}, \texttt{pandas} \citep{mckinney-proc-scipy-2010,reback2020pandas}, \texttt{Photutils} \citep{larry_bradley_2024_12585239,Bradley2022}, \texttt{SciPy} \citep{2020SciPy-NMeth}, \texttt{The Tractor} \citep{Lang2016}}, \texttt{TOPCAT}\citep{topcat1,topcat2}.

\appendix

\vspace{-1.cm}
\section{Summary of Caveats and Limitations of the UV Source Catalog}\label{sec:caveat}

While our primary goal in performing photometry on the SUMaC UV images of the Magellanic Clouds was to identify candidate stripped star binaries, the resulting source catalog will have a broad number of use cases and we provide it for the community. However, there are a number of caveats and limitations to be aware of when using this catalog. Here, we summarize a number of these issues, as well as flags that are included in the final catalog to help assess and mitigate their effects.

\textbf{\emph{Incomplete Source Coverage:}} This UV catalog is not complete, even for sources above the detection threshold. First, as described in \S \ref{sec:inputimages} we removed sources within 12" of stars bright enough to cause aberrations that would corrupt photometric estimates in their local environment. This bright source masking impacts $<$1\% of the total coverage for each galaxy. 

Second, we are likely less complete in densely clustered regions, and regions close to bright stars (which are none-the-less below the threshold for masking described above). In \S \ref{sec:qualitycuts} we describe how we remove sources with high residual flux fractions. In practice, this means we are likely less sensitive to faint sources in denser regions/near bright stars---as even small imperfections in the PSF shape can lead to a larger residual fraction for faint sources if there are sources nearby that are significantly brighter. For example, \cite{OGrady2024} used the UV catalog presented here to search for Yellow Supergiants (YSGs) in the Magellanic Clouds with evidence for excess UV light. After cross matching, they found that $\sim$15\% of their YSGs located within the SUMaC footprint were not found in the final UV catalog presented here. Of these, $\sim$27\% were excluded because they had residual flux fractions above our threshold of 0.3 (while other sources were absent either because they fell in one of the masked regions described above or because they were too faint in the UV to yield a 3$\sigma$ detection). Upon visual inspection of the images, \cite{OGrady2024} found that these sources were either in dense regions or that there was a bright object $\lesssim$6$\arcsec$ from the object of interest. Overall, we find that 86,569 sources were completely removed during our analysis (equivalent to $\sim$11\% of the number of sources in the final catalog) because they exhibited high residual fractions in all images (see \S~\ref{sec:qualitycuts}). 
Future work (Blomberg et al., in prep) will further quantify this completeness as a function of stellar density by injecting sources into the UVOT images themselves. 
Finally, we highlight that for a source from MCPS to be included in the input list for \theTractor{} to model, we require that it is brighter that 20.5 mag in either the U- or B-band \emph{and} that aperture photometry within a 5$\arcsec$ radius surrounding the source yields a count rate that is at least 1.5 times higher than the background model in the same region (see \S~\ref{InitGuess}). In practice this means that it is possible that some sources for which there is flux in the UVOT images do not appear in the final catalog because they were not modeled. In addition to a catalog of UV magnitudes, we also provide the pipeline we used to measure them, for any user who would like to examine the pipeline outputs for particular targets of interest.

\textbf{\emph{Flux Degeneracy for Close Sources:}}  As described in \S~\ref{ref:crowding}, degeneracies can occur between the count rates that \theTractor{} assigns to sources that are physically close together. In Fig \ref{fig:cluster} we illustrate for a few combinations of source magnitudes where \theTractor{} results begins to diverge from the true values as the sources are moved progressively closer together. In these examples, the divergence begins to appear once stars are closer than $\sim$2--2.5$\arcsec$ (i.e. within the FWHM of the \emph{Swift-}UVOT PSF) and is more extreme for the fainter object in the pair. For sources with roughly equal magnitudes, the maximum divergence in these examples $\sim$1.5$\sigma$ (i.e. the count rate returned by \theTractor{} is discrepant by 1.5$\sigma$ from that which was injected) while for a source that is 1.5 mag fainter than its close neighbor this increase to 2.5$\sigma$ once the sources are only $\sim$1$\arcsec$ apart. However, as described in \S~\ref{sec:morequalitycuts} there are examples of sources in our catalog that have close neighbors with more extreme magnitude differences that explicitly tested in analysis in \S~\ref{ref:crowding} (e.g 4--5 magnitudes different). As noted above, future work (Blomberg et al., in prep) will further quantify the performance of \theTractor{} on UVOT images in more complex stellar environments.
In the published catalog, we include the fraction of the flux within a 5$\arcsec$ radius attributed to a given source, the distance to the nearest neighbor, and the number of neighbors within 2.5 and 5$\arcsec$ in order to help identify when this may be problematic. Overall, we find that roughly 10\% of the stars in our final catalog have a nearest neighbor within 2$\arcsec$.

\textbf{\emph{PSF Shape Variations:}} We use a constant theoretical model of the PSF described in \S~\ref{sec:psf}, however, there are two regimes that may be negatively impacted by this. First, for bright stars coincidence narrows the PSF which could impact the magnitudes we get for those stars \citep{Breeveld2010}. Secondly, although we remove images with clear tracking issues in \S~\ref{sec:inputimages}, some images may still vary from the ideal case represented by the model PSF. We take a number of steps to account for this. In \S~\ref{sec:hea}, we add a 5\% systematic error to the measured count rates computed with \theTractor{}. In \S~\ref{sec:isolatedtest}, we compare standard \textit{Swift} routines with \theTractor{} measurements in less crowded regions, finding that for the three filters, the measurements from both methods are within the respective errors. This is also explored in \S~\ref{sec:residual} where we establish the `residual fraction' metric as an indicator of how the flux of a source could have been underestimated, therefore suggesting that the fit may not have been as robust. Figure~\ref{fig:resid} in particular shows a trend that the PSF may not be well modeled for images with low exposure times. This metric is included in the final catalog as a way to assess confidence in a source.

\textbf{\emph{UV-Optical Source Mismatches:}} We calculate UV magnitudes by performing forced-photometry at the location of MCPS sources. Hence, our final catalog contains both UV photometry from the SUMaC images, and the optical MCPS photometry for the linked source. However, in some cases, the MCPS source and the \textit{Swift}-UVOT source may be mismatched (i.e. they do not correspond to the same astronomical source). First, the MCPS catalog is not complete at bright magnitudes, which may result in a bright optical source being matched to a nearby source in the UV image. This would become apparent in the SED when a large discrepancy between the UV and optical magnitudes occurs (as described in \S~\ref{sec:SEDquality}).  Secondly, when fitting the position of a source, it is possible that \theTractor{} fits a different nearby source. To account for this, we constrain how far the source moved to within 1$\arcsec$ of the optical source. However, in densely clustered regions this could still result in a mismatch which would similarly result in an SED discontinuity.

{\textbf{\emph{Apparent Magnitude System \& Dust Extinction:}} The full SUMS UV catalog is provided in Vega magnitudes to be consistent with the original MCPS photometry, whereas the candidate catalog is given in AB magnitudes. In both cases the photometry has not been corrected for extinction. The Vega to AB conversions and dust extinction values we use in this manuscript are provided in Table~\ref{tab:conversions}, however, more detailed or alternative reddening corrections may be desired based on use case. Details of our reddening correction can be found in \S ~\ref{sec:extinction}.}

%\vspace{-0.1cm}
\section{Description of $\chi^2$ method for identifying likely foreground stars with \emph{Gaia}}\label{ap:gaia-chi2}

In Sections~\ref{sec:diverse} and \ref{sec:foreground}, we perform a $\chi^2$ analysis to compare the proper motions in right ascension and declination (designated $\mu_{\alpha}$ and $\mu_{\delta}$, respectively) for stars in our UV catalog to expectations for stars in the LMC/SMC. We follow the same procedure described in \citetalias[][]{Drout2023}, which itself was based on the methods of \citet{Gaia2018}. To describe the distribution of proper motions expected for stars in the LMC/SMC, we use the large samples of ``likely'' LMC/SMC members that were selected in \citetalias{Drout2023}. These samples consist of $\sim$1,000,000 sources in each galaxy. They were selected after rejecting stars that (i) have detected parallax (ii) lie in the yellow portion of the \emph{Gaia} CMD, which is known to have higher rates of foreground contamination, (iii) have proper motions outside 3 standard deviations from the mean and (iv) have non-zero excess noise reported by \emph{Gaia}. See \citetalias{Drout2023} for additional details. These likely members are used to construct a covariance matrix (C) based on their proper motions $\overrightarrow{\mu}$ = ($\mu_{\alpha}$, $\mu_{\delta}$). We then take measurement uncertainties into account by following the procedure described in \citet{Ogrady2023}. Specifically, we minimize the total negative log-likelihood for a set of matrices, each weighted by the measurement uncertainties of a single likely LMC/SMC member. This yields an ``optimal'' covariance matrix, C$_{*}$, which is used in the rest of our analysis. This process is done for each galaxy (LMC and SMC) separately.

We then compare the proper motions, $\overrightarrow{\mu}$, for each of our candidate stars to the distribution inferred for the galaxy that it overlaps with by computing the $\chi^2$ statistic: $\chi^{2} =$ $( \overrightarrow{\mu}-\overrightarrow{X} )^{T}\mathrm{C}_{*}^{-1}(\overrightarrow{\mu}-\overrightarrow{X})$, where $\overrightarrow{X}$ is the median proper motion of LMC or SMC (depending on the source). We consider a source to be a likely foreground star, and remove it from our candidate list, if its proper motions fall outside the region that contains 99.5\% of the ``likely'' LMC/SMC members that form our comparison samples. This corresponds to stars with $\chi^2 > 10.6$. 

\clearpage
\section{Supplementary Tables}~\label{ap:tables}
Our full UV catalog (\S~\ref{sec:catalog}) is in Vega magnitudes and our stripped star candidate catalog (\S~\ref{sec:Candidates}) is in AB magnitudes. Conversion between the two magnitude systems is described in Section~\ref{sec:Candidates}. In both cases we provide photometry in which we have not corrected for extinction. Our approach to dereddening the magnitudes for the purposes of this manuscript and our particular science case is discussed in Section~\ref{sec:extinction}. A summary of these conversions and corrections are provided in Table~\ref{tab:conversions}.

%\clearpage

\setcounter{table}{0}
\renewcommand{\thetable}{A\arabic{table}}

\vspace{-0.1cm}
\begin{table}[ht!]
\centering
\caption{Wavelength dependent conversions from Vega to AB magnitudes (\S\ref{sec:Candidates}) and extinction corrections (\S\ref{sec:extinction}).}
\label{tab:conversions}
\begin{tabular}{l@{\hskip 0.1in}|
                c@{\hskip 0.2in}
                c@{\hskip 0.2in}
                c}

\toprule\toprule

Filter (X) & $A_{X}/A_{\nu}$$^1$ (LMC)  & $A_{X}/A_{\nu}$$^1$ (SMC) & Vega to AB$^2$\\
& [mag] & [mag] & [mag] \\
\midrule
UVW2 & 2.645 & 3.523 & 1.705 \\
UVM2 & 2.723 & 3.063 & 1.659 \\
UVW1 & 2.354 & 2.744 & 1.489 \\
U & 1.563 & 1.747 & 0.761 \\
B & 1.317 & 1.407 & -0.110 \\
V & 1.033 & 1.035 & -0.001 \\
I & 0.737 & 0.591 & 0.491 \\
\bottomrule
\multicolumn{4}{l}{\footnotesize $^1$ $\textrm{Dered Mag = Mag}- A_\nu \frac{A_{X}}{A_\nu}$ where our assumed values for $A_\nu$ are 0.38 } \\
\multicolumn{4}{l}{\footnotesize \hspace{0.08in}and 0.22 mags for the LMC and SMC respectively.} \\
\multicolumn{4}{l}{\footnotesize $^2$$\textrm{Mag}_{AB} = \textrm{Mag}_{Vega} +$ Vega to AB conversion factor.} \\

\end{tabular} \end{table}

\vspace{-1.cm}
\section{Supplementary Figures}~\label{ap:figures}
\begin{figure*}[ht!]
    \centering
\includegraphics[width=0.4\textwidth]{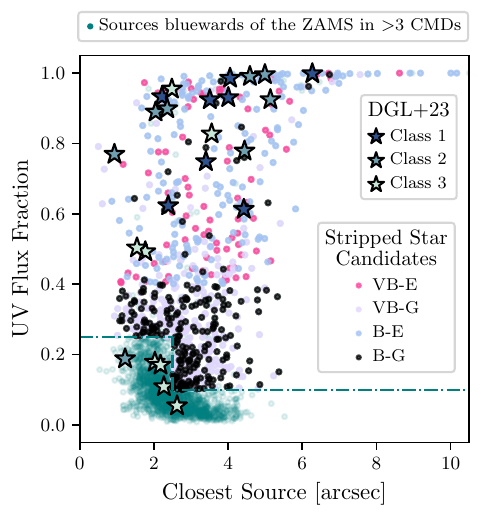}
    \caption{The average UV flux fraction versus the distance to the closest detected source in a SUMaC image for the \NAdditionalblueincmds{} sources that pass all of the SED quality cuts described in \S~\ref{sec:SEDquality} and are bluewards of the ZAMS in a at least 4 out of the 9 UV-optical CMDs examined in this manuscript (see \S~\ref{sec:bluewards} and \S~\ref{sec:morequalitycuts}). When examining these source, we found that many showed low flux fractions, typically due to a bright source near the target of interest. When selecting our stripped star candidates, we therefore impose the additional requirement that a star contribute a minimum of 10\% of the UV flux within a 5$\arcsec$ radius and that it contribute a minimum of 25\% if there is another star within 2.5$\arcsec$. These additional criteria are shown as the {teal} dashed-dot line. After imposing these cuts \NAdditionalRemain{} candidate systems (shown {colored by their rank}) remain. For comparison, we also show the location of the 25 stripped star candidates from \citetalias{Drout2023}, color coded by their spectral class: Class 1 (which appear to be dominated by a stripped star), Class 2 (which show signatures of both a hot stripped star and a B-type MS companion) and Class 3 (which are dominated by a B-type MS star). Most of the systems presented in \citetalias{Drout2023} have high UV flux fractions, indicating that they are the dominant source in their local environment. However, six of their sources (five Class 3 and one Class 2 system) do not pass the updated criteria we impose in this manuscript when selecting candidates. See \S~\ref{sec:compD23} for further details.  }
    \label{fig:ffrac_closest}
\end{figure*}

In Section~\ref{sec:Candidates} we detail the stripped star candidate selection process which includes both photometric cuts as well as cuts based on models describing the SEDs of these stars. In Figure~\ref{fig:ffrac_closest} we show one particular photometric cut, described in Section~\ref{sec:morequalitycuts}, that addresses possible contamination from nearby sources by considering both the fraction of flux within a 5$\arcsec$ radius attributed to a candidate as well as the distance to the closest neighbor. This particular cut was not included in an earlier version of our pipeline from which the \citetalias{Drout2023} sample was identified, and which is shown on the figure and discussed further in Section~\ref{sec:compD23}.

%\vspace{-1.cm}

\pagebreak
\bibliographystyle{aasjournal}
\bibliography{new.ms.bib}

\end{document}